\documentclass{jfm}
\usepackage{graphicx}
\usepackage{epstopdf, epsfig}
\usepackage{color}
\usepackage{amsmath}
\usepackage{soul}
\usepackage{comment}

\shorttitle{LagSAT for Fluid Flows}
\shortauthor{V. J. Shinde}

\title{Lagrangian Stability Analysis Technique for Fluid Flows}

\author{Vilas J. Shinde
\corresp{\email{vshinde@ae.msstate.edu}}}
\affiliation{Department of Aerospace Engineering, \\ Mississippi State University, Mississippi State, MS 39762, USA}

\begin{document}

\maketitle

\begin{abstract}
Flow transition from a stable to unstable states and eventually to turbulence is a classical fluid mechanics phenomenon with a strong practical relevance.
Conventional hydrodynamic stability analyses deal with perturbation dynamics on a steady baseflow, typically in Eulerian reference frame.
Common modal techniques, such as the linear stability theory, involve linearization of the governing flow equations in conjunction with the flow/geometrical simplifications, revealing the large time asymptotic behavior of flow perturbations in the system.
In contrast, the non-modal nonlinear stability analyses focus on a short term transient behaviour of the perturbation dynamics.
This paper presents a perturbation-free data-driven stability analysis technique by employing select elements of Lagrangian modal analysis~\citep{shinde2021lagrangian}, particularly its adjoint form in the flow direction.
The proposed Lagrangian stability analysis technique (LagSAT) is inherently spatio-temporal, in the Lagrangian sense, and forms a natural connection with the Koopman operator theory.
LagSAT is demonstrated on classical baseflows that exhibit convective/absolute instabilities, namely, the self-similar Blasius/Falkner-Skan boundary layers, a 2D compressible flow past a cylinder, and a 2D compressible lid-driven cavity flow, producing neutral stability curves, N-factor estimate, and transient energy growth.
LagSAT is innately suitable for the global stability analysis of large/complex numerical/experimental baseflows with multiphysics effects.
\end{abstract}


\section{Introduction} \label{sec:intro}

Flow transition to turbulence can profoundly affect the design and operation of air/water/space-craft.
The transition effects include flow induced unsteadiness, increased level of viscous drag, higher heat transfer, aerodynamic heating, among others.
In flow transition, an orderly/laminar (stable) flow becomes fully chaotic/turbulent (unstable) for increasing value of a critical parameter.
The early seminal work by~\cite{reynolds1883xxix} identified a non-dimensional parameter, well known as Reynolds number.
The flow remains stable for Reynolds number less than a critical value, and becomes unstable for higher values.
However, the actual process of transition is fairly intricate and comprise various stages including the receptivity of external disturbance, the growth of disturbance, and a breakdown to turbulence~\citep{reshotko1976boundary,morkovin1994transition,schneider1999flight,saric2002boundary,saric2003stability,fedorov2011transition}.
The receptivity stage generally concerns the perturbation environment and its selective acceptance by the flow~\citep{choudhari1993boundary,choudhari1996boundary}, whereas the second stage involves perturbation growth following various routes to the final stage, i.e, the breakdown to turbulence.

Flow stability deals with a stable/steady flow state - a baseflow - subjected to external perturbation, where the flow response may amplify linearly or nonlinearly via various mechanisms to cause the onset and progress of flow transition to turbulence.
The baseflow is generally a solution to the flow governing Navier-Stokes equations.
However, the choice of steady baseflow is not straightforward, particularly for flows that exhibit oscillatory nature of instability; for example, the classical case of flow past a cylinder, which results in inaccurate frequency of unstable modes~\citep{juniper2014modal}.
This issue is generally addressed by adopting time-averaging of the flowfields and selective damping of the most unstable modes~\citep{sipp2007global,aakervik2006steady}.
A detailed discussion on the validity of various baseflows for stability analysis is provided in~\citet{beneddine2016conditions}.
A typical stability ansatz comprises a baseflow along with a set of perturbation governing equations, leading to the stability characteristics of the baseflow.
The perturbation dynamics with appropriate simplifications brings about the modal/non-modal stability of the steady baseflow~\citep{tumin2001spatial,schmid2007nonmodal}.

Broadly, the flow stability mechanisms can be categorized into convective and absolute instabilities - the concepts first used in plasma physics~\citep{briggs1964electron} and later introduced to fluid dynamics~\citep{huerre1990local}.
In convective instability, the perturbation grows and sweeps away with the flow, whereas in the absolute instability it grows at the source as well.
A convectively unstable flow is also called as noise amplifier, as it convects and amplifies the perturbation, in the coordinate frame of reference.
In contrast, an absolutely unstable flow amplifies the perturbation along the flow and also at the source.
The flow is referred as an oscillator, where the absolute instability leads to an oscillatory response of flow.
Furthermore, depending on whether the stability analysis is performed on the entire flow domain or on a local flow profile, the stability properties are referred as global and local, respectively.
Together these aspects lead to the concepts of local/global convective and local/global absolute instabilities~\citep{huerre1990local}.

Sophisticated flow stability theories have been developed over the years~\citep{schmid2002stability,chomaz2005global,theofilis2011global}.
The classical Linear Stability Theory (LST) employs a linearized form of the flow equations that govern the perturbation dynamics.
In its most general form, it is referred as linear global analysis, where the baseflow is inhomogeneous in all three spatial directions.
For two/one spatial homogeneous directions, the LST simplifies to the bi-global and local analyses, respectively.
The parallel flow assumption reduces the LST formulation suitable to perform local analysis, where 1D/2D baseflow profiles can be the input to the stability analysis, as opposed to the 2D/3D baseflows in a global analysis.
The parallel flow assumption allows the normal-mode solutions, where the complex-valued normal modes are conveniently utilized to examine the stable/unstable modes of the baseflow.

As a matter of course, the parallel/local flow assumptions put limits on the generality of LST in predicting non-linear dynamics or transition front in complex flow geometries.
To relax some of the constraints under LST, Parabolized Stability Equations (PSE) allows a mild variation in the baseflow along the flow direction~\citep{herbert1997parabolized,bertolotti1992linear}. 
Although PSE builds on the LST ansatz (i.e, the initial perturbation decays for a stable baseflow or grows for an unstable baseflow), it encompasses weakly non-parallel and also the non-linear effects to some degree~\citep{esfahanian2001linear}.
Nonetheless, PSE can not capture strong non-parallel effects or streamwise gradients, which tend to occur in complex flow geometries.

To address the need to treat strongly non-parallel flows, several stability analysis frameworks have been proposed, including the Harmonic Linearized Navier-Stokes Equations~\citep{choudhari1994theoretical,streett1998direct,paredes2024harmonic}, global Input-Output analysis~\citep{jovanovic2005componentwise}, and Resolvent analysis~\citep{mckeon2010critical}.
Generally, the normal mode analyses referenced earlier in the context of local stability analyses utilize the baseflow governing equations in a linearized form before solving for an eigenvalue problem, where the numerical implementations must deal with their specific theoretical and numerical constructs.
For instance, the LST/PSE formulations in incompressible, compressible, or hypersonic flows with non-equilibrium effects are different and require a specific set of equations and treatments that govern the perturbation dynamics~\citep{mack1984boundary}.
The stability analysis techniques, such as the LST or the local analyses that consider normality of modes, fall short in dealing with the nonlinear effects as well as the short-time behaviour of the perturbation dynamics, including the transient energy amplification~\citep{schmid2007nonmodal,sipp2010dynamics}.
The optimally time-dependent framework developed by~\cite{babaee2016minimization} addresses the requirements of transient global LST for time-varying baseflows by operating in the linear tangent space instantaneously, following the time-dependent linear dynamics in non-autonomous flows where the classical stability analysis remains limited~\citep{kern2024onset}.

Majority of the prior work in flow stability has been performed in the Eulerian reference frame.
The Lagrangian description of fluid flow provides an alternative viewpoint that can address some of the difficulties in the conventional (Eulerian) approaches~\citep{friedlander2003localized}.
Even so, the Lagrangian description of fluid flow is relatively less popular, presumably due to the mathematical difficulty of dealing with the viscous terms of the Navier-Stokes equations~\citep{yakubovich2001matrix}.
An early work by~\cite{pierson1962perturbation} provides some context with regards to the Eulerian/Lagrangian approaches of linearization and solution of the Navier-Stokes equations.
From the Lagrangian flow stability standpoint, an Eulerian-equivalent Lagrangian perturbation theory has been used generally in astro/geo-physics branches of science~\citep{ehlers1997newtonian}, albeit considering the simplified flow/perturbation governing equations as compared to the full Navier-Stokes equations~\citep{fukumoto2010lagrangian,nadkarni2013modelling}.

Clearly, the state of the art flow stability frameworks have been beset by the complexities of the flow geometries and the governing equations for small amplitude perturbations in both Eulerian and Lagrangian reference frames.
This work presents a radically different approach of performing flow stability in a Lagrangian frame of reference.
The proposed LagSAT (\ul{Lag}rangian \ul{S}tability \ul{A}nalysis \ul{T}echnique) fundamentally differs from the conventional flow stability approaches in terms of the frame of reference and the absence of external perturbations.
LagSAT operates in the Lagrangian frame of reference, seeking stability features of the Eulerian baseflow that is transformed into a Lagrangian flow map.
The ansatz elicits stability features that are embedded in the Lagrangian flow map in lieu of subjecting the baseflow to external perturbation.
The Lagrangian flow maps are known to comprise various stability features including fixed points, periodic orbits, stable and unstable manifolds, and chaotic attractors.~\citep{ottino1989kinematics,wiggins2005dynamical,lekien2007lagrangian,shinde2021lagrangian}
These stability features can be unveiled from the spatio-temporal form of the Lagrangian flow map by employing Lagrangian Modal Analysis~\citep{shinde2021lagrangian}.
Among the various modal decompositions and forward/adjoint methods presented in~\citep{shinde2021lagrangian}, LagSAT conveniently adopts an adjoint Lagrangian Dynamic Mode Decomposition (LDMD) and employs it in the forward flow direction, {\it i.e.}, forward in time.

The theoretical development of LagSAT differs from LMA in terms of the treatment of Lagrangian flow map, where the latter extracts modal information that pertains to a particular modal decomposition technique, which need not be relevant to stability.
On the other hand, LagSAT is concerned about the stability of Lagrangian flow map, and thereby the Eulerian baseflow that is under consideration.
LagSAT is intrinsically connected to the Koopman operator theory~\citep{koopman1931hamiltonian,mezic2005spectral,williams2015data}, where the Lagrangian flow map is a dynamical system that comprises spacetime evolving Lagrangian trajectories.
The Koopman operator theory is known to provide the stability of a dynamical system in terms of its spectral properties (eigenvalues and eigenvectors) and eigenfunctions.
The connection between the Koopman operator and dynamic mode decomposition (DMD) has been well established in the literature~\citep{rowley2009spectral,brunton2021modern,budivsic2012applied} in that the DMD modes and eigenvalues are a linear approximation of the Koopman modes and eigenvalues, respectively.
In LagSAT framework, the spectral properties of the adjoint LDMD (in the forward time direction) are utilized to investigate the system stability, analogous to the Koopman operator ansatz.

The notion of convective and absolute instabilities inherently aligns with the formulation of LagSAT.
Furthermore, the local/global and spatio-temporal aspects of flow stability can be conveniently explored with LagSAT on complex flow geometries.
To assess these attributes, we consider three baseflows, namely, the self-similar Blasius and Falkner-Skan boundary layers, a 2D compressible flow past a cylinder, and a 2D compressible lid-driven cavity.
These baseflows are known to exhibit, respectively, only convective instability, both the convective/absolute instabilities, and only absolute instability.
Analogous to the local and global analysis in LST, LagSAT adopts the local/global and spatial/temporal forms of stability analysis.
Notably, LagSAT does not require revisiting the flow governing equations for the analysis.
In this sense, LagSAT is a data driven method that can be employed directly on numerical/experimental flow fields in pre/post-critical flow regimes.

The paper is organized as follows: Sec.~\ref{sec:lagsat} presents the theoretical and mathematical construct of LagSAT.
LagSAT on the self-similar boundary layers, namely, the Blasius boundary layer and the Falkner-Skan boundary layer with pressure gradient effects is discussed in Sec.~\ref{sec:lagsat_BlaFalSka}.
Section~\ref{sec:lagsat_cyl} employs LagSAT on flow past a cylinder in 2D compressible setting, exploring various transition points and convective/absolute instabilities of the flow.
Section~\ref{sec:lagsat_ldc} examines the absolute instability of a 2D lid-driven cavity flow at Mach $M=0.5$.
Lastly, Sec.~\ref{sec:concl} provides discussion and concluding remarks on LagSAT.
The paper also includes two appendices. 
Appendix~\ref{sec:Blasius_baseflow} provides details on the modeling and simulation of the self-similar boundary layers, while Appendix~\ref{sec:Num_Sense} presents the numerical sensitivity of LagSAT results.

\section{\ul{Lag}rangian \ul{S}tability \ul{A}nalysis \ul{T}echnique - LagSAT}\label{sec:lagsat}

In this section, we recall the core elements of LST to establish connections with the elements of LagSAT in the context local/global stability analyses.
An essential first step of LagSAT is the formation of Lagrangian flow map using an Eulerian baseflow.
The Lagrangian flow map comprises a set of Lagrangian flow states that evolve in the selected time or the flow direction.
This is achieved by employing the adjoint LMA approaches presented in~\citet{shinde2021lagrangian}, constructing a Lagrangian flow map that begins at a negative time and progresses toward the identity map ($t=0$).
The second step involves subjecting the spatio-temporal Lagrangian flow map to LMA, namely, adjoint LDMD in the forward time setting.
Thus, LagSAT leads to a set of stability modes with growth/decay rates, as well as it provides the time evolution of the flow energy, in Lagrangian sense, via Lagrangian Proper Orthogonal Decomposition (LPOD).

The flow directionality construct of LagSAT is convenient to establish the notion of flow stability.
It aligns well with the aspects of convective/absolute instability as well as with the conventional Parabolized Stability Equation (PSE) formulation, which progresses along the flow direction.
The standard forward and adjoint formulations of LMA employ a time progress that moves the flow away from the current flow state, whereas reversing the time direction leads to the current flow state, which we consider for the stability analysis.
This provides two possibilities of performing LMA in the context of flow stability: 1. a forward LMA with reversed time direction, 2. a backward (adjoint) LMA with the forward time direction.
We adopt the latter approach, since it aligns with the flow direction, leading to the fate/stability at the end state, which remains at the intersection of the Lagrangian and Eulerian frames of reference.

\subsection{Flow stability in the Eulerian reference frame}\label{sec:LST}

Consider a real Euclidean vector space $\mathsf{E}$ of dimension $d$, with the inner product $\langle \pmb{x}, \pmb{x} \rangle > 0$ for non-zero $\pmb{x}$, and real norm $\| \pmb{x} \|=\sqrt{\langle \pmb{x}, \pmb{x} \rangle}$.
Here, for convenience we consider $\textsf{E}$ as a $d$-dimensional point space with a coordinate system and a frame of reference, on which the Euclidean space with $\mathbb{R}^{d=3}$ can be realized by considering an orthonormal basis.
A flow in a suitable closed domain $\mathsf{D} \subseteq \mathsf{E} = \mathbb{R}^3$ may be represented in terms of a vector field $\pmb{u}$ through the mapping
\[ \pmb{u}:\mathsf{D}\times [0,\mathsf{T}] \rightarrow \mathbb{R}^3:(\pmb{x},t)\mapsto \pmb{u}(\pmb{x},t)\text{ with } \pmb{x}\in \mathsf{D},\]
where $t$ is an instant from the total time $\mathsf{T}\subset \mathbb{R}$.
In the Eulerian description, all physical quantities (scalar, vector or tensor) are expressed at each instant and at every fixed spatial location with respect to the frame of reference.
Thus, the fixed spatial coordinates $x_i$ of vector $\pmb{x}$ and time $t$ constitute the Eulerian coordinates with respect to a fixed (Eulerian) frame of reference of $\mathsf{E}$.
The Eulerian description refers to  flow fields at an instant $t$ mapping on another time $t+dt$, where $dt$ is the time differential.
Thus, a steady flow in the Eulerian reference frame remains invariant with respect to the time change.

In LST, the solution state variables are decomposed into steady baseflow ($\pmb{s}$) and the flow disturbances ($\pmb{s}'$), where the solution state vector can be $\pmb{s}=[\pmb{u},p,\rho]=[u,v,w,p,\rho]$.
Here $u,v,w$ are the velocity components and $p, \rho$ are the pressure and the flow density, respectively.
For a known baseflow, the flow governing equations are developed for nonlinear disturbance equations, which are simplified (linearized) by eliminating the non-linear terms assuming small amplitude disturbances.
In general, the linearized governing equations of the fluctuation dynamics can be expressed as,
\begin{equation}\label{eq:LST_gov}
    \frac{\partial \pmb{s}'}{\partial t} = \pmb{A}\pmb{s}',
\end{equation}
where matrix $\pmb{A}$ is a linear operator.
It is a function of the baseflow and other flow parameters, such as the Reynolds number and Mach number.
To solve the linearized flow equations, the disturbance can be described in normal modes as
\begin{equation}\label{eq:LST_dis}
    \pmb{s}'(\pmb{x},t)=a \pmb{\phi}(\pmb{x})\exp({\omega}t),
\end{equation}
where $\pmb{s}'=[u',v',w',\rho',p']$ is the perturbation vector of flow variables over a three-dimensional space $\pmb{x}=(x,y,z) \in \mathbb{R}^{d=3}$.
$\pmb{\phi}(\pmb{x})$ is a vector of the associated shape functions~\citep{herbert1997parabolized}, while $a$ and $\omega$ are, respectively, the corresponding amplitude and angular frequency.

The stability equations are represented in a compact eigenvalue form:
\begin{equation}\label{eq:LST_eig}
    \mathcal{L}\pmb{\phi}=\{\pmb{J}-{\omega}\pmb{B}\}\pmb{\phi}=0,
\end{equation}
for homogeneous boundary conditions on the shape functions, where $\mathcal{L}$ is an operator comprising the Jacobian ($\pmb{J}$) and mass matrix ($\pmb{B}$) of the flow governing equations.
The eigenvalues ${\omega}\in \mathbb{C}$ in complex form are given as ${\omega}={\omega}_r+i{\omega}_i$, with the real (${\omega}_r$) and imaginary (${\omega}_i$) parts.
The complex eigenvectors $\pmb{\phi}$ represent the shape functions of Eq.~\ref{eq:LST_dis}.
The normal mode solution of Eq.~\ref{eq:LST_dis} can be expanded as,
\begin{equation}\label{eq:LST_dis_glob}
    \pmb{s}'(\pmb{x},t)=a\pmb{\phi}(\pmb{x})\exp({\omega}_rt)\exp(i{\omega}_it).
\end{equation}
The real part of the eigen value (${\omega}_r$) dictates the growth rate of the wave perturbation, whereas imaginary part provides the frequency of the perturbation wave.
For a positive value of the real part ${\omega}_r > 0$ the perturbation grows exponentially, indicating an unstable baseflow.
On the other hand, the real part ${\omega}_r < 0$ indicates a stable baseflow, whereas ${\omega}_r = 0$ for a neutrally stable baseflow.
The imaginary part of the eigenvalue (${\omega}_i$) provides the information about the modal frequency of the perturbation wave.

The formulation of LST where the baseflow is inhomogeneous in all/more-than-one directions (involving the flow direction) is known as global stability analysis.
From a practical viewpoint, the global stability analysis involves a huge number of degrees of freedom.
Thus, solving a generalized eigenvalue problem becomes cumbersome using direct eigenvalue solvers, and generally iterative solvers are employed partially accessing the eigenspectrum~\citep{arnoldi1951principle}.
Nonetheless, the advent of computing resources and advanced algorithms~\citep{demmel2012communication,shinde2025distributed} have made the global stability analysis of fully inhomogeneous, 3D baseflows a routine~\citep{theofilis2011global}.

A computationally tractable alternative of local LST has been enabled by the parallel flow assumption.
This includes an invariant baseflow in, say, the streamwise ($x$) and spanwise ($z$) directions and a negligible baseflow velocity $v$ in $y$ direction.
For a two dimensional baseflow invariant in $z$ direction,  Eq.~\ref{eq:LST_dis} takes the following form:
\begin{equation}\label{eq:LST_dis_biglobal}
    \pmb{s}'(x,y,t)=a\pmb{\phi}(x,y)\exp(\beta z+{\omega}t),
\end{equation}
which further simplifies to
\begin{equation}\label{eq:LST_dis_parallel}
    \pmb{s}'(y,t)=a\pmb{\phi}(y)\exp({k}x+{\beta}z+{\omega}t),
\end{equation}
employing the parallel-flow assumptions.
The exponents ${k}$, ${\beta}$ are the wavenumbers in the $x$ and $z$ directions, respectively.
The normal-mode solutions provide the exponents of Eq.~\ref{eq:LST_dis_parallel}, where for spatially developing disturbances, the spatial wavenumbers (\textit{e.g.}, ${k}={k_r}+i {k_i} \in \mathbb{C}$) are complex and the frequency ${\omega} \in \mathbb{R}$ is real.
This is referred as the spatial LST.
In contrast, the temporal LST considers a complex frequency (\textit{e.g.}, ${\omega}={\omega}_r+i{\omega}_i \in \mathbb{C}$) and real wavenumbers ($k, \beta\in \mathbb{R}$) in space.
In a spatio-temporal stability analysis, both the spatial wavenumber and temporal angular frequency are complex, where the real parts dictate the growth rate of the perturbations and the imaginary part providing wavenumber/angular frequency of the spatio-temporal wave.

\subsection{Lagrangian flow map through the lens of Koopman operator} \label{sec:lfm}

Lagrangian flow map constitutes an essential step of LagSAT, where a given Eulerian baseflow is transformed into a flow map that is suitable to perform local/global stability analyses, in Lagrangian sense.
A uniform baseflow typically results in a rank deficient Lagrangian flow map, from discrete viewpoint; thus, the non-uniformity of baseflow is required for the meaningful analysis.
A Lagrangian description of fluid flow provides instantaneous flow states that are observed in a time dependent reference frame.
The flow field, $\pmb{\mathcal{U}}$, over a closed domain $\mathcal{D} \subseteq \mathsf{E} = \mathbb{R}^3$ and time interval $[-\mathsf{T},0] \subset \mathbb{R}$ may be mapped as:
\[ \pmb{\mathcal{U}}:\mathcal{D}\times [-\mathsf{T},0] \rightarrow \mathbb{R}^3:(\pmb{\chi},t)\mapsto \pmb{\mathcal{U}}(\pmb{\chi},t)\text{ with } \pmb{\chi}\in \mathcal{D}, t\in [-\mathsf{T},0].\]
The flow evolves to a reference state, mapping on a deformed geometrical configuration.
As opposed to the adjoint/backward LMA~\citep{shinde2021lagrangian}, the reference state is the end state, evolving from an arbitrary Lagrangian flow state from a past time $t=-\mathsf{T}$.
This is in accordance with the flow directionality consideration noted before (at the beginning of Sec.~\ref{sec:lagsat}).
Thus, a reference configuration, $\Omega_0\in \mathcal{D}$ at $t=t_0$ is utilized to compute an \textit{orientation and measure-preserving diffeomorphism} of $\mathcal{D}$ in the negative time direction, leading to a Lagrangian flow map that evolves from $\Omega_{-\mathsf{T}} \in \textit{SDiff}(\mathcal{D})$ at $t=-\mathsf{T}$ to the reference state at $t=0$.
Mathematically, the flow map can be expressed as
\begin{eqnarray} \label{eq:mapping}
\mathcal{M}:\mathcal{D}\times [-\mathsf{T},0] \rightarrow \textit{SDiff}(\mathcal{D})\subseteq \mathsf{E}=\mathbb{R}^3: (\pmb{\chi},t) \mapsto \mathcal{M}(\pmb{\chi},t)=(\pmb{x},t), \text{ and }  \nonumber \\
\mathcal{M}(\pmb{\chi}_0,t_0)=\text{identity map}.
\end{eqnarray}
The triple components $\chi_i$ of vector $\pmb{\chi}$ and time $t$ comprise the Lagrangian coordinates, which can be explicitly expressed using the Eulerian frame of reference as
\begin{equation} \label{eq:x=chi}
(x_i,t) = \mathcal{M}^i(\pmb{\chi},t) = \mathcal{M}^i(\chi_1,\chi_2,\chi_3,t).
\end{equation}

The Lagrangian flow mapping from an initial configuration $\Omega_0$ to a current configuration $\Omega$ must meet the regularity conditions of the transformation, mainly that it be injective and $\mathcal{M}$ be a bijection.
The inverse $\mathcal{M}^{-1}$ exists due to the regularity conditions, and by considering the existence of the inverse at any instant $t$, we can state
\begin{equation} \label{eq:M=M-1}
(\pmb{\chi},t) = \mathcal{M}^{-1}(\pmb{x},t) \Leftrightarrow (\pmb{x},t)=\mathcal{M}(\pmb{\chi},t).
\end{equation}
Consequently, the Jacobian matrix $\mathcal{J}=\partial (\pmb{\chi},t)/\partial (\pmb{x},t)$ is invertible, which plays an important role in the domain deformations.
The vector fields of the Lagrangian and Eulerian frame of references are related as
\begin{equation}\label{eq:ul-ue}
\pmb{\mathcal{U}}(\pmb{\chi},t)=\pmb{u}(\mathcal{M}(\pmb{\chi},t)) \Leftrightarrow \pmb{u}(\pmb{x},t)=\pmb{\mathcal{U}}(\mathcal{M}^{-1}(\pmb{x},t)),
\end{equation}
which also applies to each physical quantity of the flow.

The total or material derivative of a quantity, \textit{e.g.,} flow velocity vector, in the Lagrangian frame of reference is simply its partial derivative with respect to time $t$, written as
\begin{equation}\label{eq:tot_Lag}
\frac{D \pmb{\mathcal{U}}}{Dt}=\frac{\partial \pmb{\mathcal{U}}}{\partial t}\Big|_{\pmb{\chi}}.
\end{equation}
On the other hand, the Eulerian frame of reference accounts for the local and convective rates of change of a quantity.
The total derivative from Eq.~\ref{eq:ul-ue} is then:
\begin{eqnarray}
\frac{D\pmb{u}}{Dt}&=&\frac{\partial \pmb{u}}{\partial t}\Big|_{\pmb{x}} + \frac{\partial \pmb{u}}{\partial \pmb{x}}\cdot \frac{\partial \pmb{x}}{\partial t}\Big|_{\pmb{\chi}} \\
&=&\underbrace{\frac{\partial \pmb{u}}{\partial t}}_{\text{local rate of range}}+\underbrace{ (\pmb{u}_{\pmb{\chi}}\cdot\nabla) \pmb{u}}_{\text{convective rate of change}}. \label{eq:tot_Eul}
\end{eqnarray}
The Eulerian convective flow velocity $\pmb{u}_{\pmb{\chi}}$ is the mapping of the vector field from the fixed spatial coordinates $\pmb{x}$ to $\pmb{x}+d\pmb{x}$, where $d\pmb{x}$ is the differential of space.
The flow velocity vector fields can be expressed in terms of total derivative of the space vector fields in the Eulerian and Lagrangian approaches, respectively, as 
\begin{equation}\label{eq:Eul_Lag_vel}
\underbrace{\pmb{u}(\pmb{x},t)=\frac{D \pmb{x}}{Dt}}_{Eulerian} \text{ and } \underbrace{\pmb{\mathcal{U}}(\pmb{\chi},t)=\frac{\partial \pmb{\chi}}{\partial t}\Big|_{\pmb{\chi}}}_{Lagrangian}.
\end{equation}
Note that the velocity field in the Lagrangian frame of reference is always a function of time for a non-uniform flow - a key assumption in Lagrangian Modal Analysis (LMA) ansatz~\citep{shinde2021lagrangian}.

The Lagrangian flow map can be deemed as a dynamical system that evolves spatio-temporally.
The embedded Lagrangian dynamics can be described using Koopman operator theory~\citep{koopman1931hamiltonian,mezic2005spectral}.
Irrespective of the linearity (or non-linearity) of the Lagrangian dynamics, its stability can be studied by employing a linear Koopman operator~\citep{mezic2005spectral,rowley2009spectral,budivsic2012applied}, where a set of Koopman operators act on linearly independent observable functions of the system.
The Lagrangian flow dynamics for a vector flowfield $\pmb{\mathcal{S}}$ on $\pmb{\mathcal{\chi}}$ can be expressed as
\begin{equation}\label{eq:koop_dyn_sys}
     \pmb{\mathcal{S}}(\pmb{\mathcal{\chi}},t)=\pmb{\mathsf{F}}^t\left( \pmb{\mathcal{S}}(\pmb{\mathcal{\chi}},t=-\mathsf{T}) \right),
\end{equation}
where the flow map operator, $\pmb{\mathsf{F}}^t: \pmb{\chi} \rightarrow \pmb{\chi}$, maps the Lagrangian flowfield from an initial state at $t=-\mathsf{T}$ to $t$ in a time forward direction toward the final state $t=0$.
A semigroup of Koopman operators $\pmb{\mathcal{K}}^t: \mathcal{F}\rightarrow \mathcal{F}$, where $\mathcal{F}$ is a linear vector space of functions, act on vector-valued (observable) functions $\pmb{\phi}:\pmb{\chi}\rightarrow\pmb{\chi}$ on the same manifold.
It gives
\begin{equation} \label{eq:koop_obs}
    \left[\pmb{\mathcal{K}}^t\pmb{\phi}\right](\pmb{\mathcal{S}})=(\pmb{\phi}\circ\pmb{\mathsf{F}}^t)(\pmb{\mathcal{S}}),
\end{equation}
where the symbol $\circ$ indicates a sequential function operation (or a composition).
Notably, the Koopman operator mapping of the system observables (Eq.~\ref{eq:koop_obs}) is linear, regardless of the linearity of the Lagrangian flow map (Eq.~\ref{eq:koop_dyn_sys}).

The key mathematical properties of Koopman operator from the viewpoint of system dynamics, stability, and control include the linearity, and the spectral decomposition, where the latter inherits linearity from the eigenfunctions of the Koopman operator~\citep{budivsic2012applied,bevanda2021koopman,brunton2021modern}.
The Koopman operator acts on an observable function as
\begin{equation}\label{eq:koop_eig}
    \pmb{\mathcal{K}}^t\pmb{\phi}=\sum_{l=1}^\infty \pmb{v}_l\lambda_l^t\pmb{\psi}_l,
\end{equation}
where the $\pmb{\psi}_l$ are the eigenfunctions and $\pmb{v}_l$ are the Koopman operator modes. 
The eigenvalues $\lambda^t_l$ comprise the transient/oscillatory dynamics of the observable function, $\pmb{\phi}$, via the set of eigenfunctions, pertaining to the stability of the system (Lagrangian flow map).
The Koopman modes and eigenvalues can be approximated by means of the data-driven modal decomposition techniques.
For instance, \cite{rowley2009spectral} established a connection between the Koopman operator and DMD procedure~\citep{schmid2009dynamic,schmid2010dynamic}.
The DMD operates directly on the state space variables, $\pmb{\mathcal{S}}$, as opposed to the observable functions, $\pmb{\phi}$, making a linear approximation for the Koopman modes and eigenvalues.
Further, it approximates the Koopman eigenfunctions, $\pmb{\psi}$, in terms of the DMD modal amplitude at the initial condition, here $t=-\mathsf{T}$.
In the context of LagSAT, the Eulerian connections between the Koopman operator and DMD framework are extended to the Lagrangian reference frame, by means of Lagrangian modal analysis, specifically the LDMD and LPOD.
The method is described in the next sections.

\subsection{Lagrangian modal analysis} \label{sec:lma}

Modal/non-modal analysis techniques, particularly concerning the spatio-temporal system dynamics, have been developed predominantly by considering a fixed/Eulerian frame of reference.
For instance, the more popular Proper Orthogonal Decomposition (POD) and DMD typically operate in a time-independent space domain.
LMA provides a formal mathematical framework to employ these techniques in a moving/Lagrangian reference frame, which is generally useful in problems with time-dependent space domains, e.g., fluid-structure interactions (FSI).
\citet{shinde2021lagrangian} also demonstrate the application of LMA in various flow configurations, including an Eulerian steady flow, anticipating Lagrangian coherent structures pertinent to the flow stability.
Among various modal decomposition techniques, DMD closely aligns with the theory of flow stability.
For instance, for a linearized flow about a steady state (in Eulerian frame of reference), the DMD modes are equivalent to global stability modes~\citep{schmid2010dynamic}.
In the context of Lagrangian flow maps, the Lagrangian DMD can be employed to extract spatio-temporally coherent structures evolving with unique wavenumbers/frequencies and growth/decay rates.

For completeness, here we present the LDMD formulation based on~\citet{shinde2021lagrangian}.
Typically, DMD derives a mapping between suitably constructed sequences of flow states.
It then solves for the basis functions (eigenvectors) of a reduced-order representation of the mapping.
In the context of LagSAT, these flow states are extracted from the Lagrangian flow map in a suitable direction.
The application of LDMD in the negative flow direction produces adjoint (backward in time) modal features.
However, as noted before, a set of flow states evolving in the flow direction and leading up to the identity map is of particular interest from the flow stability point of view.
In this manner, we can analyze the past events on the Lagrangian flow map leading to the present state, {\it i.e.}, the identity map $\mathcal{M}(\pmb{{\chi}_0},t_0)$.

Here, we consider $\pmb{X}$ and $\pmb{Y}$ as tensors whose elements are the Lagrangian flow fields, \textit{e.g.}, the velocity vector $\pmb{\mathcal{U}}(\pmb{\chi},t)$, where time $t \in [-\mathsf{T},0]$, such that $t=\{t_n, t_{n-1},\cdots,t_2 \}$ for $\pmb{X}$ and $t=\{t_{n-1}, t_{n-2},\cdots,t_1 \}$ for $\pmb{Y}$.
Similarly, the Lagrangian space coordinate vector $\pmb{\chi}$ is considered to be discrete of size $m$, thus $\pmb{X}, \pmb{Y}\in \mathbb{R}^{m\times n}$.
The aim of the DMD procedure is to find $\pmb{A} \in \mathbb{R}^{m\times m}$ such that
\begin{equation}\label{eq:A_operator}
\pmb{A}\pmb{X} = \pmb{Y}\text{ or } \pmb{A}=\pmb{Y}\pmb{X}^+,
\end{equation}
where $\pmb{X}^+$ is the Moore-Penrose pseudoinverse of $\pmb{X}$.
It can be obtained using the compact singular value decomposition of $\pmb{X}$ as,
\[
\pmb{X}^+=\pmb{V}\pmb{\Sigma}^{-1}\pmb{U}^{T} \hspace{5mm}\text{with}\hspace{5mm} \pmb{X}=\pmb{U}\pmb{\Sigma}\pmb{V}^{T},
\]
where $\pmb{U}\in \mathbb{R}^{m \times n}$ and $\pmb{V}\in \mathbb{R}^{n\times n}$ are orthogonal matrices, while $\pmb{\Sigma}$ is a diagonal matrix of size $n\times n$ with non-zero real singular values.
The matrix $\pmb{A}$ can be then expressed as,
\begin{equation}\label{eq:A}
    \pmb{A}=\pmb{Y}\pmb{V}\pmb{\Sigma}^{-1}\pmb{U}^{T} \in \mathbb{R}^{m \times m}.
\end{equation}
As in the regular Eulerian approach, in practice $m \gg n$ which complicates the use of Eq.~\ref{eq:A_operator}.
A low-order representation of $\pmb{A}$ is sought through the compact singular value decomposition of $\pmb{X}$. 
This leads to an approximate representation of $\pmb{A}$ as,
\begin{equation}\label{eq:A_tilde}
\pmb{\tilde{A}} = \pmb{U}^{T}\pmb{A}\pmb{U} = \pmb{U}^{T}\pmb{Y}\pmb{V}\pmb{\Sigma}^{-1} \in \mathbb{R}^{n \times n},
\end{equation}
Lastly, the Lagrangian DMD modes, $\pmb{\Phi}_l \in \mathbb{C}^m$, are obtained by
\begin{equation}\label{eq:dmd_modes}
\pmb{\Phi}_l = \pmb{U}\pmb{v}_l,
\end{equation}
where the $l$th eigenvector $\pmb{v}_l \in \mathbb{C}^m$ (or $\pmb{v}_l \in \mathbb{C}^n$ for $m \gg n$) is a solution of the eigenvalue problem: 
\begin{equation}\label{eq:evp_t}
\pmb{A}\pmb{v}_l={\lambda}_l \pmb{v}_l \hspace{5mm}\text{or}\hspace{5mm}\pmb{\tilde{A}}\pmb{v}_l={\lambda}_l \pmb{v}_l \hspace{3mm}\text{for}\hspace{3mm} m \gg n.
\end{equation}
The corresponding eigenvalue ${\lambda}_l \in \mathbb{C}$.
The temporal growth rate and angular frequency of the LDMD mode are $\mathcal{R}(\omega_l)=\ln |{\lambda}_l|/\delta t$ and $\mathcal{I}(\omega_l)=\arg({\lambda}_l)/\delta t$, respectively, where $\delta t$ is the Lagrangian uniform time discretization.

The spatial form of the Lagrangian DMD formulation comprises the Lagrangian flow field tensors $\pmb{X}$ and $\pmb{Y}$ constructed by spatially shifting the Lagrangian flow fields, $\pmb{\mathcal{U}}(\pmb{\chi},t)$.
Thus, the elements of $\pmb{X}$ comprise the Lagrangian flow fields at $\pmb{\chi}=\{\pmb{\chi}_1, \pmb{\chi}_2,\cdots,\pmb{\chi}_{m-1} \}$, while the elements of $\pmb{Y}$ are the Lagrangian flow fields at $\pmb{\chi}=\{\pmb{\chi}_2, \pmb{\chi}_3,\cdots,\pmb{\chi}_m\}$ for all discrete times, $n$.
The eigenvalue problem of Eq.~\ref{eq:evp_t} can expressed in terms of the spatial wavenumber ${k}\in\mathbb{C}$ as,
\begin{equation}\label{eq:evp_x}
\pmb{A}\pmb{v}_l={\lambda}_l \pmb{v}_l \hspace{5mm}\text{or}\hspace{5mm}\pmb{\tilde{A}}\pmb{v}_l={\lambda}_l \pmb{v}_l \hspace{3mm}\text{for}\hspace{3mm} m \gg n.
\end{equation}
The spatial growth rate and wavenumber of the LDMD mode are $\mathcal{R}(k_l)=\ln |{\lambda}_l|/\delta x$ and $\mathcal{I}(k_l)=\arg({\lambda}_l)/\delta x$, respectively, where $\delta x$ is the Lagrangian uniform space discretization, accounting for the diffeomorphism of $\mathcal{D}$.

The Lagrangian stability analysis framework seeks to extract the baseflow stability features that are embedded in the Lagrangian flow map.
Some of the features include fixed points, periodic orbits, stable and unstable manifolds, and chaotic attractors~\citep{ottino1989kinematics,wiggins2005dynamical,lekien2007lagrangian,shadden2005definition,haller2015lagrangian}.
Indeed, the dynamic mapping formed by matrix $\pmb{A}$ and its approximation $\pmb{\tilde{A}}$ for $m \gg n$ comprise the properties of the Lagrangian flow map $\mathcal{M}(\pmb{\chi},t)$ (of Eq.~\ref{eq:mapping}) in terms of the Lagrangian flow fields $\pmb{X}$.
In particular, these include eigenvalues, eigenvectors, energy amplification, and resonance behavior~\citep{schmid2010dynamic}, which reveal the dynamic characteristics of the process that is governing the flow map.
The Lagrangian flow fields at any time instant $t$ can be expressed as,
\begin{equation} \label{eq:ldmd_recT}
\pmb{X}=\pmb{\Phi}\exp{\left( \pmb{\omega}t \right)}\pmb{a} \hspace{5mm}\text{with}\hspace{5mm}\pmb{a}=\pmb{\Phi}^+\pmb{X}_{t_n}.
\end{equation}
Here $\pmb{\phi}=\{\pmb{\phi}_l\}_{l\in\{1,...,n\}}\in \mathbb{C}^{m\times n}$ is a set of complex LDMD modes, while $\pmb{\omega}=\text{diag}\{\omega_1,...,\omega_n\}$ are the eigenvalues (also Ritz values).
The initial conditions $\pmb{a}\in \mathbb{C}^{n}$ are obtained by means of the pseudoinverse $\pmb{\phi}^+$ and the last Lagrangian flow field in the negative time direction $\pmb{X}_{t_n}=\mathcal{M}(\pmb{\chi}_n,t_n)$.
In the proposed LagSAT formulation, $\pmb{X}_{t_n}$ becomes the first flow field of the dataset $\pmb{X}$ in the flow direction (or positive time direction), progressing toward the identity map $\pmb{X}_{t_0}=\mathcal{M}(\pmb{\chi}_0,t_0)$.

Similarly in the spatial form of LDMD, we can reconstruct the Lagrangian flow fields as,
\begin{equation} \label{eq:ldmd_recX}
    \pmb{X}=\pmb{\Phi}\exp{\left( \pmb{k}\chi \right)}\pmb{a} \hspace{5mm}\text{with}\hspace{5mm}\pmb{a}=\pmb{\Phi}^+\pmb{X}_{\chi_1},
\end{equation}
where $\pmb{\phi}=\{\pmb{\phi}_l\}_{l\in\{1,...,m\}}\in \mathbb{C}^{m\times m}$ is a complex set of LDMD modes, while $\pmb{k}=\text{diag}\{k_1,...,k_n\}$ are the eigenvalues (also Ritz values).
The initial amplitudes associated with the spatial modes are estimated at the upwind end of the spatial domain, here $\chi_1$, to remain consistent in terms of the flow direction.

\subsection{Flow stability in Lagrangian reference frame} \label{sec:lagsat_subsec}

The elements of stability theory, Lagrangian flow map as a dynamical system, and LMA are consolidated here to form LagSAT.
Clearly, LagSAT framework operates directly on the Lagrangian spatio-temporal features of a baseflow, as opposed to the conventional approach of solving for the perturbation dynamics on a baseflow.
Nonetheless, we present LagSAT in a form that is analogous to the Eulerian stability framework of Sec.~\ref{sec:LST} and also suitable for the local/global stability analyses.
Following the Eulerian stability equations of Sec.~\ref{sec:LST}, Eq.~\ref{eq:ldmd_recT} can be expressed in a functional form as,
\begin{equation}\label{eq:ldmd_func}
        \pmb{\mathcal{S}}(\pmb{\chi},t)=a\pmb{\phi}(\pmb{\chi})\exp{({\omega} t)},
\end{equation}
where $\pmb{\mathcal{S}}$ is the Lagrangian flow map obtained using the mapping $\mathcal{M}$ of Eq.~\ref{eq:mapping}.
The eigenvalues of Eq.~\ref{eq:ldmd_func} are complex numbers, where the real (${\omega}_r$) and imaginary (${\omega}_i$) parts provide, respectively, the growth rate and angular frequency of the associated shape functions/modes, $\pmb{\phi}(\pmb{\chi})$.
Equation~\ref{eq:ldmd_func} can be further expanded as,
\begin{equation}\label{eq:ldmd_func_realimag}
        \pmb{\mathcal{S}}(\pmb{\chi},t)=a\pmb{\phi}(\pmb{\chi})\exp{({\omega}_r t)}\exp{(i{\omega}_i t)}.
\end{equation}
For a positive value of the real part ${\omega}_r > 0$, the specific mode grows exponentially.
The baseflow in this case is unstable even for a single such mode.
On the other hand, the real part ${\omega}_r < 0$ indicates a decaying mode, where the baseflow remains stable only if all modes decay.
In addition, the baseflow is neutrally stable for ${\omega}_r = 0$.
Similar to the Eulerian linear stability framework, the imaginary part of the eigenvalue (${\omega}_i$) provides the modal frequency.

To perform local analysis, the simplifying assumptions on baseflow can be invoked in LagSAT framework as well, subjected to the non-uniformity of the baseflow.
For a 2D non-parallel baseflow, Eq.\ref{eq:ldmd_func_realimag} can be written as,
\begin{equation}
    \pmb{\mathcal{S}}(\chi_1,\chi_2,t)=a\pmb{\phi}(\chi_1,\chi_2)\exp{({\omega}_r t)}\exp{(i{\omega}_i t)}.
\end{equation}
In this case, one can perform LagSAT with $\chi_1$ as the flow direction and $\chi_3$ as the homogeneous flow direction; this corresponds to the bi-global approach of Eulerian stability.
On the other hand, for a flow with $\chi_3$ as the flow direction and nominal variation of the flow in that direction, correspond to the PSE approach of Eulerian flow stability~\citep{bertolotti1992linear}.
To perform LagSAT in either case, there is no restriction on the spatial variation of the baseflow, as long as there is some.

Next, for a 1D non-parallel/parallel baseflows, Eq.~\ref{eq:ldmd_func_realimag} can be expressed as,
\begin{equation} \label{eq:x2}
    \pmb{\mathcal{S}}(\chi_2,t)=a\pmb{\phi}(\chi_2)\exp{({\omega}_r t)}\exp{(i{\omega}_i t)},
\end{equation}
assuming $\chi_2$ as the wall normal direction.
The 1D non-parallel flow assumption requires a non-zero wall-normal flow velocity $\mathcal{V}\neq 0$, whereas 1D-parallel flow assumption employs a zero ($\mathcal{V}=0$) wall-normal velocity.
To conveniently retain the non-uniformity of the base flow, alongside the 1D-parallel flow assumption, one can also employ the following form of the Eq.~\ref{eq:x2},
\begin{equation}\label{eq:x1}
    \pmb{\mathcal{S}}(\chi_1,t)=a\pmb{\phi}(\chi_1)\exp{({\omega}_r t)}\exp{(i{\omega}_i t)},
\end{equation}
where the non-uniformity of flow variables in the streamwise direction can be exploited.
The wall-normal velocity still remains zero ($\mathcal{V}=0$) due the parallel flow assumption.
When subjected to the spatial form of LagSAT, the above equation becomes,
\begin{equation}
        \pmb{\mathcal{S}}(\chi_1,t)=a\pmb{\phi}(t)\exp{({k}_r \chi_1)}\exp{(i{k}_i \chi_1)}.
\end{equation}
The temporal and spatial forms of the presented LagSAT framework provide access to the phase velocities ${c}$ of the interference between the temporal, ${\phi}(t)$ and spatial ${\phi}(\chi)$.
The phase velocities for a solution state variable can be given as,
${c}={\omega}_i/{k}_i$.

\subsection{N-factor and total energy estimation}

Flow transition to turbulence is a non-linear process, where the linearized formulations of flow stability generally remain inadequate in predicting the onset of transition.
More commonly, n-factor is used in the transition prediction by measuring the actual growth of perturbation amplitude in the flow.
It is an exponential factor obtained using the initial and final amplitudes of perturbation over a distance.
It is defined as,
\begin{equation}\label{eq:N-fac1}
    n=\ln{\frac{\mathcal{A}}{\mathcal{A}_0}},
\end{equation}
where $n$ is the n-factor of a particular mode, and $\mathcal{A}_0$ and $\mathcal{A}$ stand for the initial and final amplitudes of the perturbation, respectively.
In the context of LagSAT, where no external perturbation is present, the modal/non-modal time evolution of the Lagrangian flow map can be utilized to estimate the n-factor.
For an initial modal amplitude, $a_{-\mathsf{T}}$, of a LagSAT mode, $$\mathcal{A}_0=a_{-\mathsf{T}}\exp{(\omega_r t_{-\mathsf{T}})} \text{ and } \mathcal{A}=a_{-\mathsf{T}}\exp{(\omega_r t_0)}.$$
The n-factor can be given by
\begin{equation}\label{eq:N-fac2}
    n=\omega_r(t_0-t_{-\mathsf{T}})=\omega_r(0+\mathsf{T})=\omega_r\mathsf{T},
\end{equation}
where $\omega_r$ is the growth rate associated with the mode, and $\mathsf{T}$ is the time-duration of the Lagrangian flow map.
For all stable/unstable modes, different values of modal growth rate are expected.
Furthermore, for a constant modal growth rate ($\omega_r$) and a fixed $\mathsf{T}$, a local n-factor can be estimated using a local time duration $T_{local}$ that corresponds to a fraction of the spatial extent of the Lagrangian flow map as 
\begin{equation}\label{eq:N-fac3}
    n=\omega_rT_{local}=\omega_r\frac{L_{local}}{U_{ref}}.
\end{equation}
Lastly, the global N-factor envelope can be obtained using the local/individual mode n-factors.

The total energy growth associated with the Lagrangian flow map can be directly estimated by subtracting the time-mean Lagrangian flow map.
Alternatively, the energy associated with the Lagrangian flow map can be expressed in a functional form in terms of the LPOD time coefficients as~\citep{shinde2021lagrangian}: 
\begin{equation}\label{eq:trans-energy}
    {E}(t)=\sum_{l=1}^3\sum_{n=1}^\infty \lambda_n^{l} ||{\psi}^{l}_n(t)||^2,
\end{equation}
where the $\lambda$ and $\psi$ are the eigenvalue and eigenvector of the LPOD.
The index $l$ stands for the three components of velocity.
The total energy of Eq.~\ref{eq:trans-energy} is an optimal form for the energy associated with LPOD modes for a given number of modes~\citep{berkooz1993proper,shinde2020proper}, however, each LPOD mode may comprise multiple spectral/LagSAT modes.
Notably, this energy estimate does not participate in/affect the LagSAT procedure presented before.
Thus, a rather simple definition of the total energy is considered here (Eq.~\ref{eq:trans-energy}); however, one could adopt a commonly used Chou norm~\citep{chu1965energy}, which accounts for the other flow variables and compressibility effects.

Similar variational formulations to estimate energy have been used in the formulation of linearized stability ansatzes, for instance in~\cite{paredes2016transient,paredes2016nonlinear} in the context of linear PSE.
Although LPOD spatio-temporal modes are orthonormal, they retain the non-linearity of the input~\citep{lumley67,berkooz1993proper,shinde2021lagrangian}.
The total energy estimate $E(t)$ remains conserved at all times and it is equal to the direct estimate of the total kinetic energy, in Lagrangian sense.
Thus, the effects of non-normality/non-linearity of LagSAT modes, and thus the transient energy growth~\citep{schmid2007nonmodal}, can be alluded via the total energy estimate of Eq.~\ref{eq:trans-energy}.

In the following sections, LagSAT is applied to three baseflows in local/global forms, examining convective/absolute aspects of instabilities.
The global stability analysis is generally restrictive in terms of computational resources, and requires efficient parallel algorithms~\citep{rodriguez2009massively}.
Several advanced libraries exist for singular/eigen-value problems in parallel, for instance, PETSc and SLEPc with specific packages for parallel implementations~\citep{balay2020petsc,amestoy2001fully,roman2016slepc}.
The parallel formulation of LagSAT employs an in-house Fortran-based set of parallel algorithms to solve singular/eigen-value decompositions~\citep{shinde2025distributed}.

\section{LagSAT on Blasius Boundary Layer} \label{sec:lagsat_BlaFalSka}

The self similar derivation of a boundary layer by~\cite{blasius1907grenzschichten} is a classical baseflow to investigate flow transition.
It is primarily because one can extract baseflow profiles to perform local/quasi-local/global analyses.
In this section, we will perform a local LagSAT analysis, considering the parallel flow assumption alongside a global stability analysis of self-similar boundary layers under various pressure gradients~\citep{falkneb1931lxxxv}.
\begin{figure}
    \centering
    \begin{minipage}{0.32\textwidth}
    \centering
    \includegraphics[width=1.0\linewidth]{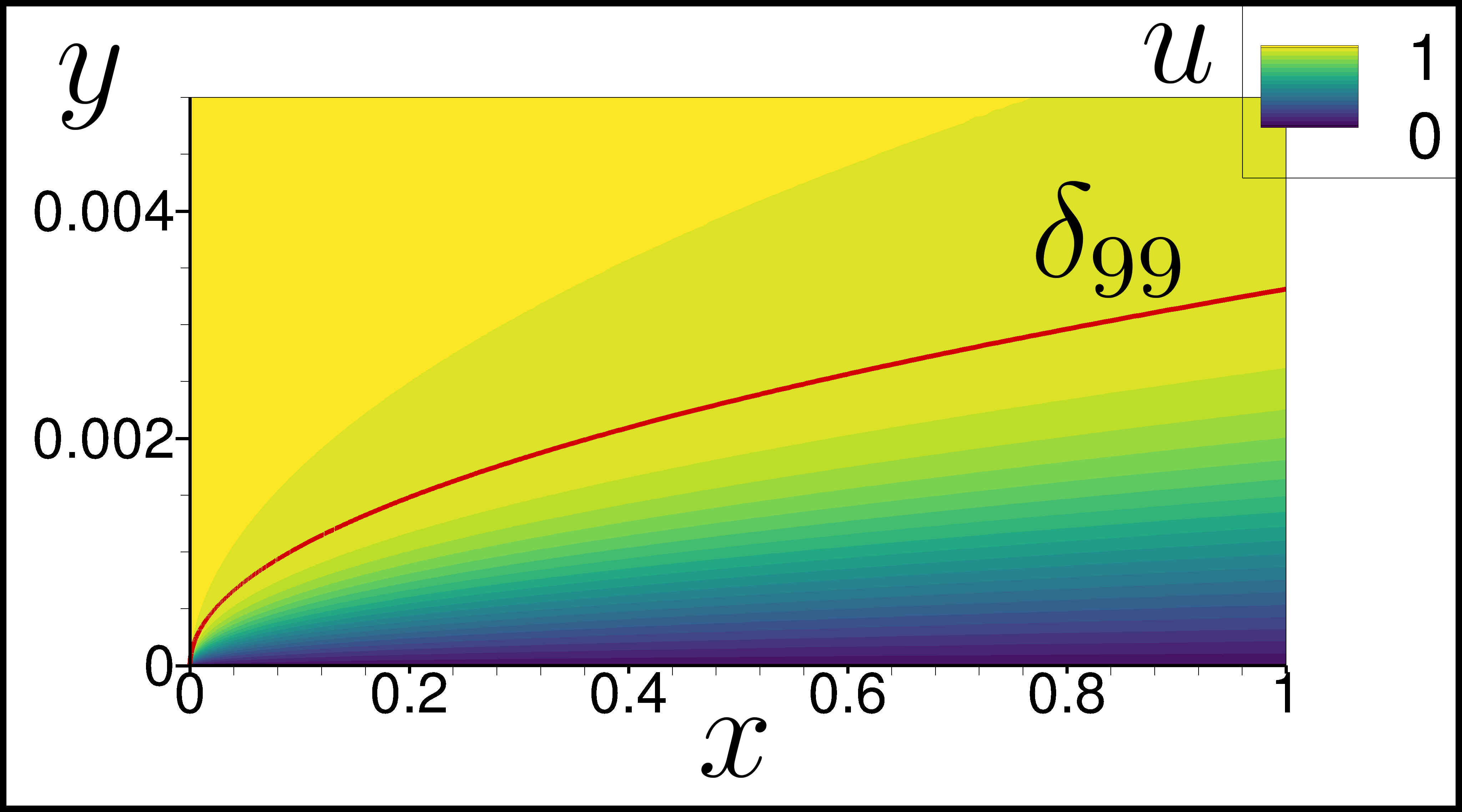}\\(a) 2D $u(x,y)$ profile
    \end{minipage}  
    \begin{minipage}{0.32\textwidth}
    \centering
    \includegraphics[width=1.0\linewidth]{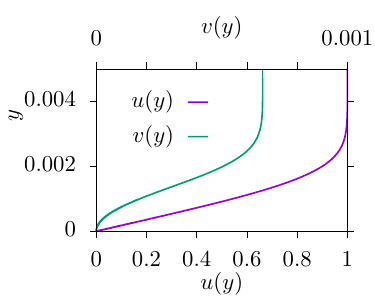}\\(b) 1D $u(y)$, $v(y)$ profiles at $x=0.77$
    \end{minipage}  
    \begin{minipage}{0.32\textwidth}
    \centering
    \includegraphics[width=1.0\linewidth]{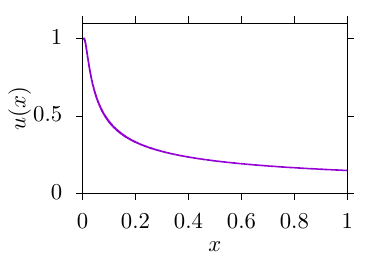}\\(b) 1D $u(x)$ profile at $y=\delta_{99}$ at $x=0.77$
    \end{minipage}
    \caption{Incompressible Blasius boundary layer profile of $u$ with $Re_L=2{,}197{,}134$ and $\delta^\ast=0.00067$ ($\delta_{99}=3.494\times 10^{-3}L$) at the end of the computational domain ($x=1.0$). The edge of the boundary layer is shown in using a pink curve.}
    \label{fig:Blasius_xy}    
\end{figure}
A 2D incompressible Blasius boundary layer is shown in Fig.~\ref{fig:Blasius_xy}(a), where the freestream velocity $u_\infty=1$ and streamwise domain extent $L=1$.
The corresponding Reynolds number is $Re_L={u_\infty L}/{\nu}=2{,}197{,}134$, where $\nu$ is the kinematic viscosity.
The local velocity profiles are displayed in Fig.~\ref{fig:Blasius_xy}(b) and (c), where Fig.~\ref{fig:Blasius_xy}(b) are the conventional streamwise ($u(y)$) and wall-normal ($v(y)$) velocity profiles as a function of the wall-normal distance ($y$) at a streamwise location $x=0.77L$.
Fig.~\ref{fig:Blasius_xy}(c) also displays a local streamwise velocity ($u(x)$) profile as a function of the streamwise distance from the leading edge at a $y$ location matching the boundary layer thickness at $x=0.77L$.

More details of self-similar boundary layers, Blasius and Falkner-Skan, associated equations, and solution methodology are presented in Appendix~\ref{sec:Blasius_baseflow}.
Among various length scales of the self-similar boundary layers, such as the boundary layer thickness, the displacement thickness, the inverse of unit Reynolds number, a general length scale for self similar boundary layers is defined as, $l_x=\sqrt{\frac{\nu x}{u_e}}.$
For the local boundary layer edge velocity, $u_e$, and the local streamwise length from the leading edge, ($x$), a non-dimensional Reynolds number parameter ($R$) is defined as $R=\frac{u_el_x}{\nu}=\sqrt{Re_x}=\sqrt{\frac{u_e x}{\nu}},$
where $Re_x$ is the local Reynolds number.
The modal angular frequencies are also non-dimensionalized in terms of the local variables as, $\omega=\frac{\omega_x l_x}{u_e},$
leading to a non-dimensional frequency parameter given as, 
\begin{equation} \label{eq:F}
F=\frac{\omega}{R}=\frac{2\pi f \nu}{u_e^2},
\end{equation}
where $f$ is a dimensional frequency.

\subsection{Local stability analysis using LagSAT: 1D parallel Blasius boundary layer}\label{sec:lagsat_1dp}

In a local stability analysis, typically the flow profile at a streamwise station is extracted and subjected to stability analysis.
In the parallel flow assumption, the wall-normal mean flow ($v(y)$) is zero; this leads to a uniform streamwise velocity ($u(y)$) profile in the streamwise direction ($x$).
Thus, to perform LagSAT analysis in the local and parallel flow settings, we can utilize $u(x)$ at a $y$ location that corresponds to a streamwise location with the same boundary layer thickness or the local Reynolds number.
In this case, the wall normal velocity $v(x)$ is set to zero due to the parallel flow setting.
Nevertheless, one can employ LagSAT in a local and non-parallel setting, considering both velocity profiles of Fig.~\ref{fig:Blasius_xy}(b), since these velocity profiles constitute a non-uniform flow, albeit in a weak form.
The weak form is due to the relatively small magnitude of the wall-normal component of velocity compared to the streamwise velocity in the boundary layer, as shown in Fig.~\ref{fig:Blasius_xy}(b).

\begin{figure}
    \centering
    \begin{minipage}{0.32\textwidth}
    \centering
    \includegraphics[width=1.0\linewidth]{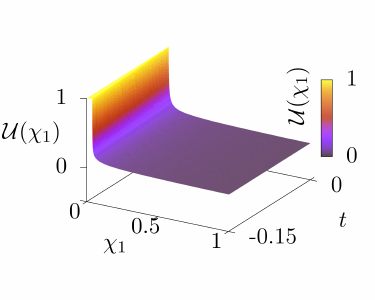}\\(a) $R=134$
    \end{minipage}
    \begin{minipage}{0.32\textwidth}
    \centering
    \includegraphics[width=1.0\linewidth]{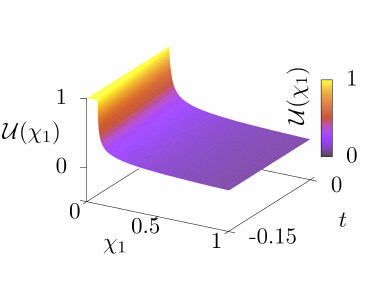}\\(b) $R=402$
    \end{minipage}
    \begin{minipage}{0.32\textwidth}
    \centering
    \includegraphics[width=1.0\linewidth]{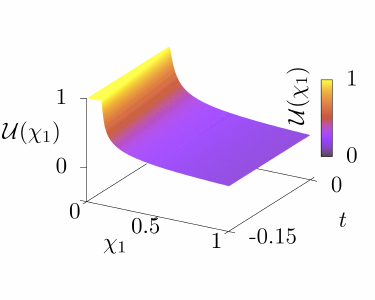}\\(c) $R=670$
    \end{minipage}
    \caption{Lagrangian flow maps of 1D parallel $u(x)$ profiles for (a) $R=134$ (b) $R=402$ (c) $R=670$}
    \label{fig:lagmap_1dp}
\end{figure}

To perform the local LagSAT on 1D velocity profile of Blasius, a Lagrangian flow map of the streamwise velocity $u(x)$ is constructed at a given Reynolds number parameter $R$.
The velocity field is essential in constructing a Lagrangian flow map, as stated in Eq.~\ref{eq:mapping}, where starting with an identity map ($\mathcal{M}(\chi_0,t_0)$) the flow states in the past are gathered by employing an Euler integration with an appropriate size of the time step.
The adjoint form of the Lagrangian modal analysis utilizes the same approach for constructing the Lagrangian flow map~\citep{shinde2021lagrangian}.
The Eulerian baseflow velocity profile $u(x)$ at $t=0$ evolves into the Lagrangian flow map in terms of the Lagrangian velocity field $\mathcal{U}(\chi_1,t)$.
The Lagrangian flow maps are shown in Fig.~\ref{fig:lagmap_1dp} for three values of the local Reynolds number parameter, namely $R=134$, $R=402$, and $R=670$.
The $y$ distance, where the 1D Blasius velocity profile $u(x)$ is extracted, increases with the local Reynolds number.
The Lagrangian flow map flattens (Fig.~\ref{fig:lagmap_1dp}) with increasing $R$ due to a higher velocity ($u(x)$) away from the wall as well as a better alignment of the boundary layer edge with the inflow as $y$ increases.
A total of $150$ time instances were selected with a time-step of $0.001$ in order to build the Lagrangian flow maps of Fig.~\ref{fig:lagmap_1dp}, leading to $\chi_1\in[0,1]$ and $t\in [-0.15,0]$.

\begin{figure}
    \centering
    \begin{minipage}{0.192\textwidth}
    \centering
    \includegraphics[width=1.0\linewidth]{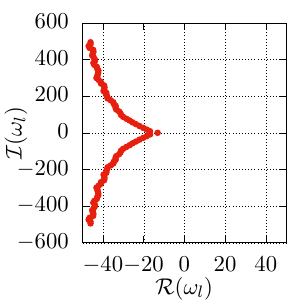}\\(a) $R=134$
    \end{minipage}
    \begin{minipage}{0.192\textwidth}
    \centering
    \includegraphics[width=1.0\linewidth]{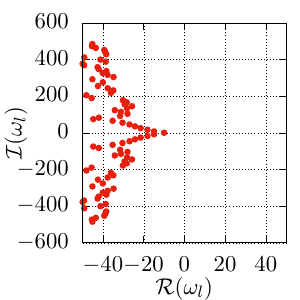}\\(b) $R=268$
    \end{minipage}
    \begin{minipage}{0.192\textwidth}
    \centering
    \includegraphics[width=1.0\linewidth]{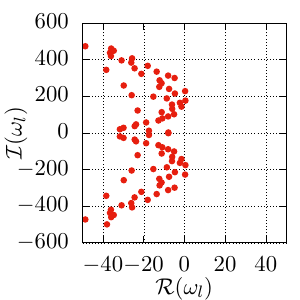}\\(c) $R=402$
    \end{minipage}
    \begin{minipage}{0.192\textwidth}
    \centering
    \includegraphics[width=1.0\linewidth]{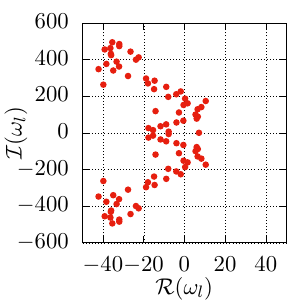}\\(d) $R=536$
    \end{minipage}
    \begin{minipage}{0.192\textwidth}
    \centering
    \includegraphics[width=1.0\linewidth]{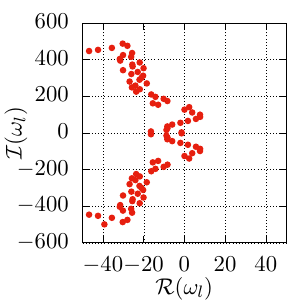}\\(e) $R=670$
    \end{minipage}
    \caption{Eigen spectra and stability of 1D parallel $u(x)$ profiles for (a) $R=134$ (b) $R=268$ (c) $R=402$ (d) $R=536$ (e) $R=670$.}
    \label{fig:R_1dp}
\end{figure}

As noted before, the Lagrangian flow maps are subjected to LDMD in the adjoint setting with the progress of time toward the identity map, i.e., the Eulerian baseflow, in order to obtain the LagSAT stability modes and associated dynamics.
In addition, the flow direction plays a key role in extracting the stability dynamics in an unambiguous manner.
For instance, the eigen spectra of Fig.~\ref{fig:R_1dp}(a) display the modal growth rates and associated frequencies for a lower value of $R$.
The growth rates are all negative, indicating all decaying modes implying a stable flow configuration at $R=134$.
For increasing values of the local Reynolds number, the modal growth rates increase, which is evident if we compare Fig.~\ref{fig:R_1dp}(a) and Fig.~\ref{fig:R_1dp}(b).
Further increase of the Reynolds number to $R=402$ leads to near-zero growth rates of the stability modes (Fig.~\ref{fig:R_1dp}c), where the modes with zero growth rate ($\mathcal{R}(\omega_l)=\omega_r=0$) are called neutrally stable modes.
At further higher values of $R$, increasing number of modes exhibit positive growth rates, see Fig.~\ref{fig:R_1dp}(d) and Fig.~\ref{fig:R_1dp}(e), indicating unstable baseflows.

The local LagSAT on Blasius boundary layer reproduces some of the salient features of the well known neutral stability curve given by LST.
In Fig.~\ref{fig:R_1dp}, the local Blasius boundary layer becomes unstable for $R\approx 402$, where the non-dimensional frequency of the unstable mode $St=fL/u_\infty$ is $\approx 200$ (or in terms of local length scale $fl_x/u_e$ $\approx 55$).
The corresponding frequency in the non-dimensional form of Eq.~\ref{eq:F} is $F=159\times 10^{-6}$, which is in a good agreement with the literature~\cite{mack1984boundary}.
Notably, the lower and higher frequency modes are stable, Fig.~\ref{fig:R_1dp}(c).
For higher Reynolds numbers, the range of unstable modal frequencies shifts toward lower values; for instance, the unstable range is centered around $F=135\times 10^{-6}$ at $R=536$ and $F=98\times 10^{-6}$ at $R=670$.
The higher frequency modes become stable as Reynolds number increases, emphasizing the role of viscosity on the instability~\citep{tumin2010flow}.
The corresponding unstable modes are well known as Tollmien-Schlichting (TS) modes.

\begin{figure}
    \centering
    \begin{minipage}{0.64\textwidth}
    \centering
    \includegraphics[width=1.0\linewidth]{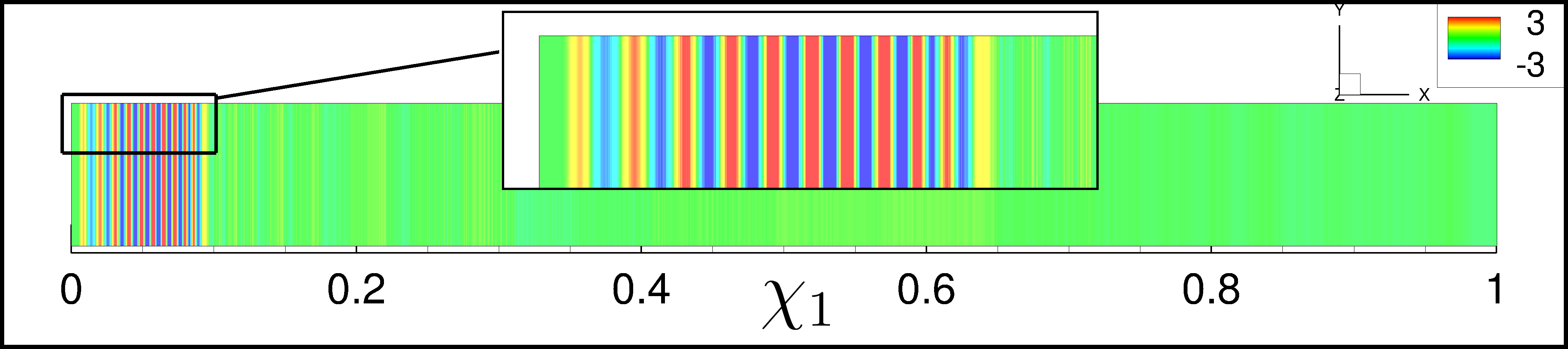}\\(a) Most unstable TS mode
    \end{minipage}
    \begin{minipage}{0.32\textwidth}
    \centering
    \includegraphics[width=1.0\linewidth]{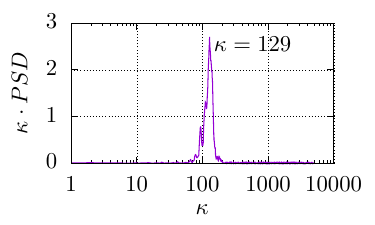}\\(b) TS-mode spatial spectrum
    \end{minipage}
    \caption{Spatial form of the most unstable TS mode, $\mathcal{I}(\omega_l)=74$. (a) TS spatial mode shape for $y=45\% \delta_{99}$ (or $R=670$). (b) Pre-multiplied PSD of the spatial TS mode, providing the spatial wavenumber $\kappa=129$.}
    \label{fig:TS-wave}
\end{figure}
The leading unstable TS mode of the Blasius boundary layer is displayed in Fig.~\ref{fig:TS-wave}.
The mode has an associated non-dimensional frequency of $\mathcal{I}(\omega_l)=75$, which remains consistent through the unstable cases shown in Fig.~\ref{fig:R_1dp}, namely, $R=402$, $R=536$, and $R=670$.
The spatial form of the TS wave clearly exhibits the characteristics wavy pattern, whose amplitude increases downstream toward the identity map, from the LagSAT viewpoint.
For the considered time duration ($t\in [-0.15,0]$) of the Lagrangian flow map and the reference velocity, the LagSAT modes extend toward the identity map over $0 \leq \chi_1 \lessapprox 0.15$.
The spatial wavenumber of the TS wave of Fig.~\ref{fig:TS-wave}(a) is $\kappa\approx 129$, given by the pre-multiplied power spectral density (PSD) of Fig.~\ref{fig:TS-wave}(b).
The phase velocity, i.e., the ratio of the angular frequency and the wavenumber can be estimated as $c=\mathcal{I}(\omega_l)/\kappa$.
Among the spectrum of phase velocity values associated with the LagSAT modes, the leading unstable mode exhibits a phase velocity of $c \approx 0.58$.
In general, the unstable modes at a particular wavenumber display different angular velocities, and therefore the different phase velocities, indicating that this is a dispersive medium~\citep{huerre1990local,juniper2014modal}.

As noted before, the modal growth rate, $\mathcal{R}(\omega_l)=\omega_r$, of the most unstable TS wave increases with the increase of local Reynolds number (Fig.~\ref{fig:R_1dp}).
The n-factor of Eq.~\ref{eq:N-fac3} is essentially the non-dimensional modal growth $n=\omega_rT_x=\omega_rl_x/u_e$.
The local modal growth rates of the most unstable TS mode for the local Reynolds numbers $R=402$, $R=536$, and $R=670$ are $\omega_r=1.5$, $\omega_r=4.7$, and $\omega_r=6.2$, respectively.
For the corresponding local lengths of $l_x=0.27L$, $l_x=0.36L$, and $l_x=0.46L$, the n-factor values are $0.4$, $1.7$, and $2.8$, respectively.
The most unstable TS modal frequencies at these three stations are comparable with the LST results, see for example~\citet{mack1984boundary}, where the maximum n-factor values (or the N-factor envelope values) at the local Reynolds numbers $R=402$, $R=536$, and $R=670$ are represented by the frequency parameter $F\approx 200\times 10^{-6}$, $F\approx 150\times 10^{-6}$, and $F\approx 105\times 10^{-6}$, respectively.
The LST modal growth rates (N-factors) for these frequencies at the corresponding local Reynolds numbers are, respectively, $\approx 0.25$, $\approx 1$, and $\approx 1.5$.
Broadly, the most unstable modal frequencies of LagSAT are marginally lower compared those predicted by the LST, whereas the corresponding N-factor values are moderately higher in LagSAT compared to the LST results.

\begin{figure}
    \centering
    \begin{minipage}{0.48\textwidth}
        \centering
        \includegraphics[width=1.0\linewidth]{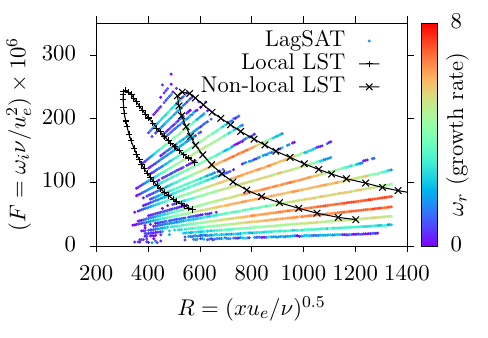}\\(a)
    \end{minipage}
    \begin{minipage}{0.48\textwidth}
        \centering
        \includegraphics[width=1.0\linewidth]{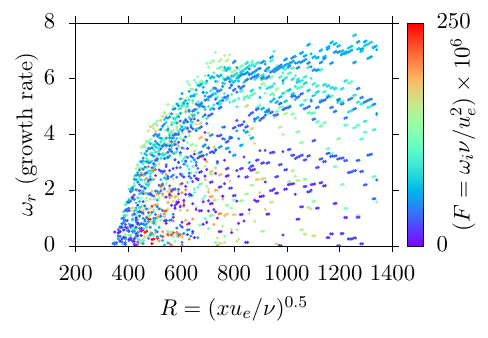}\\(b) 
    \end{minipage}\\
    \begin{minipage}{0.48\textwidth}
        \centering
        \includegraphics[width=1.0\linewidth]{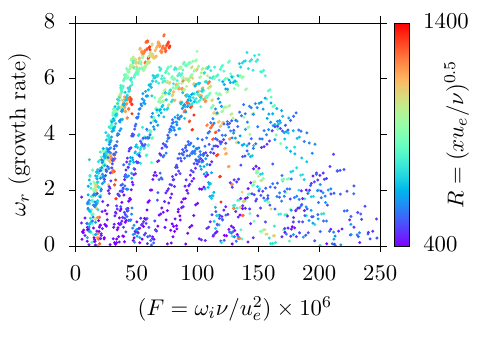}\\(c)
    \end{minipage}
    \begin{minipage}{0.48\textwidth}
        \centering
        \includegraphics[width=1.0\linewidth]{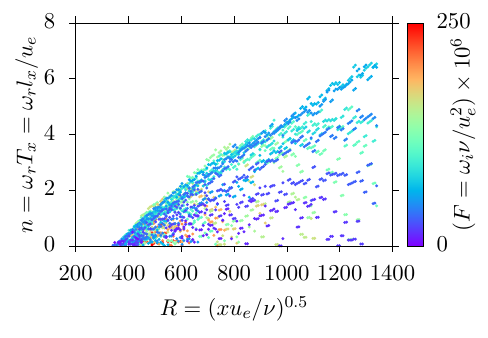}\\(d)
    \end{minipage}
    \caption{Stability map of Blasius boundary layer in 1D parallel flow configuration. (a) Neutrally stable or unstable modal frequencies for increasing Reynolds number parameter $R$ colored with the modal growth rate ($\omega_r$). (b) Growth rate for increasing Reynolds number parameter colored with the frequency parameter ($F$), (c) Growth rate versus modal frequency colored with the Reynolds number parameter, and (d) n-factor values for increasing Reynolds number colored by modal frequency.}
    \label{fig:nsc-nfac}
\end{figure}

To further assess the local stability results, the local velocity profiles are subjected to LagSAT over a wide range of Reynolds numbers.
The neutrally stable ($\omega_r=0$) and unstable ($\omega_r>0$) LagSAT modes are assimilated in Fig.~\ref{fig:nsc-nfac}(a), generating a discrete neutral/unstable region of the Reynolds and frequency parameters.
At lower Reynolds numbers that is less than a critical value, all modes of LagSAT on Blasius boundary layer remain stable with negative growth rates.
The critical Reynolds number is $R_c\approx 350$, at $R$ higher than that ($350 \lessapprox R \lessapprox 600$) the LagSAT modes with a higher frequency range, $50\times 10^{-6} \lessapprox F \lessapprox 250\times 10^{-6}$ become unstable.
For further increase of Reynolds number, the range of unstable frequencies shortens and shifts toward the lower values of the frequency, producing the typical shape of the neutral stability curve~\citep{mack1984boundary}.
The LagSAT unstable region, in a discrete sense, colored by means of the modal growth rates, where the growth rates increase toward the center of the region and tends to zero in the outer region (Fig.~\ref{fig:nsc-nfac}a).
The outer region with zero growth rate represents the neutral stability curve, albeit discretely.

The neutral stability curves of local LST~\citep{juniper2014modal} and PSE (non-local LST)~\citep{gaster1974effects} analyses are regenerated and are also shown in Fig.~\ref{fig:nsc-nfac}(a)
Characteristically, the neutral stability curves of LST/PSE exhibit similarities with LagSAT results, yet the critical Reynolds number and the modal frequency ranges show some quantitative differences.
The critical Reynolds number of LagSAT is marginally higher than the local LST, indicating non-zero growth rates at lower frequencies ($50 \lessapprox F \lessapprox 150$) compared to the local LST frequencies $F\approx 200$.
The LagSAT prediction of the upper limit of the unstable frequencies for all Reynolds numbers (the upper branch of the neutral stability curve) compare well with the PSE (non-local LST) curve.
However, at the lower branch of the stability curve, the LagSAT exhibits unstable modes at much lower Reynolds numbers compared to the non-local LST results, particularly for $F \lessapprox 150$.
Overall, the LagSAT results are consistent with LST in terms of the near-zero frequency stable region for all Reynolds numbers considered.
Furthermore, the LagSAT and LST show similarity in terms of the stability mechanism, where the flow becomes stable at higher Reynolds number, in particular for higher frequency modes.

The local LagSAT results are alternatively presented in Fig.~\ref{fig:nsc-nfac}(b), displaying the modal growth rate versus Reynolds number eigen spectra colored by the modal frequency, and in and Fig.~\ref{fig:nsc-nfac}(c), which displays the growth rate versus modal frequency colored by the Reynolds number.
Fig.~\ref{fig:nsc-nfac}(b) shows the growth rate evolution of LagSAT stability modes for increasing Reynolds number. 
The unstable modes form near-parabolic distributions with some plateauing around $600 \lessapprox R \lessapprox 800$, this is consistent with the linear/non-linear PSE results presented in~\cite{esfahanian2001linear,bertolotti1992linear}.
Clearly, the unstable modes of Fig.~\ref{fig:nsc-nfac} correspond to the first mode or TS waves of instability.
Consistent with the local LST results~\citep{mack1984boundary}, the growth rate as a function of Reynolds number of Fig.~\ref{fig:nsc-nfac}(b) exhibits higher frequencies toward the lower Reynolds number values, whereas the lower modal frequencies persist in the higher Reynolds number range.
Alternatively, the growth rate versus modal frequency of Fig.~\ref{fig:nsc-nfac}(c) displays higher growth rate at higher Reynolds numbers, with the peak growth rates, $\omega_r\approx 7$ attained for the frequency range of $50 \lessapprox F \lessapprox 75$ as compared to the LST results of $30 \lessapprox F \lessapprox 60$~\cite{mack1984boundary}.

The modal amplification is often defined in terms of N-factor (Eq.~\ref{eq:N-fac1}) envelope, which indicates the maximum amplification of individual modes (n-factor) at a Reynolds number.
Early works by~\citet{smith1956transition} and~\citet{jaffe1970determination} suggest N-factor values $\approx 10$ for the transition to occur; however, it depends on several factors including the perturbation environment.
In the present perturbation-free LagSAT ansatz, the n-factor estimator, Eq.~\ref{eq:N-fac2}, relates the logarithmic amplification rate $\ln{(\mathcal{A}/\mathcal{A}_0)}$ to the LagSAT modal growth rate $\omega_r$ and a representative time/length scale of the Lagrangian flow map over which the modal growth occurs.
For instance, the maximum N-factor at $R=600$ is approximately $N=\omega_r\times l_x/u_e=\approx 5\times 0.41=2$, while the N-factor value of $\approx 10$ is obtained for $R > 1472$ - at higher Reynolds number that considered here.
Nonetheless, the maximum amplification (N-factor) values as a function of increasing Reynolds number that are predicted using LagSAT (Fig.~\ref{fig:nsc-nfac}d) are in a good agreement with the literature~\cite{mack1984boundary}.
In the proposed model of N-factor (Eq.~\ref{eq:N-fac2}, Eq.~\ref{eq:N-fac3}), the time-duration $\mathsf{T}$ is a physical rather than numerical parameter, relating to spatio-temporal extent of the Lagrangian flow map.
The numerical uncertainties in the N-factor estimate are primarily associated with the modal growth rates and also depend on the time-resolution (or the number of snapshots) employed in LagSAT.
A broader discussion on the numerical sensitivity of LagSAT results is provided in Appendix~\ref{sec:Num_Sense}.

\subsection{Global stability analysis using LagSAT: 2D Blasius/Falkner-Skan boundary layer}\label{sec:lagsat_1dnp}

For baseflows with two/three inhomogeneous directions, we resort to global analysis, specifically these analyses are referred as bi/tri-global LST, respectively.
The presence of directional inhomogeneity, particularly in steady baseflows, works favorably for LagSAT as it seeks the non-uniformity of baseflow.
In this section, LagSAT is directly applied to a 2D Blasius boundary layer that is displayed in Fig.~\ref{fig:Blasius_xy}(a).
Similar to the local LagSAT procedure, at first, the Eulerian Blasius boundary layer is transformed into a Lagrangian flow map, which is essentially a three-dimensional diffeomorphism of the flow with computational domain.
The spatio-temporal Lagrangian flow map of the 2D Blasius boundary layer is then subjected to Lagrangian modal analysis (LDMD/LPOD) in an adjoint setting but time-progressing toward the identity map, {\it i.e.}, the Eulerian baseflow.

The baseflow of 2D Blasius boundary layer of Fig.~\ref{fig:Blasius_xy}(a) is at a local Reynolds number of $R=1472$ (or the global/free-stream Reynolds number $Re_L=2{,}197{,}134$), which is higher than the critical Reynolds number of $R_c=402$ in anticipation of unstable global modes.
The computational domain is discretized in $1000\times 1000$ mesh points that are non-uniformly distributed in $x$ and $y$ directions to capture the velocity gradients, particularly near the leading edge.
More details of the computational domain and flow solution method are provided in Appendix~\ref{sec:Blasius_baseflow}.
The flow conditions upstream of the inflow boundary are assumed to be the same as at the inflow boundary in order to accommodate the time-evolution of the Lagrangian flow map in the upstream region due to the adjoint setting.
A total of $500$ time instances with a time-step of $0.002$ are utilized to form the Lagrangian flow map.

The 2D Blasius boundary layer comprises convective instability waves, commonly known as Tollmien-Schlichting waves~\citep{schubauer1947laminar,lin1955theory,schubauer1956contributions}.
The linear global LST investigations~\citep{huerre1985absolute,brevdo1995convectively,alizard2007spatially} have shown that the open wall bounded shear layers including Blasius boundary layer are convectively unstable but absolutely stable.
The global LST modes and their genesis in the boundary layers have been extensively studied~\citep{cossu1997global,alizard2007spatially,aakervik2008global,rodriguez2008instability,ehrenstein2008two}, elucidating the mode shapes and potential of transient mechanisms.

\begin{figure}
    \centering
    \begin{minipage}{1.0\textwidth}
    \centering
    \includegraphics[width=1.0\linewidth]{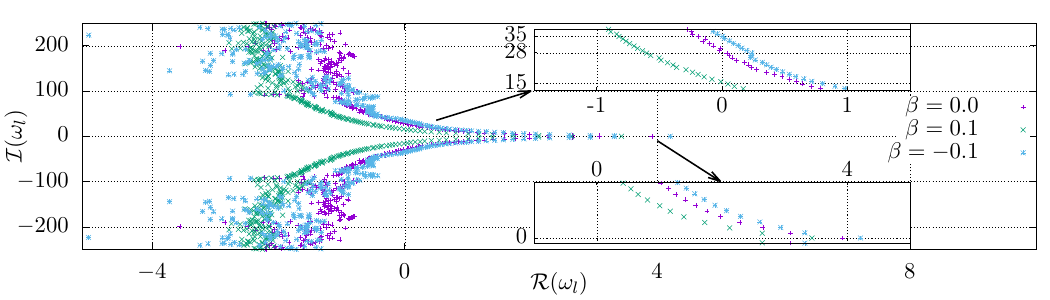}
    \end{minipage}
    \caption{LagSAT eigen spectra of 2D Blasius/Falkner-Skan profile under pressure gradient: zero ($\beta=0$), favorable ($\beta=0.1$), and adverse ($\beta=0.1$) pressure conditions.} 
    \label{fig:2d_phi1}
\end{figure}

In the context of LagSAT for global stability analysis, we subject the 2D Blasius boundary layer to LagSAT anticipating convectively unstable modes as well as their tendency under pressure gradient.
The pressure gradient is imposed by means of the parameter $\beta$ of the Falkner-Skan generalized boundary layer flow (see Appendix~\ref{sec:Blasius_baseflow} for more details), where $\beta=0$ corresponds to the Blasius boundary layer.
A favorable pressure gradient with $\beta=0.1$ and an adverse pressure gradient with $\beta=-0.1$ are considered.
The eigen spectra of the LagSAT on three boundary layers (Fig.~\ref{fig:2d_phi1}) indicate the sets of unstable modes at lower frequencies, namely, $St \lessapprox 35$, $St \lessapprox 28$, and $St \lessapprox 15$ for $\beta=0.1$, $\beta=0.0$, and $\beta-0.1$, respectively.
The insets of the figure display the near the zero frequency (bottom inset) and near-zero growth rate (top inset) modal behaviors.
Clearly, the favorable pressure gradient tends to stabilize the Blasius boundary layer, whereas the adverse pressure gradient has an opposite effect of destabilization.
This is consistent with the prior stability analyses results that were collectively discussed in~\cite{drazin1981hydrodynamic,obremski1969portfolio}.
This effect of pressure gradient on the stability of modes does not seem to persist for higher frequency modes, $St\gtrapprox 100$ (see Fig.~\ref{fig:2d_phi1}).

\begin{figure}
    \centering
    \begin{minipage}{1.0\textwidth}
    \centering
    \includegraphics[width=0.24\linewidth]{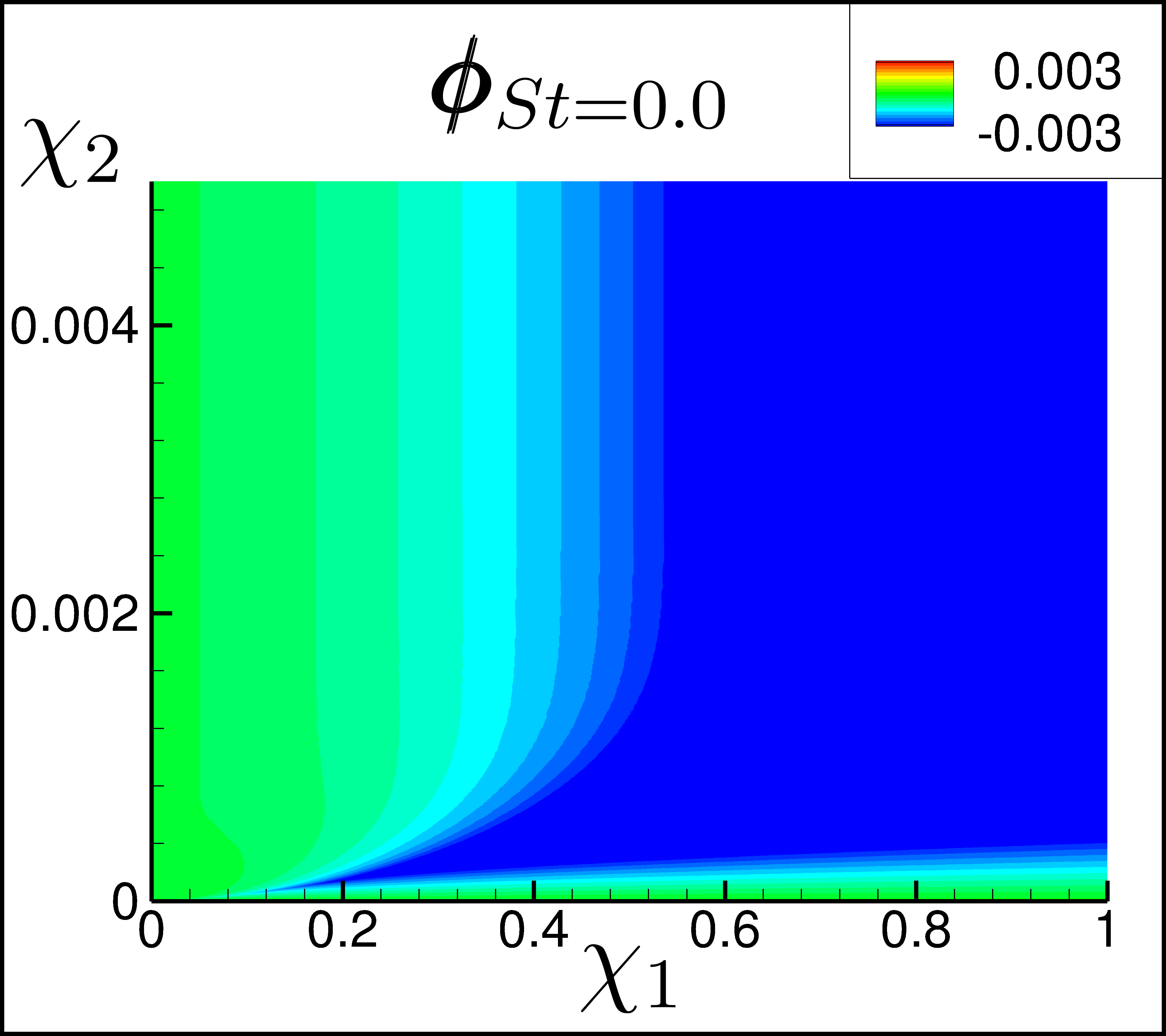}
    \includegraphics[width=0.24\linewidth]{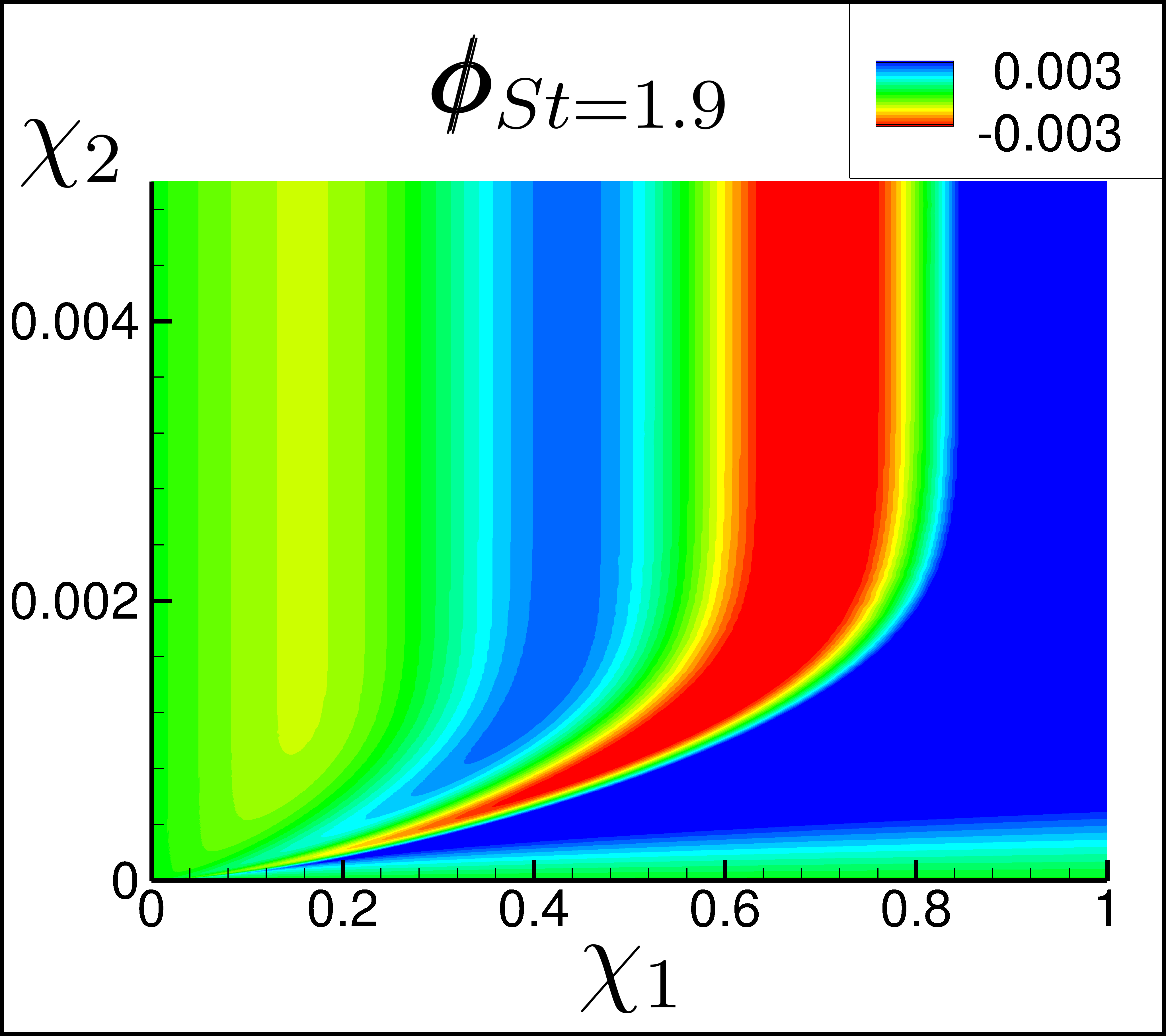}
    \includegraphics[width=0.24\linewidth]{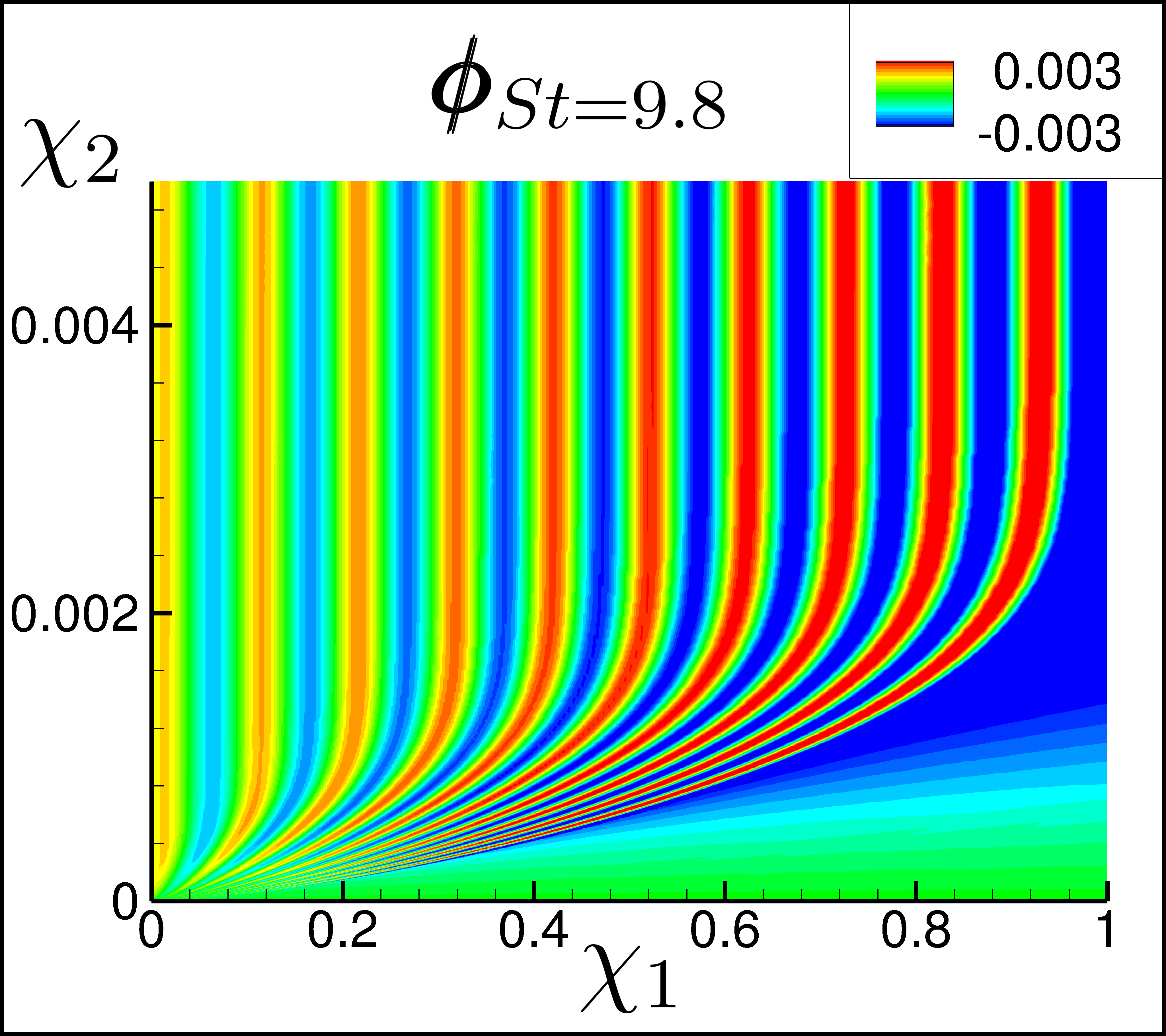}
    \includegraphics[width=0.24\linewidth]{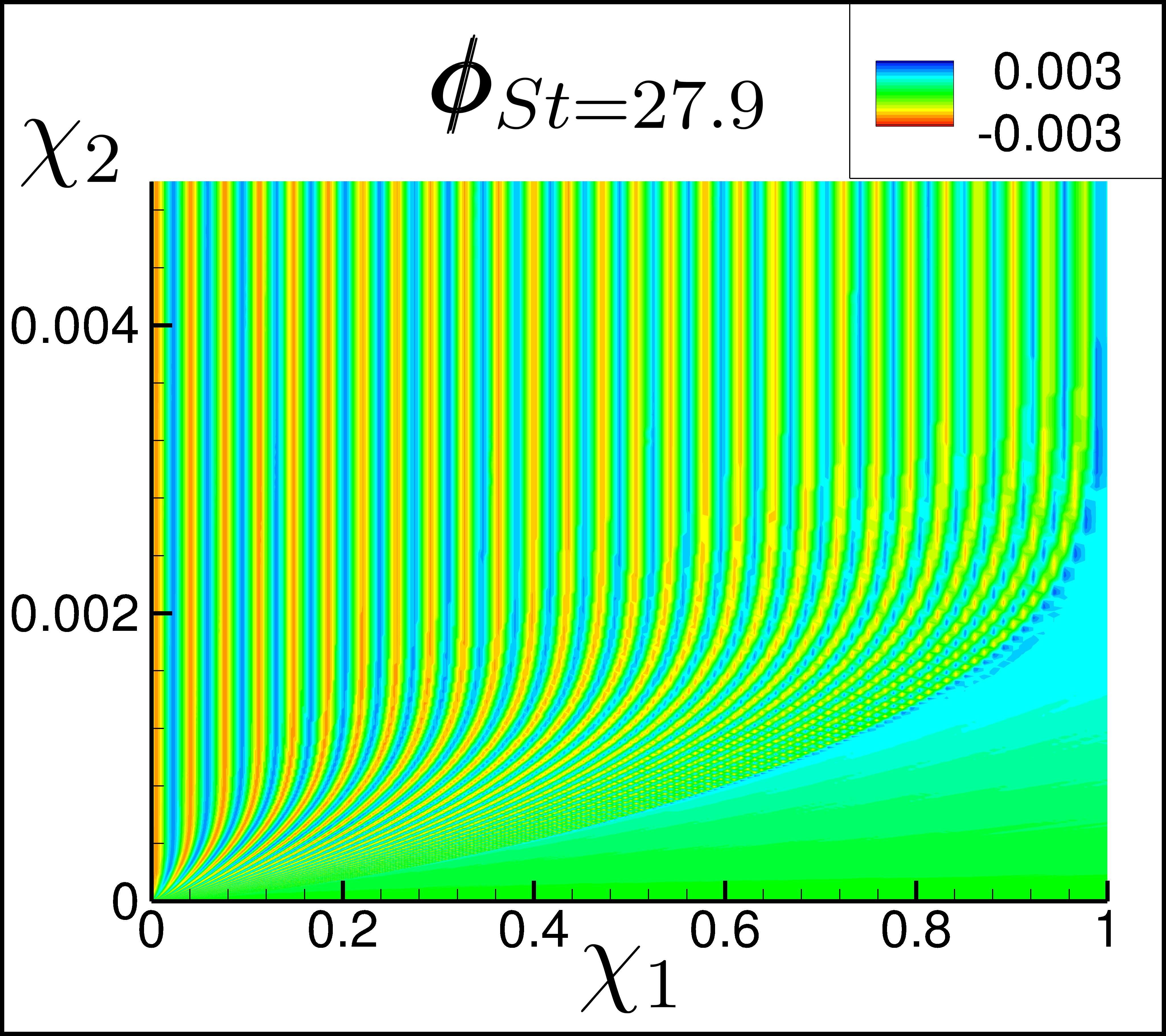}\\(a) Unstable modes for $\beta=0$
    \end{minipage}
    \begin{minipage}{1.0\textwidth}
    \centering
    \includegraphics[width=0.24\linewidth]{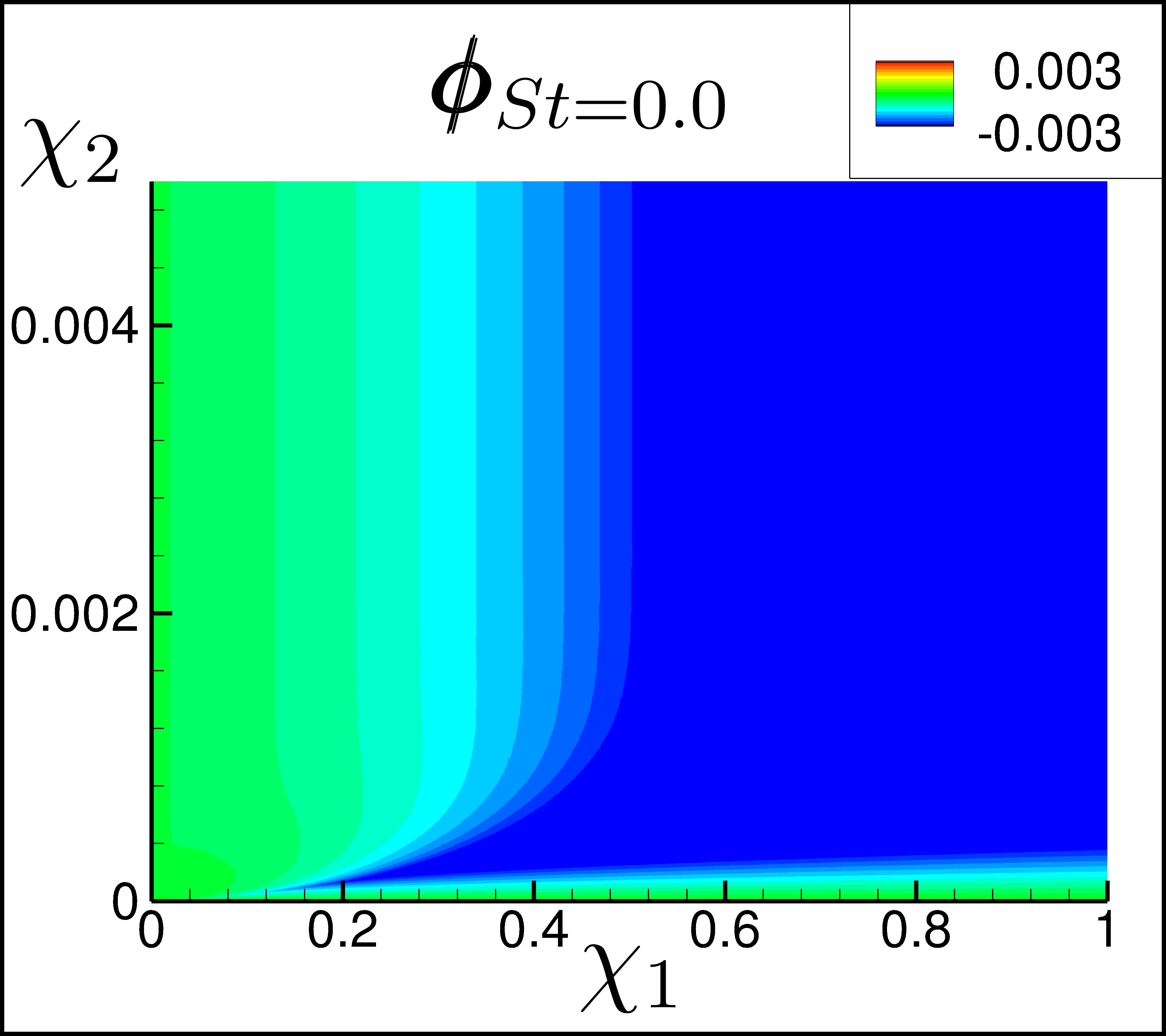}
    \includegraphics[width=0.24\linewidth]{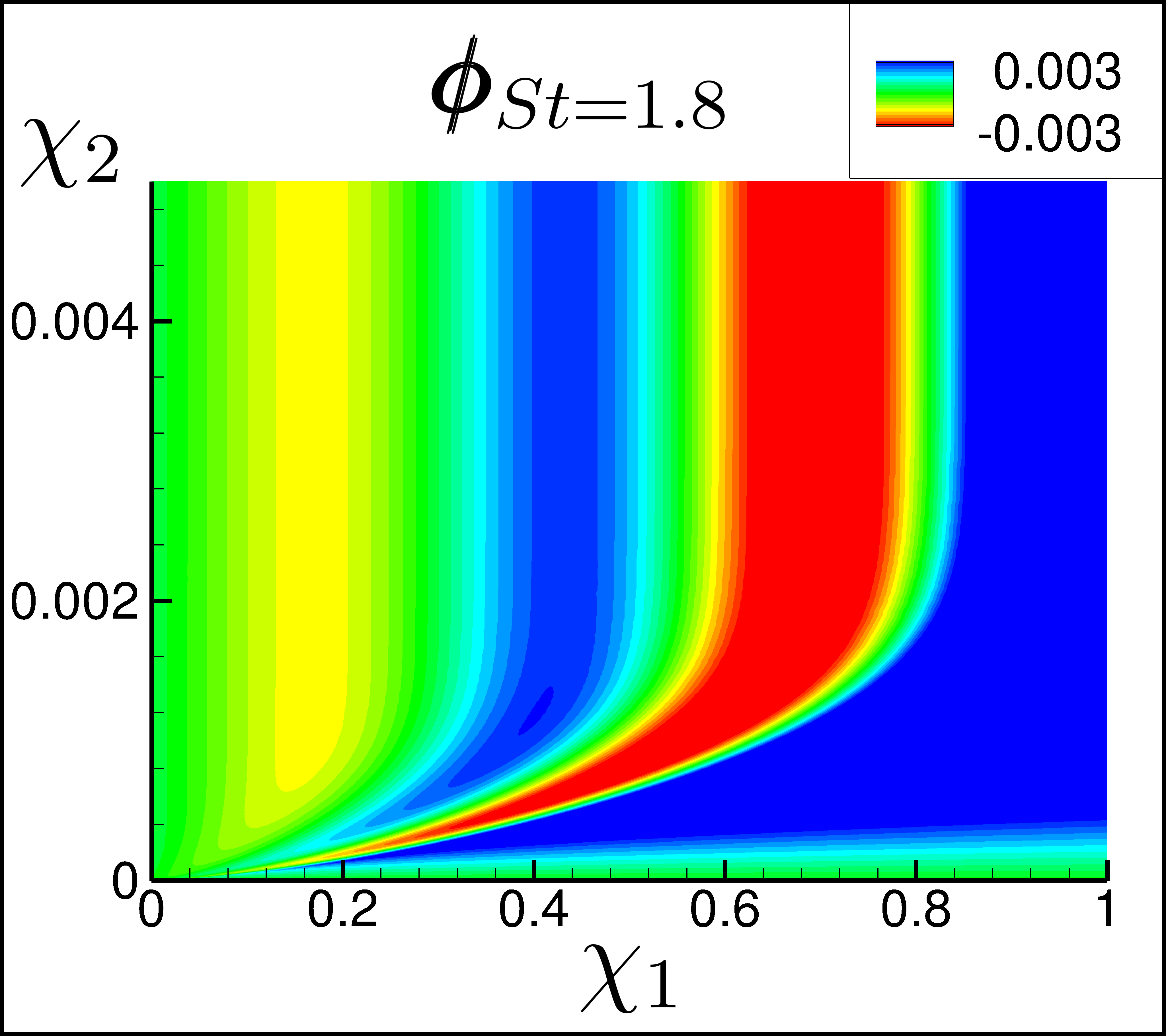}
    \includegraphics[width=0.24\linewidth]{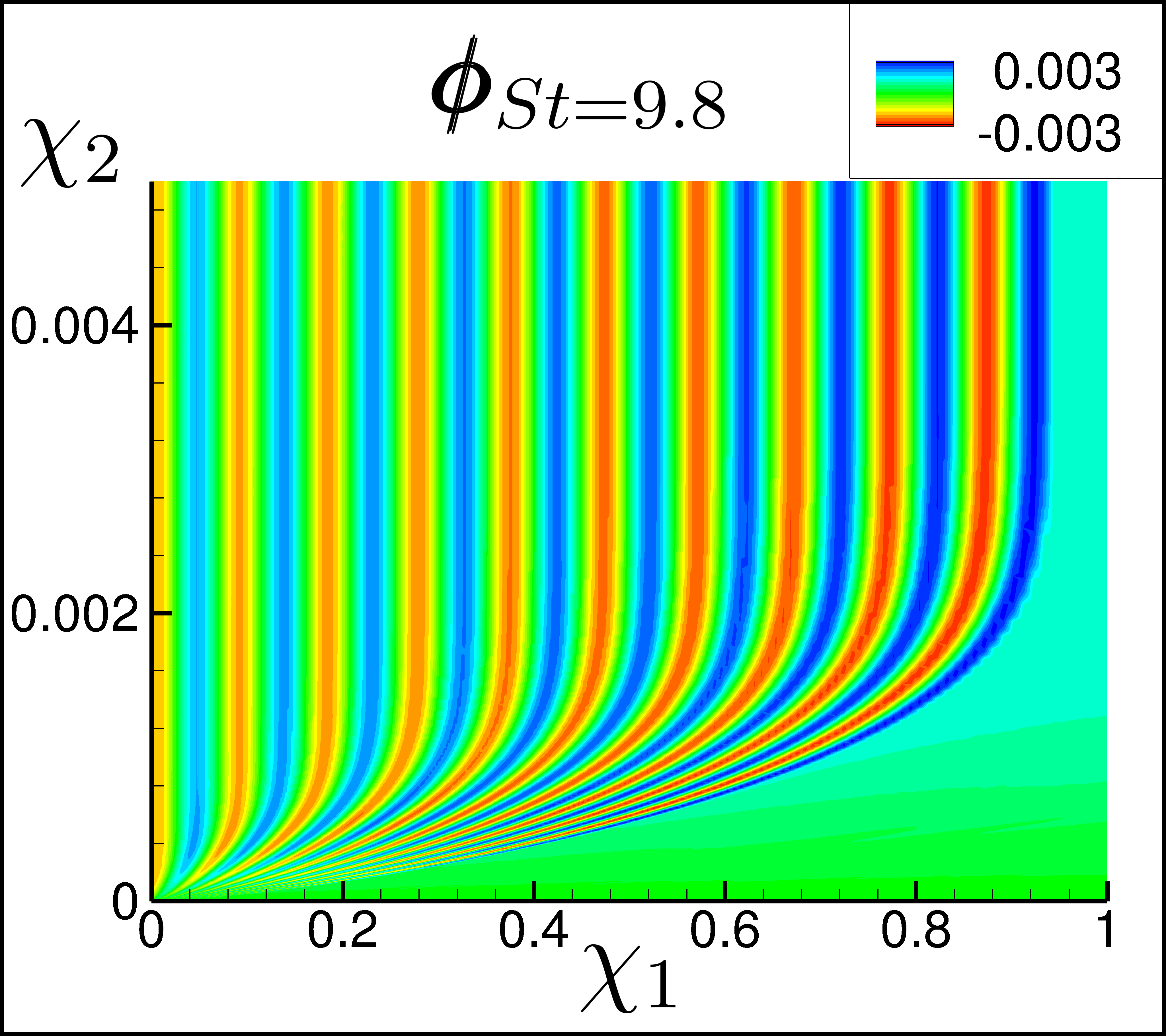}
    \includegraphics[width=0.24\linewidth]{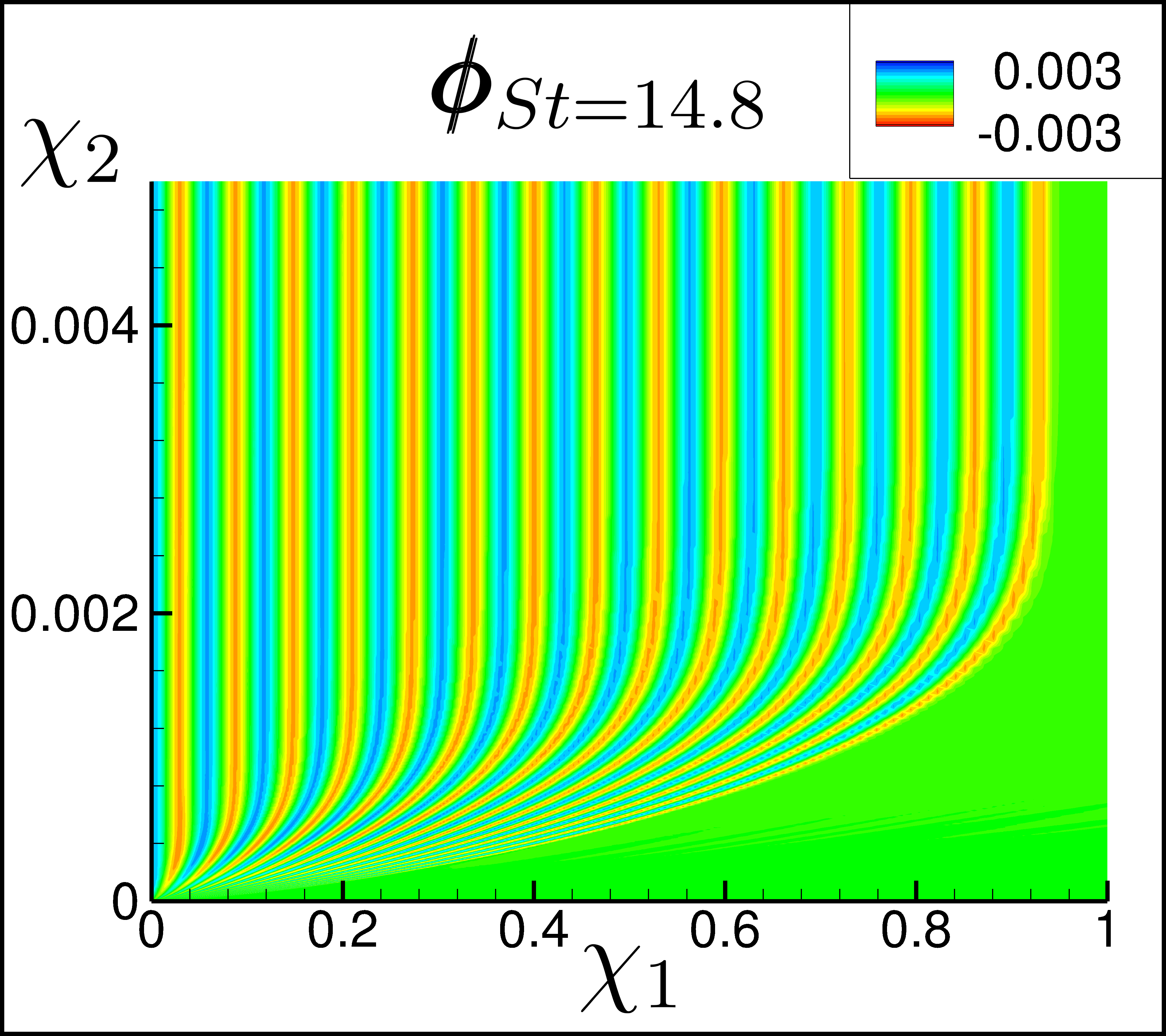}\\(b) Unstable unstable modes for $\beta=0.1$
    \end{minipage}
    \begin{minipage}{1.0\textwidth}
    \centering
    \includegraphics[width=0.24\linewidth]{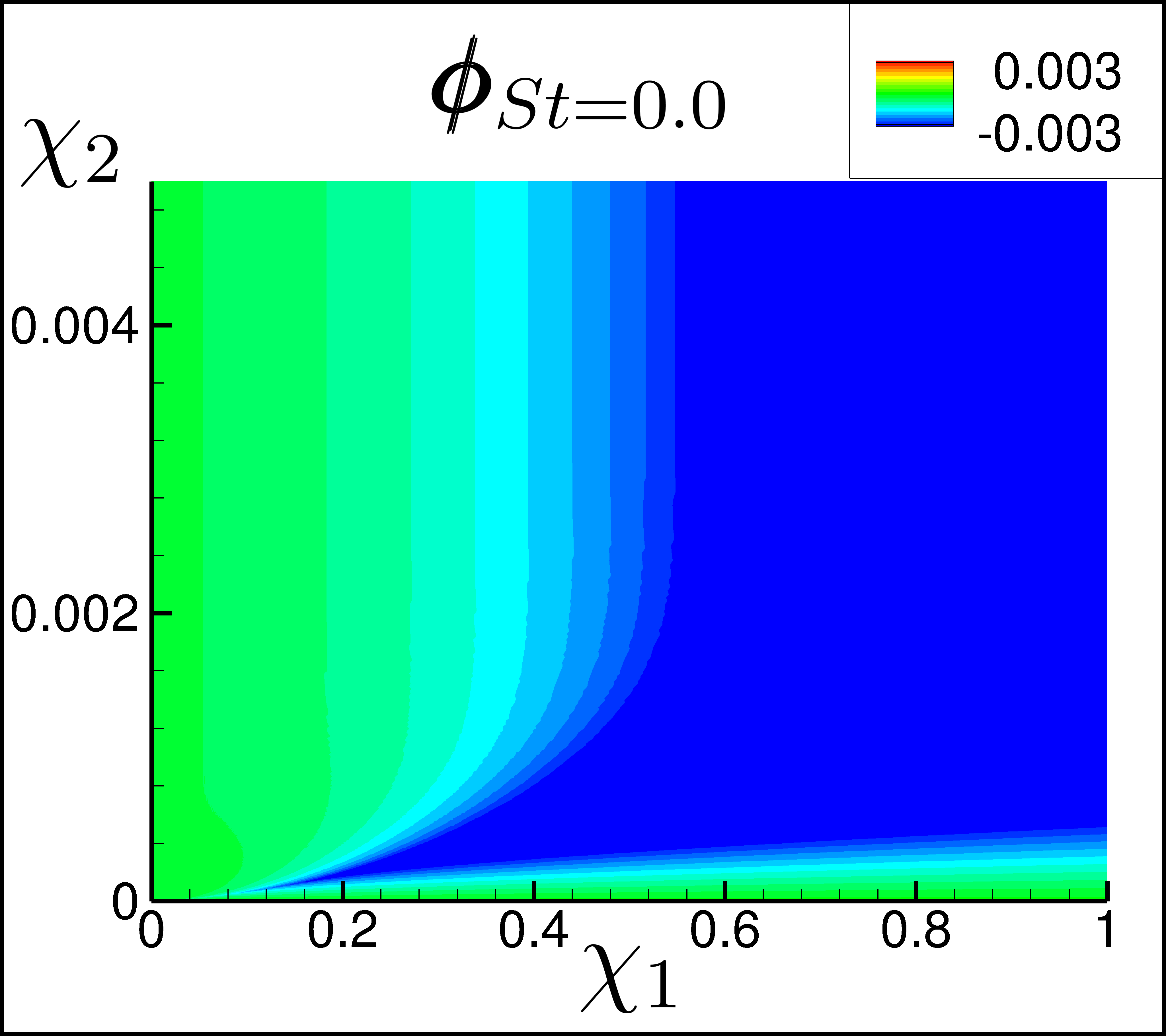}
    \includegraphics[width=0.24\linewidth]{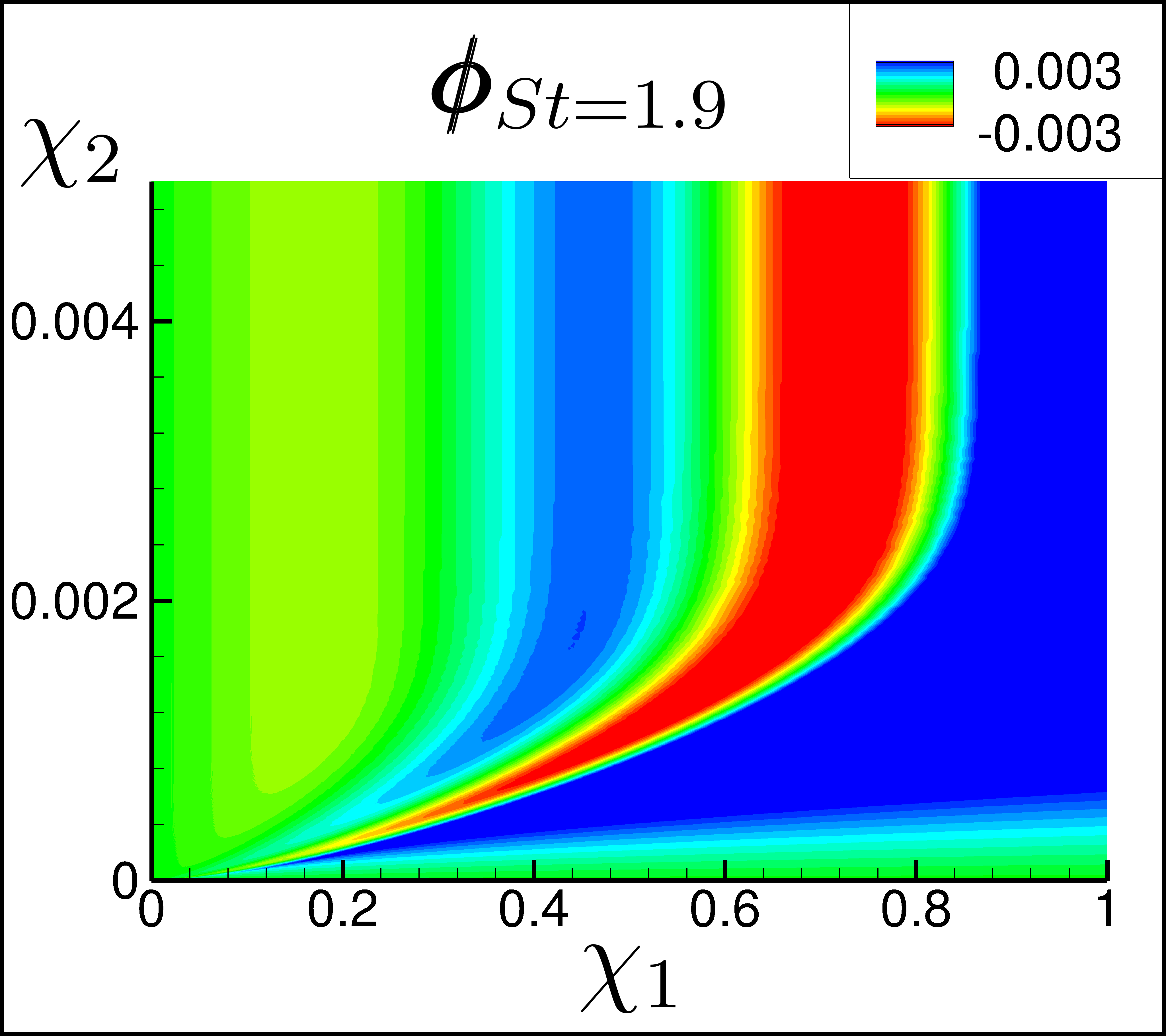}
    \includegraphics[width=0.24\linewidth]{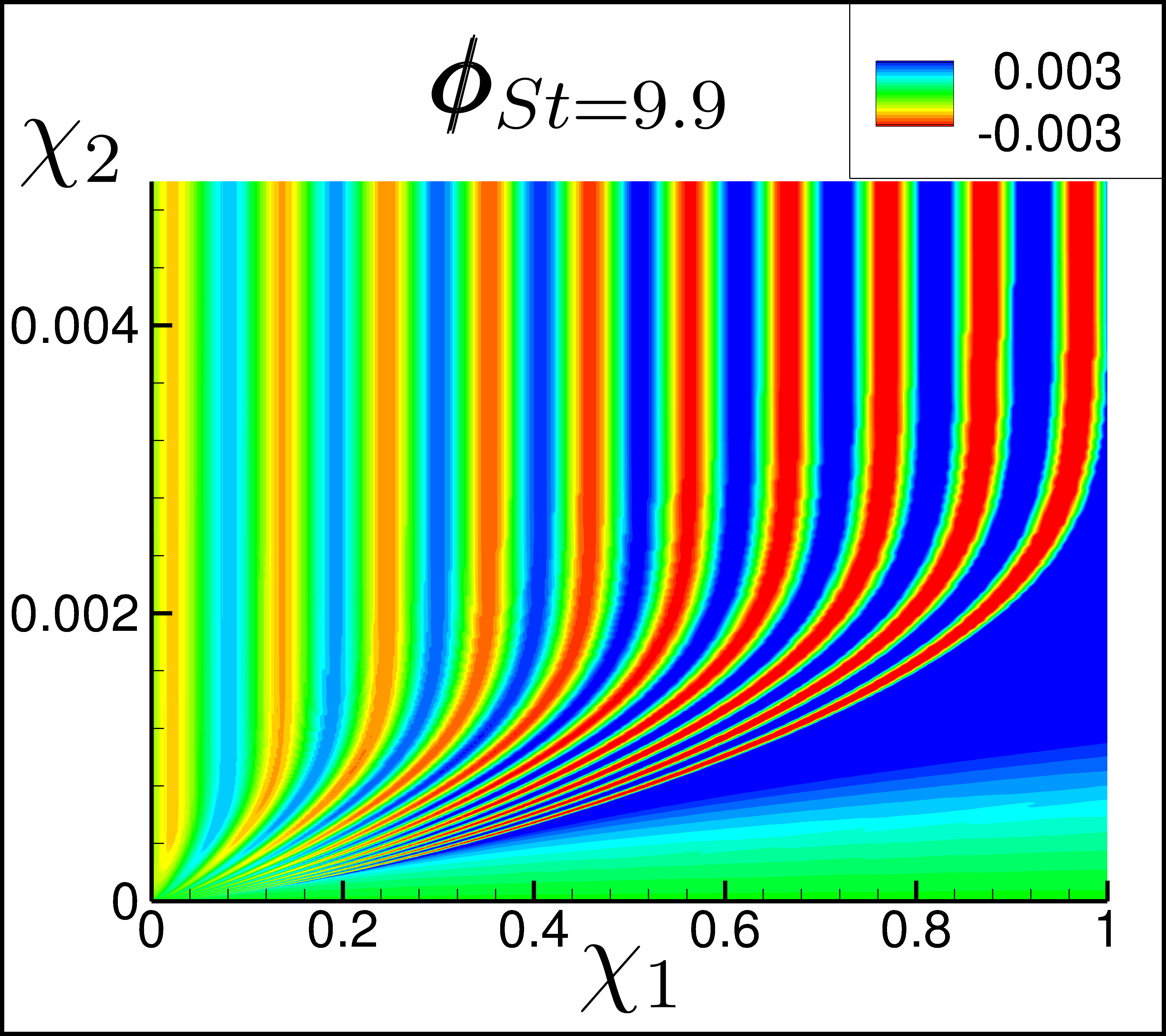}
    \includegraphics[width=0.24\linewidth]{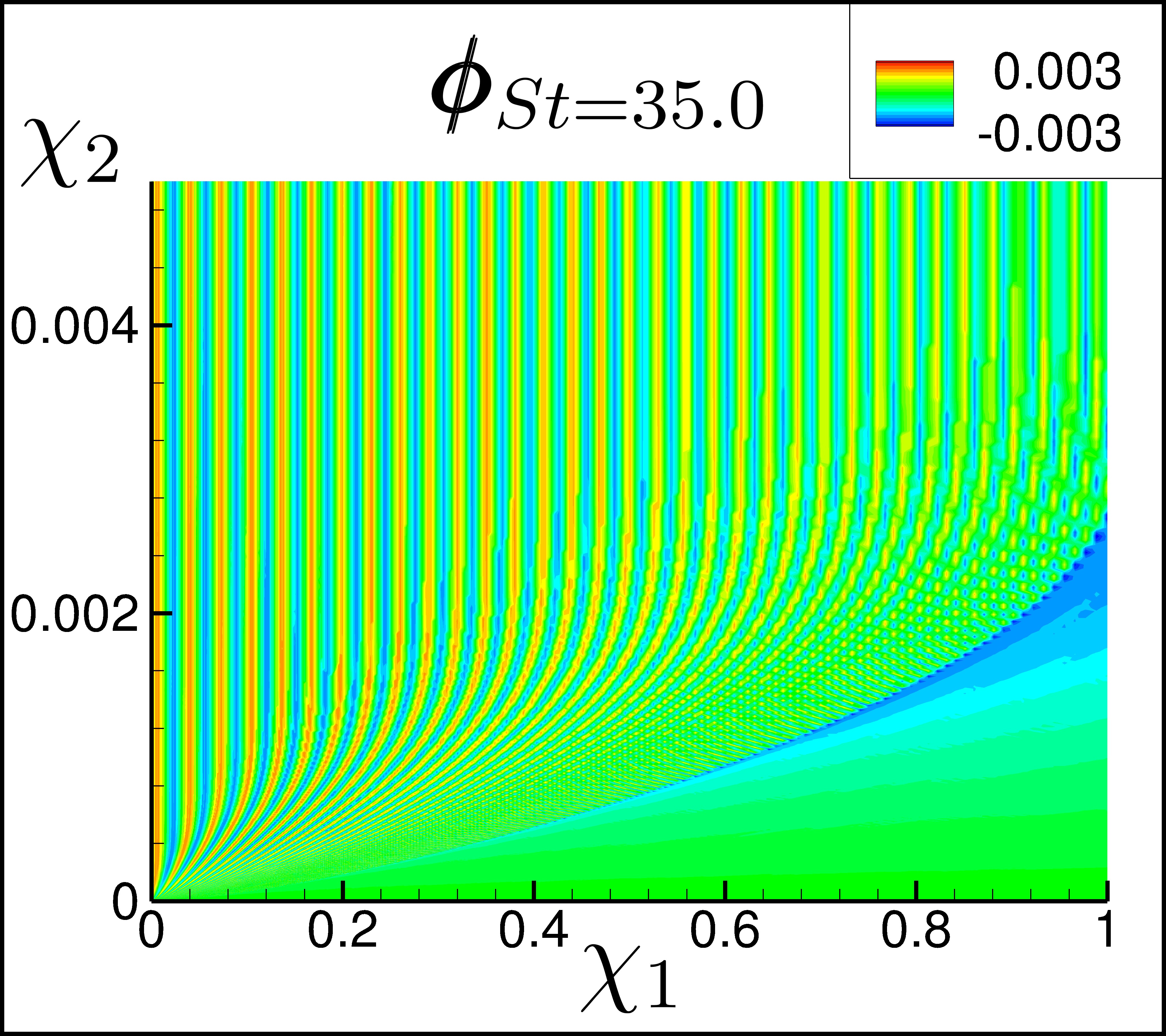}\\(c) Unstable modes for $\beta=-0.1$
    \end{minipage}
    \caption{LagSAT representative instability modes of 2D Blasius/Falkner-Skan boundary layer under zero ($\beta=0$), favorable ($\beta=0.1$), and adverse ($\beta=0.1$) pressure conditions. (a) Unstable modes under zero pressure gradient, (b) Unstable modes under adverse pressure gradient, and (c) Unstable modes under adverse pressure gradient.} 
    \label{fig:2d_phi2}
\end{figure}

The spatial modal shapes of representative LagSAT modes of the Blasius boundary layer are shown in Fig.~\ref{fig:2d_phi2}(a).
Fig.~\ref{fig:2d_phi2}(b) and Fig.~\ref{fig:2d_phi2}(c) display the same set of modes, except the highest frequency mode on the right, under the favorable and adverse pressure gradients, respectively.
The highest frequency mode (on the extreme right) is the least unstable mode of the respective set.
Naturally, the cut-off modal frequency for the least unstable mode due to the stabilizing effect under favorable pressure gradient becomes $St=14.8$ compared to the $St=27.9$ for $\beta=0$, whereas it increases for the destabilizing effect under the adverse pressure gradient ($\beta=-0.1$) to become $St=35$.
The lower frequency modes, including the displayed stationary mode ($St=0.0$) and modes at $St=1.9$ and $St=9.8$, exhibit marginal effect of the pressure gradient, in particular on the stationary mode.
However, the dynamic modes, $St=1.9$ and $St=9.8$, show some attenuation/amplification of the spatial growth for, respectively, favorable/adverse pressure gradient values.

\begin{figure}
    \centering
    \begin{minipage}{1.0\textwidth}
    \centering
    \includegraphics[width=1.0\linewidth]{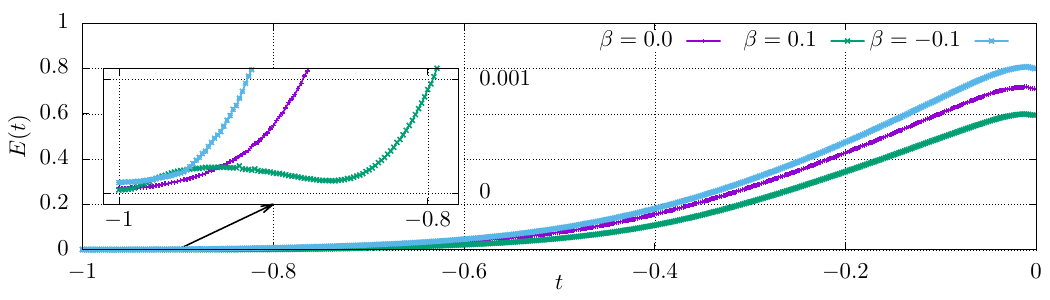}
    \end{minipage}
    \caption{Transient energy growth in LagSAT on 2D Blasius/Falkner-Skan boundary layer under zero ($\beta=0$), favorable ($\beta=0.1$), and adverse ($\beta=0.1$) pressure conditions.} 
    \label{fig:2d_phi3}
\end{figure}

As expected, the global LagSAT modes and global LST modes (see for instance~\cite{alizard2007spatially,rodriguez2008instability}) have considerable differences, in particular in their growth rates and mode shapes.
However, there are some salient features pertaining to the convective form of the TS waves where the global modes exhibit similarities.
These include the spatially growing wave form along the flow direction as well as the relevancy of the least unstable/most stable modes, since these are characteristically similar modes.
Nonetheless, the LagSAT does not invoke any assumptions on the linearity of the baseflow, and the analysis is inherently transient.
In an effort to relate the linear local and global convective properties of a boundary layer, ~\cite{ehrenstein2008two} examined a transient growth mechanism triggered by the streamwise non-normality of the modes.
With regard to LagSAT, where the variance of the flow map constitutes the fluctuation energy, the total energy is given by Eq.~\ref{eq:trans-energy}.
The total energy growth for the three cases is displayed in Fig.~\ref{fig:2d_phi3}, where the favorable pressure gradient $\beta=0.1$ on the Blasius boundary layer leads to an early transient peak of energy (see the inset of Fig.~\ref{fig:2d_phi3}) before turning into a relatively stable boundary layer.
The exact roles of the modal non-normality and non-linearity in the transient growth, in the LagSAT context, needs further investigation.

\section{LagSAT on 2D Flow Past a Cylinder} \label{sec:lagsat_cyl}

Flow past a cylinder is a well studied canonical configuration, where the flow exhibits unsteady oscillations for Reynolds number, based on the freestream velocity ($u_\infty$) and cylinder diameter ($D$), higher than $Re\approx 47$.
The steady flow at lower Reynolds number becomes unsteady via an absolute instability of the wake, undergoing a supercritical Hopf bifurcation~\citep{mathis1984benard,mittal2007stabilized}.
However, starting with $Re=0$, the flow past a cylinder undergoes several critical points, exhibiting changes in the flow pattern alongside the nature of instabilities, namely convective/absolute.
The flow essentially creeps over the cylinder for $Re<4$, while forming a pair of symmetric steady recirculation regions with increasing $Re$ until loosing its steadiness at $Re=47$~\citep{mathis1984benard} to engender alternating vortex shedding, well known as Karman vortex street.
Here, we examine the flow past a cylinder through the lens of LagSAT, exploring the nature of instabilities across the transition points.

To demonstrate the use of LagSAT for compressible flows, the 2D flow past a cylinder is compressible with Mach number of $M=0.5$.
In general, the flow compressibility at Mach numbers below $0.5$ marginally affects the instabilities, in particular near the critical points~\citep{canuto2015two}.
The compressibility of the flow (at these Mach numbers) has a stabilizing effect on the flow instabilities, leading to higher values of the critical Reynolds number~\citep{canuto2015two}.
The baseflow of the compressible 2D flow past a cylinder at $M=0.5$ is obtained using previously computed Direct Numerical Simulations (DNS) dataset~\citep{shinde2021lagrangian}.
Details of the mesh and computational domain settings are provided in Appendix~C of~\citet{shinde2021lagrangian}.
Multiple simulations were performed in the Reynolds number range of $1\le Re \le 60$.
Each simulation was conducted for over $500$ time units using a time step of $0.0001$, ensuring a converged baseflow.

\begin{figure}
    \centering
    \begin{minipage}{0.48\textwidth}
    \centering    
    \includegraphics[width=0.48\linewidth]{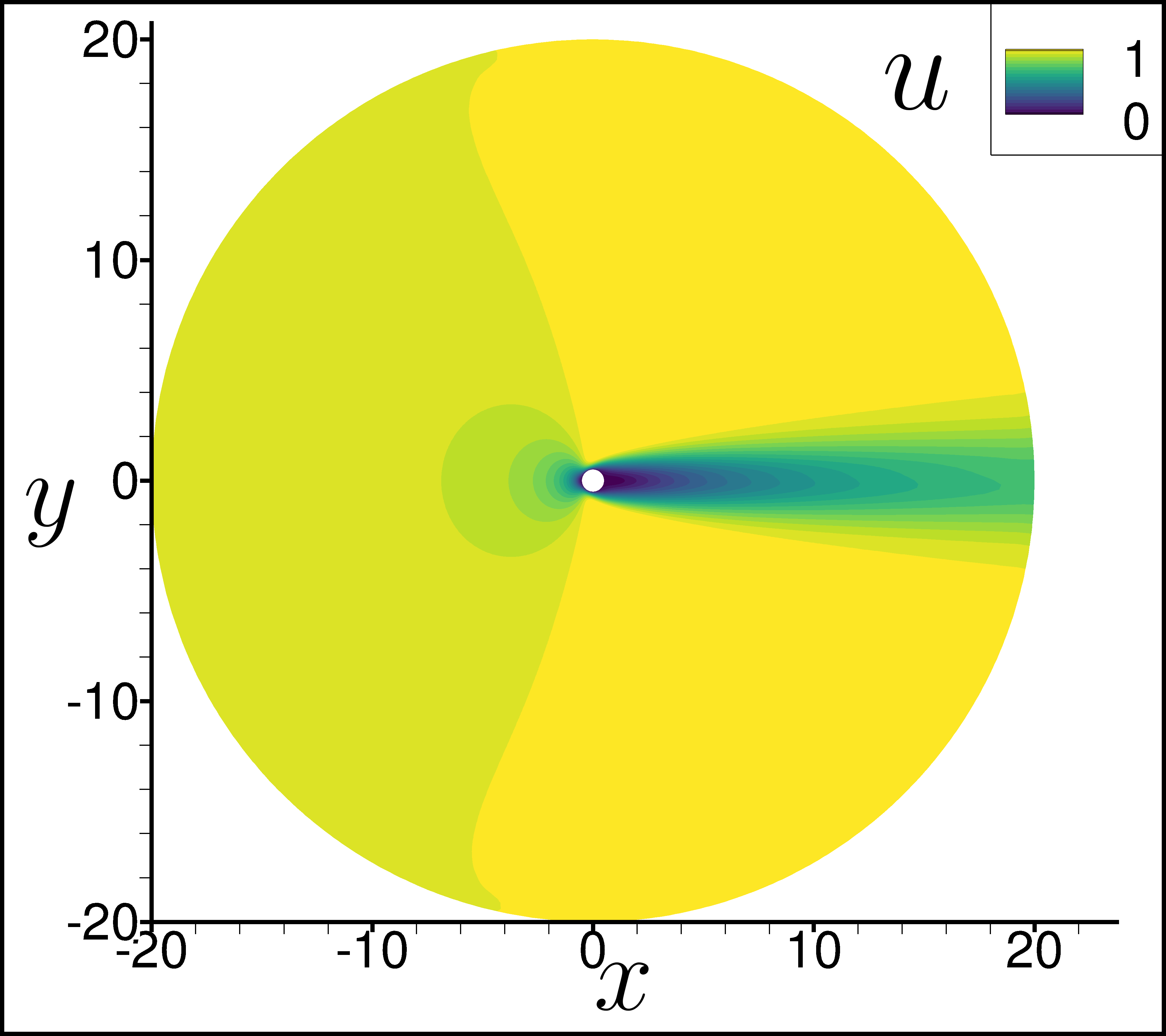} 
    \includegraphics[width=0.48\linewidth]{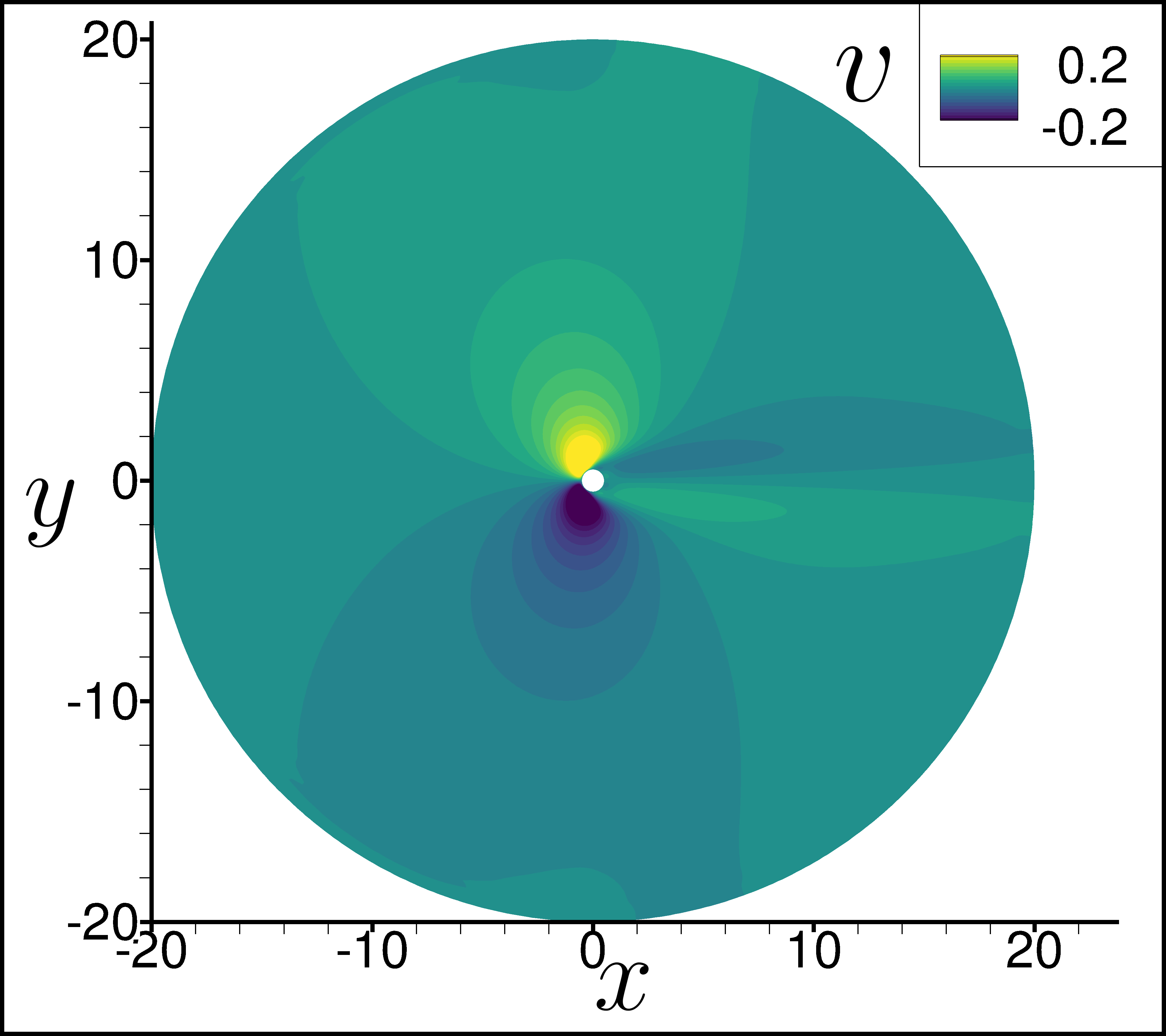}\\ (a) $u(\pmb{x},t=0); v(\pmb{x},t=0); Re_D=20$
    \end{minipage}
    \begin{minipage}{0.48\textwidth}
    \centering    
    \includegraphics[width=0.48\linewidth]{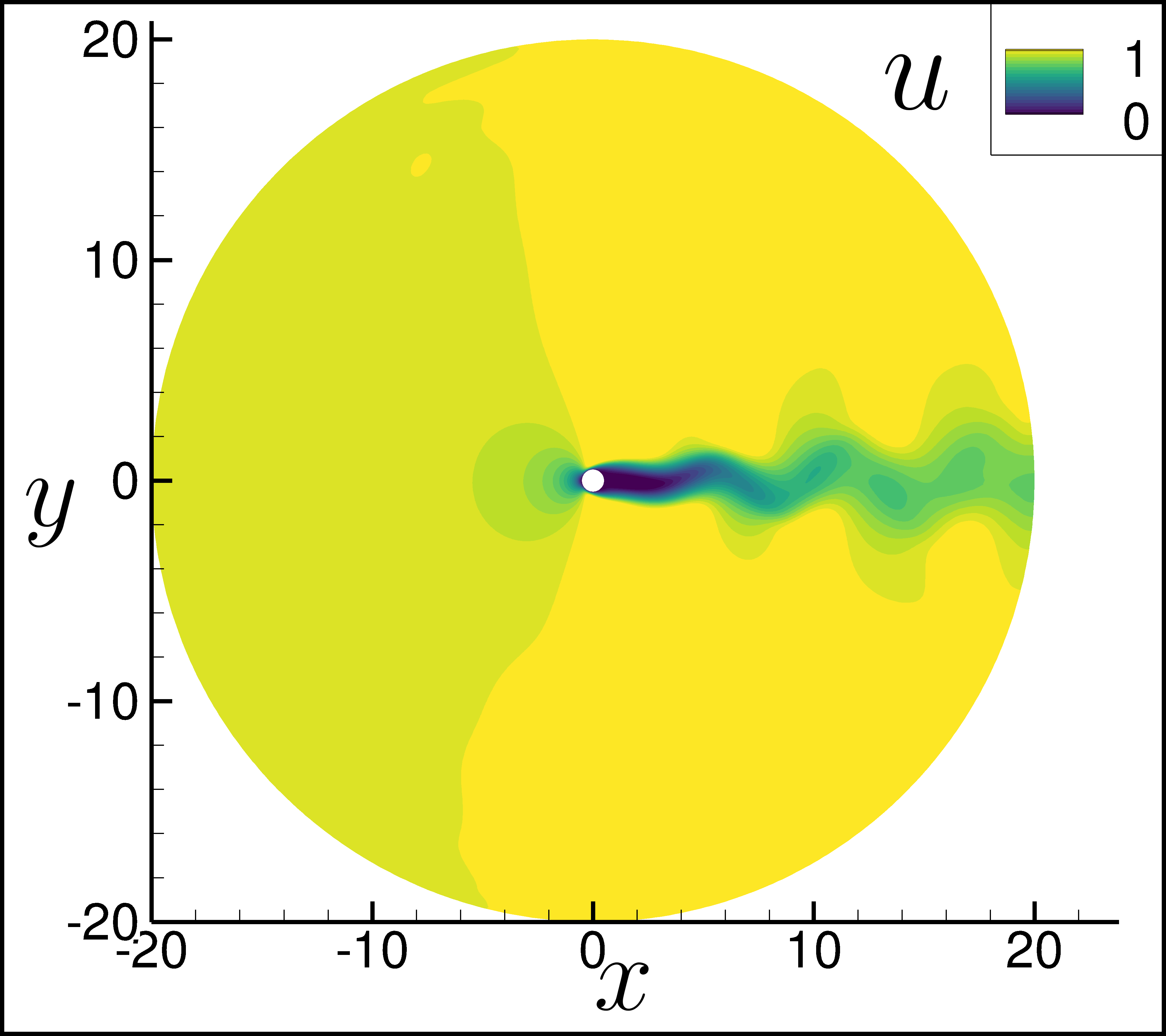}
    \includegraphics[width=0.48\linewidth]{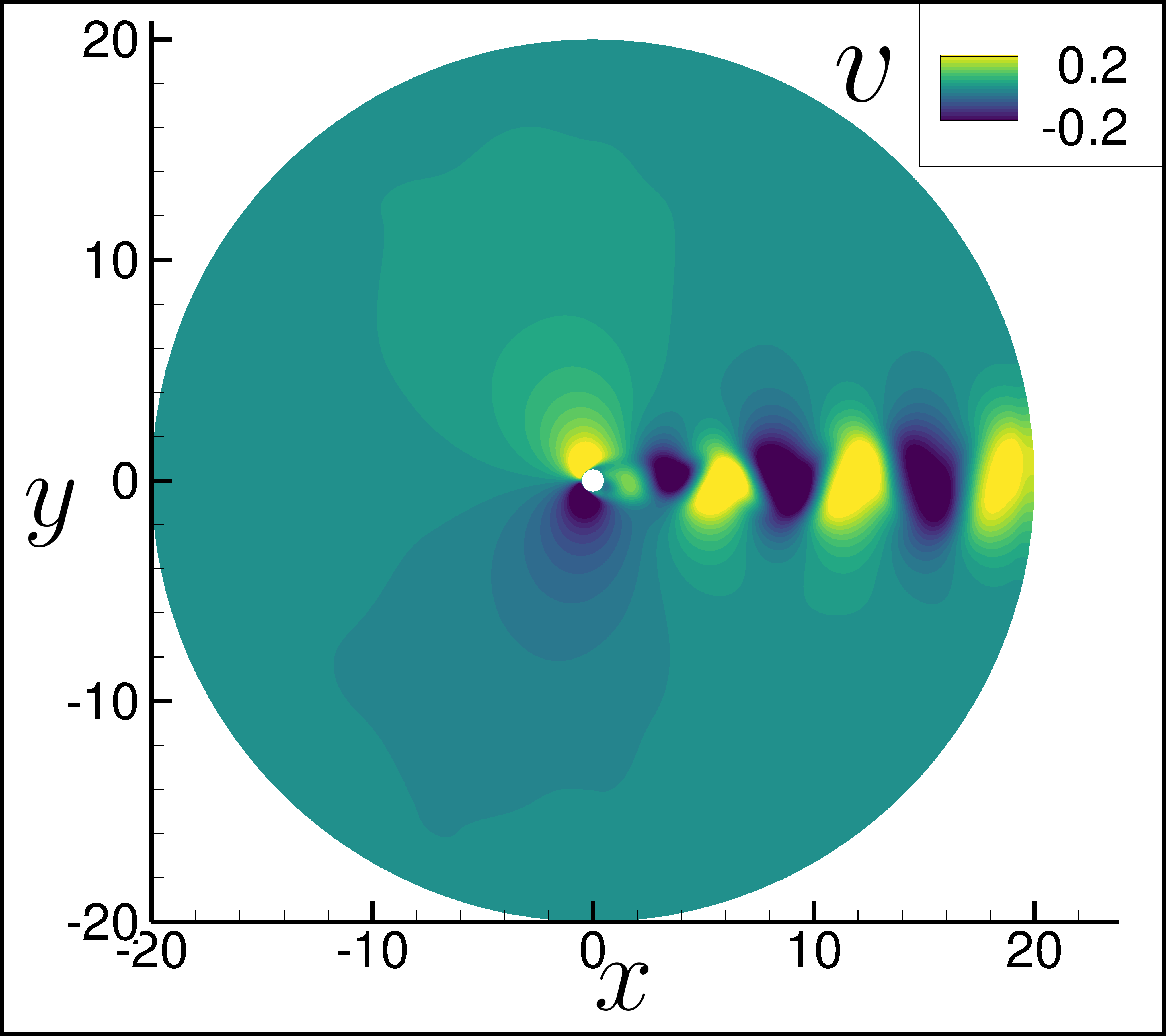} \\ (b) $u(\pmb{x},t=0); v(\pmb{x},t=0); Re_D=60$
    \end{minipage}
    \caption{Compressible flow past a cylinder at Mach $M_\infty=0.5$. Eulerian steady and unsteady baseflows at (a) $Re_D=20$ and (b) $Re_D=60$.}
    \label{fig:cyl_base}
\end{figure}

The compressible flow past a cylinder at Mach $0.5$ becomes unsteady at $Re=53$ in our simulations with a resolution of $\Delta Re=1$, exhibiting the von Karman street of vortices.
The baseflows in terms of the streamwise and cross flow velocities at $Re=20$ and $Re=60$ are displayed in Fig.~\ref{fig:cyl_base}(a) Fig.~\ref{fig:cyl_base}(b), respectively.
For a post-critical Reynolds numbers with an unsteady flow, the construction of Lagrangian flow maps requires the precomputed unsteady flow snapshots of the DNS.
This is one of the salient features of LMA~\citep{shinde2021lagrangian}, where the analysis can be extended across the transition points to trace the specific Lagrangian coherent features of the flow.
LagSAT inherits this feature which allows stability analysis of the post-critical flow states, tracing the genesis of instability/unsteadiness in the pre-critical regime.

\begin{figure}
    \centering
    \begin{minipage}{0.24\textwidth}
    \centering    
    \includegraphics[width=1.0\linewidth]{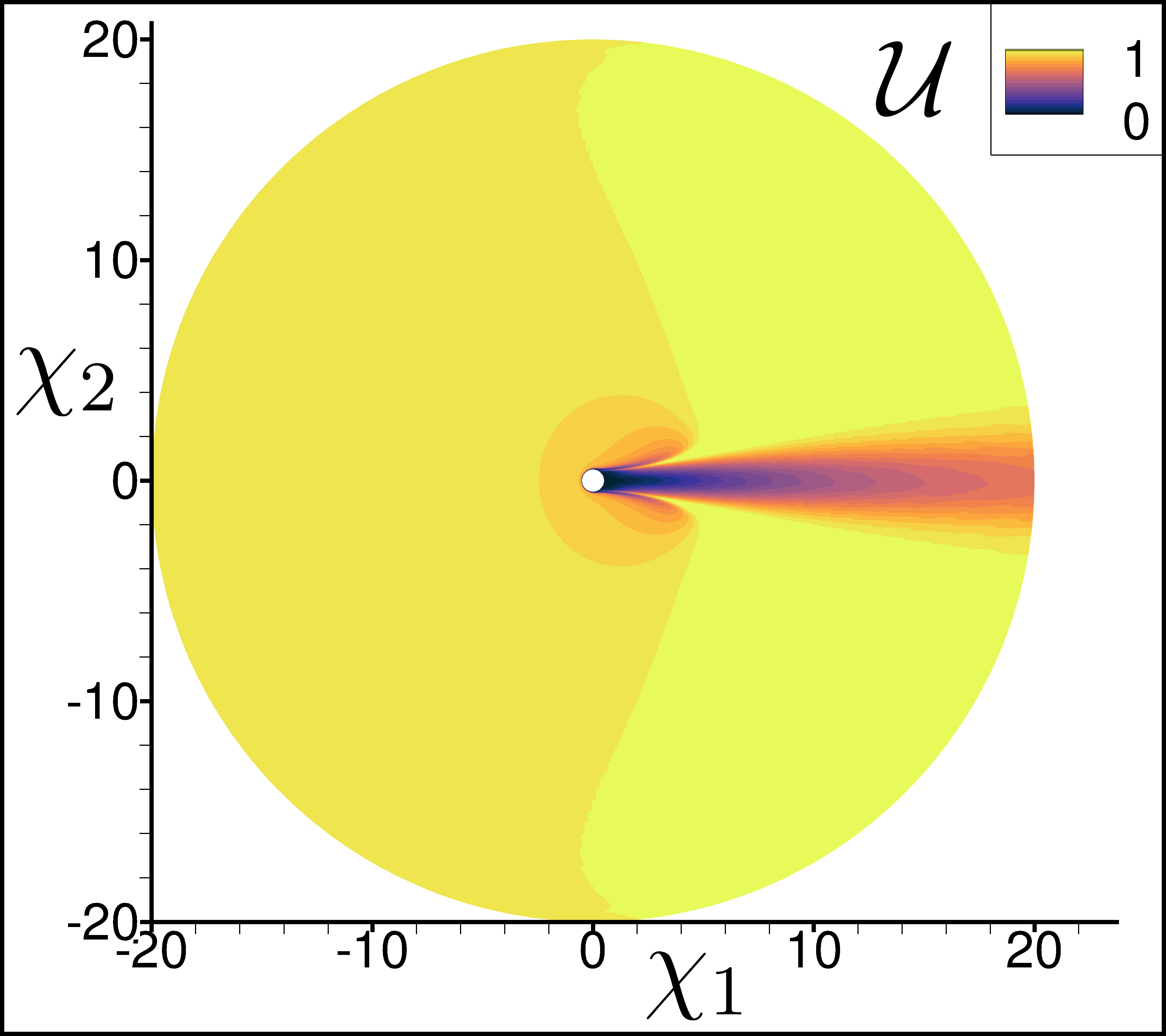} \\ (a) $\mathcal{U}(\pmb{\chi},t=-5)$
    \end{minipage}
    \begin{minipage}{0.24\textwidth}
    \centering    
    \includegraphics[width=1.0\linewidth]{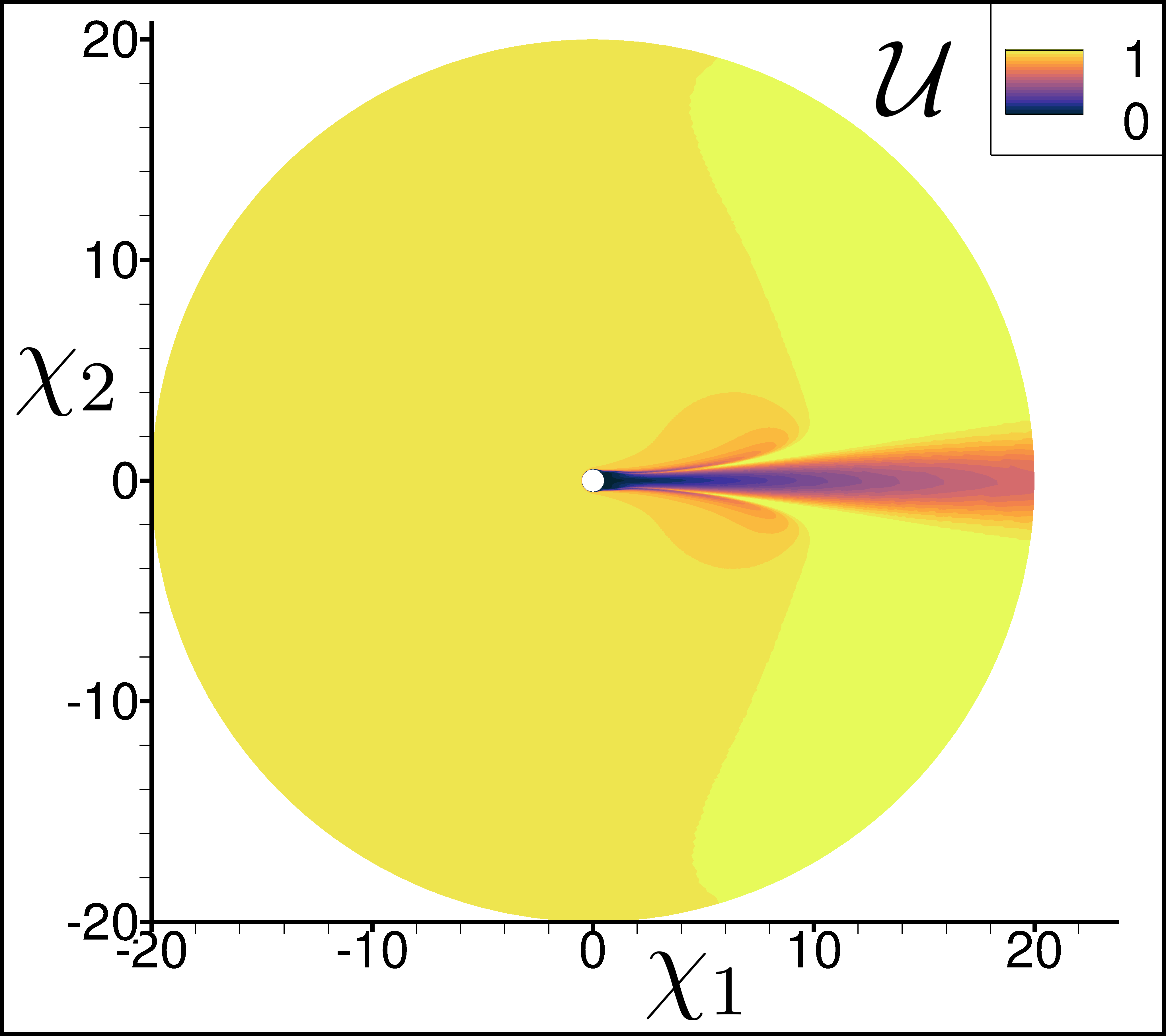} \\ (b) $\mathcal{U}(\pmb{\chi},t=-10)$
    \end{minipage}
    \begin{minipage}{0.24\textwidth}
    \centering    
    \includegraphics[width=1.0\linewidth]{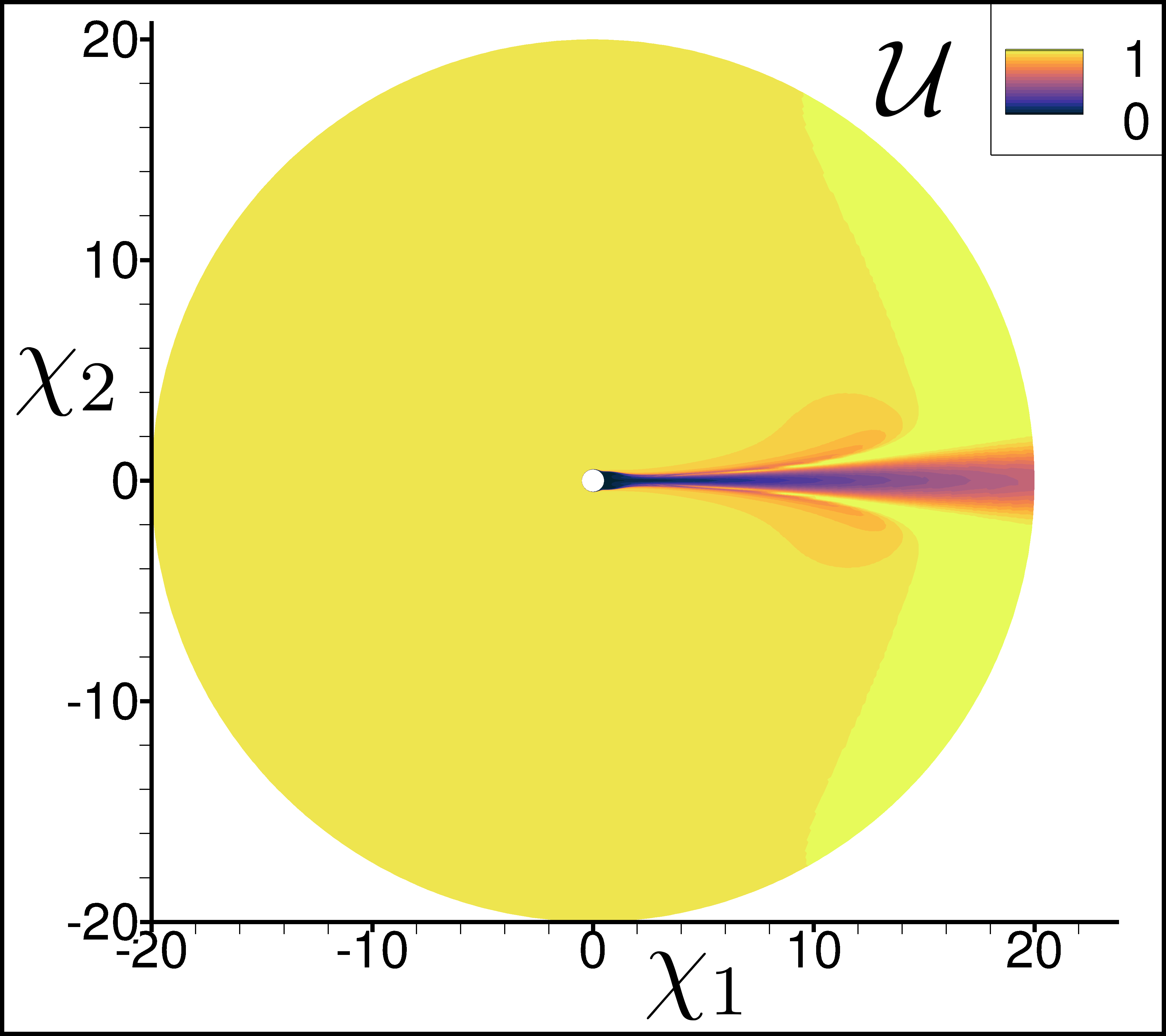} \\ (c) $\mathcal{U}(\pmb{\chi},t=-15)$
    \end{minipage}
    \begin{minipage}{0.24\textwidth}
    \centering    
    \includegraphics[width=1.0\linewidth]{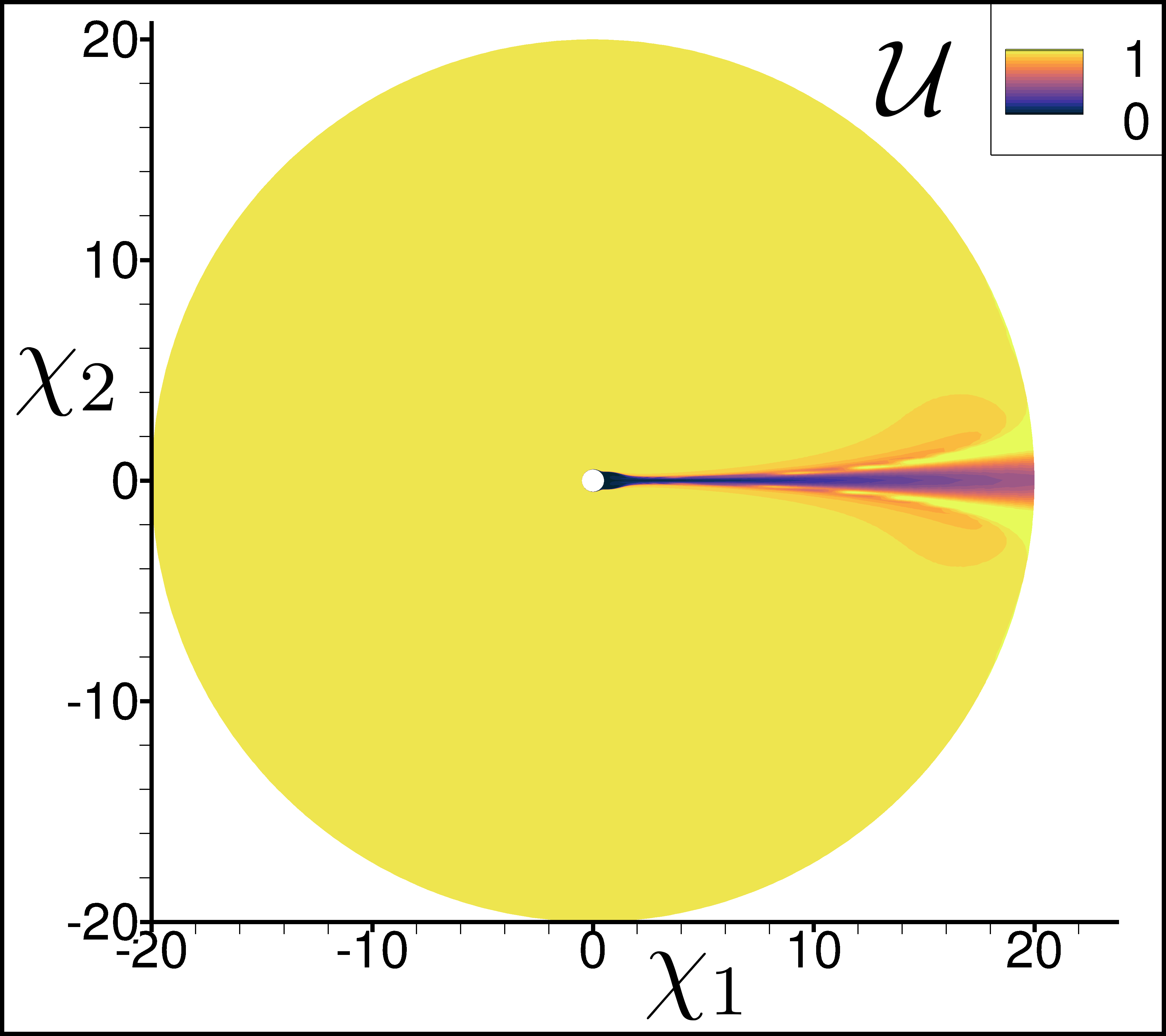} \\ (d) $\mathcal{U}(\pmb{\chi},t=-20)$
    \end{minipage}
    \begin{minipage}{0.24\textwidth}
    \centering    
    \includegraphics[width=1.0\linewidth]{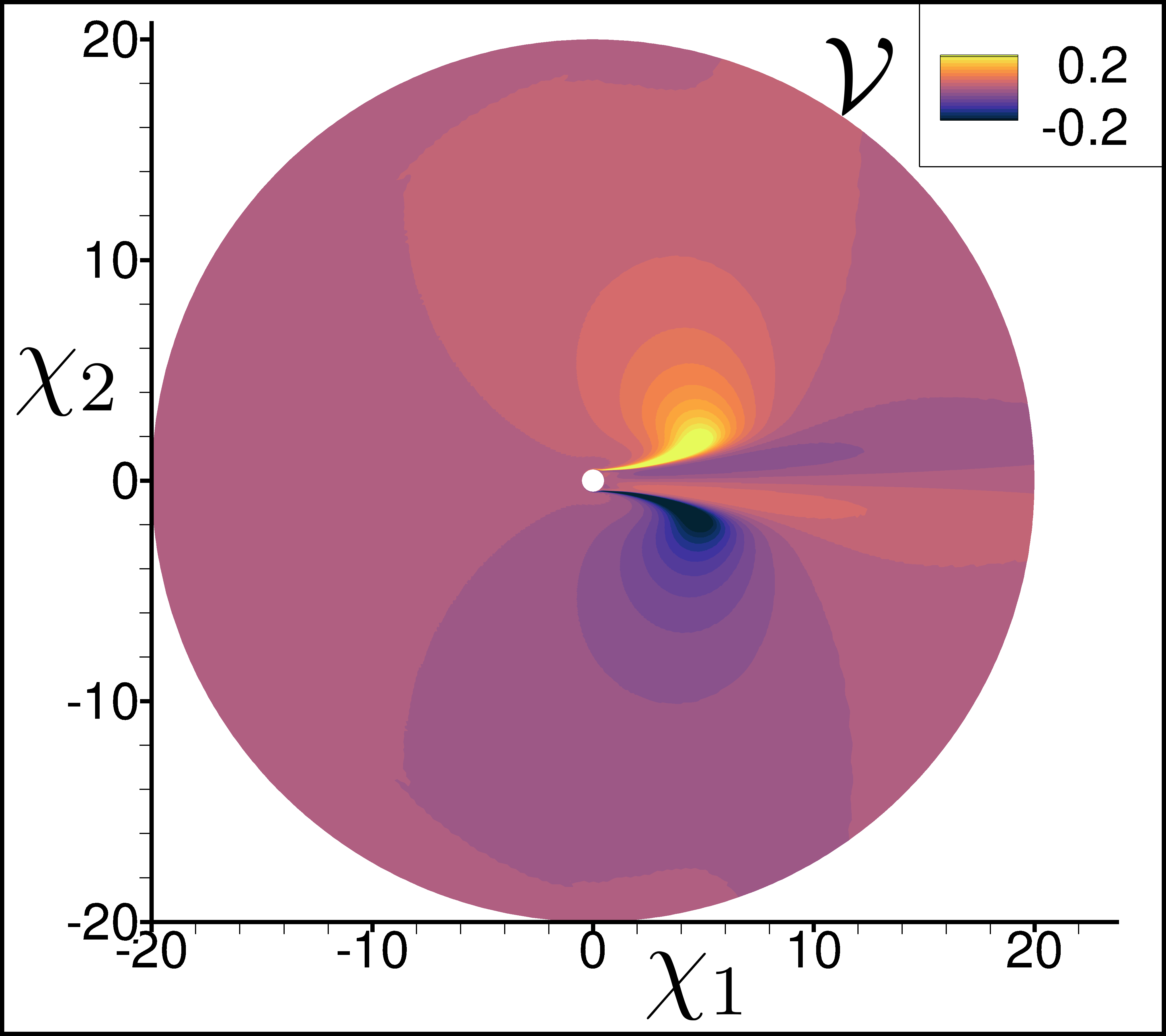} \\ (e) $\mathcal{V}(\pmb{\chi},t=-5)$
    \end{minipage}
    \begin{minipage}{0.24\textwidth}
    \centering    
    \includegraphics[width=1.0\linewidth]{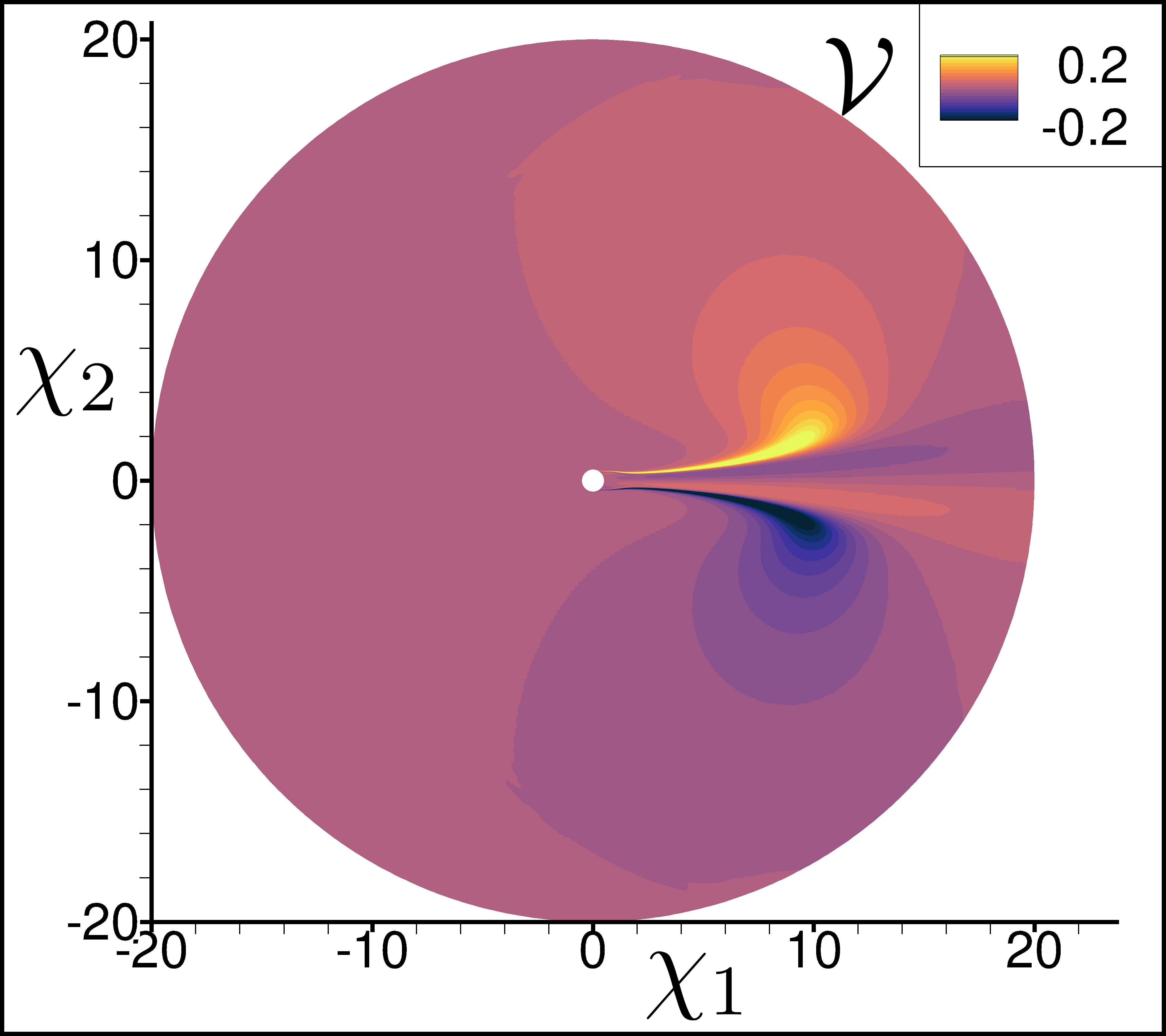} \\ (f) $\mathcal{V}(\pmb{\chi},t=-10)$
    \end{minipage}
    \begin{minipage}{0.24\textwidth}
    \centering    
    \includegraphics[width=1.0\linewidth]{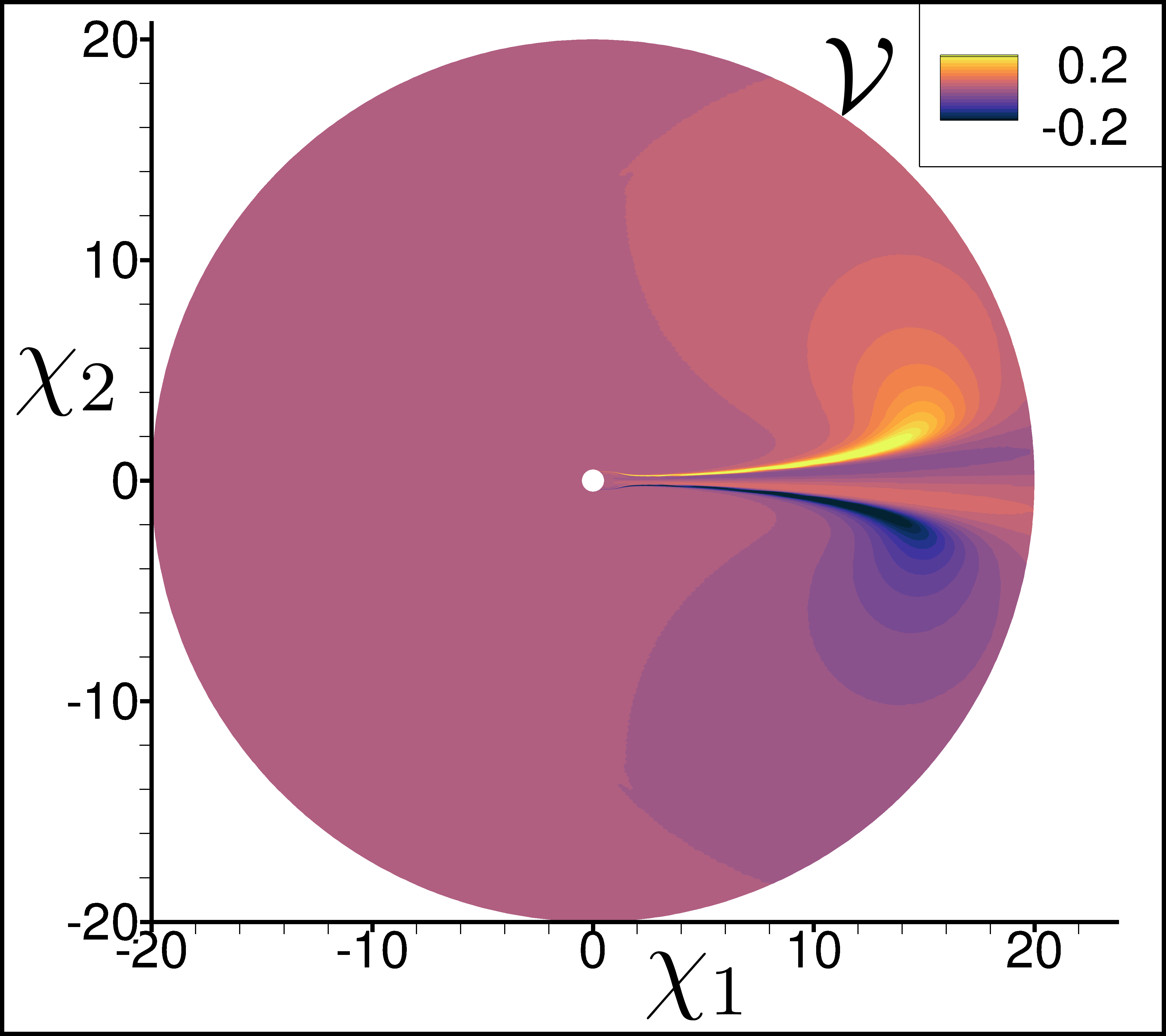} \\ (g) $\mathcal{V}(\pmb{\chi},t=-15)$
    \end{minipage}
    \begin{minipage}{0.24\textwidth}
    \centering    
    \includegraphics[width=1.0\linewidth]{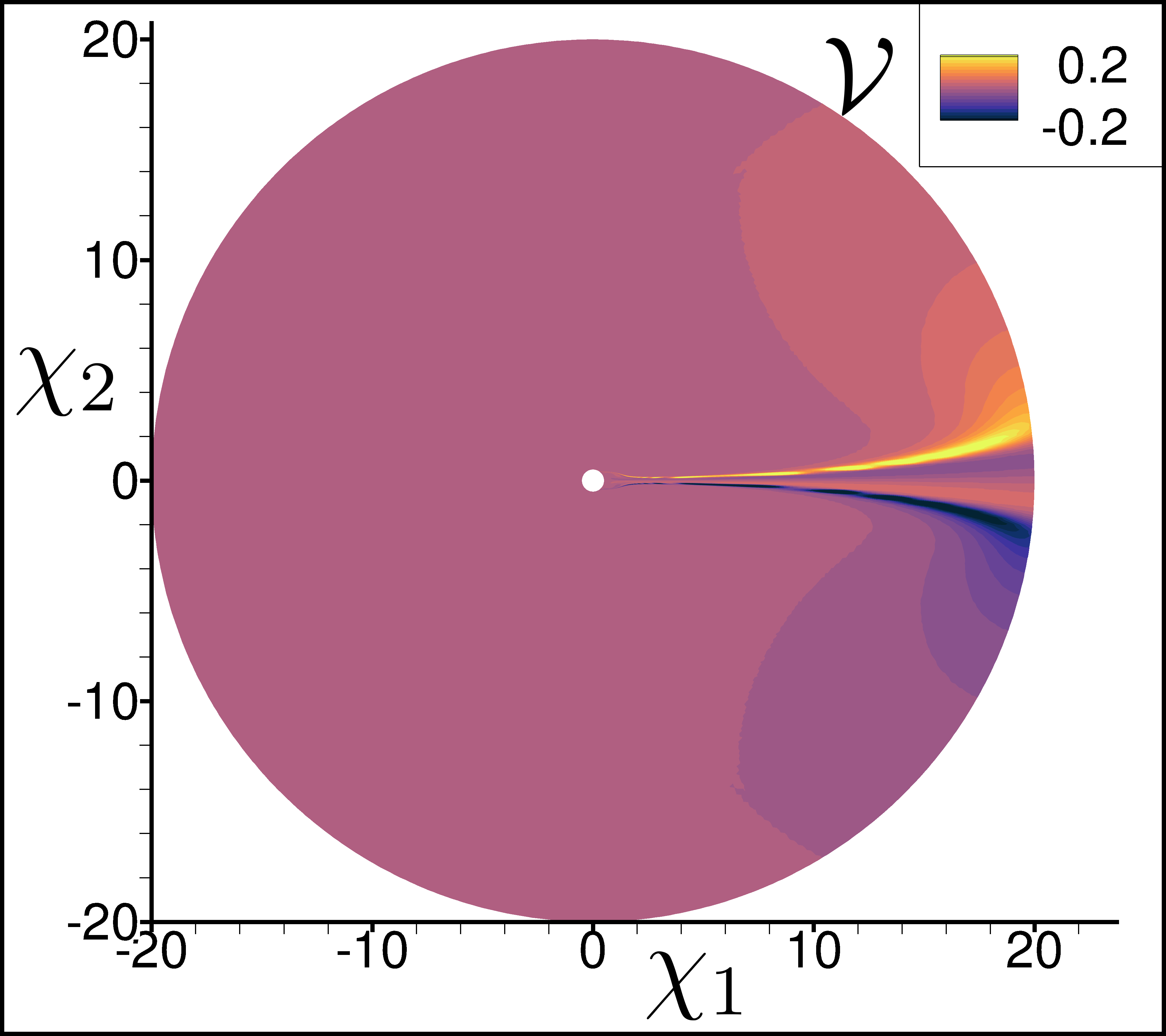} \\ (h) $\mathcal{V}(\pmb{\chi},t=-20)$
    \end{minipage}
    \caption{Lagrangian flow map (top row - streamwise velocity; bottom row - crossflow velocity) of compressible flow past a cylinder at Mach $M_\infty=0.5$ and $Re=20$.}
    \label{fig:cyl_lagmap}
\end{figure}

The Lagrangian flow map of the flow past a cylinder at $M=0.5$ and $Re=20$ is displayed in Fig.~\ref{fig:cyl_lagmap}, in terms of the Lagrangian streamwise (top row) and crossflow (bottom row) velocities.
A total of $400$ time instances at $0.1$ time step constitute the flow map, including the identity map of Fig.~\ref{fig:cyl_base}(a).
The choice of number of snapshots and the time step accounts for the computational domain extent in the flow direction (or the streamwise $x$ diction), which is $40$.
Despite this is an open flow, the uniform inflow boundary condition and the adjoint form of LagSAT do not pose any difficulty in accessing the flow outside the inflow boundary location.
Figure~\ref{fig:cyl_lagmap} displays four instances of the Lagrangian flow map at $t=-20$ (Fig.~\ref{fig:cyl_lagmap}d, h), $t=-15$ (Fig.~\ref{fig:cyl_lagmap}c, g), $t=-10$ (Fig.~\ref{fig:cyl_lagmap}b, f), and $t=-5$ (Fig.~\ref{fig:cyl_lagmap}a, e), depicting the time evolution of the flow map into the identity map at $t=0$ (Fig.~\ref{fig:cyl_base}a).
In addition to the flow velocities, the Lagrangian flow maps can be build for the co-variables, namely, the density, the pressure, and the temperature or for the passive-variables, such as the species concentration~\citep{shinde2021lagrangian}.
The LagSAT can be employed on the co/passive-variables to extract the stability characteristics of their dynamics.

\begin{figure}
    \centering
    \begin{minipage}{1.0\textwidth}
    \centering
    \includegraphics[width=1.0\linewidth]{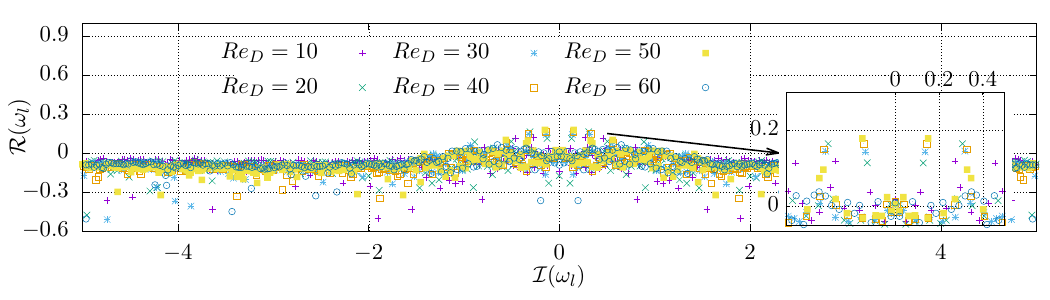}\\(a) Eigen spectra of $\mathcal{U}(\pmb{\chi})$
    \end{minipage}
    \begin{minipage}{1.0\textwidth}
    \centering
    \includegraphics[width=1.0\linewidth]{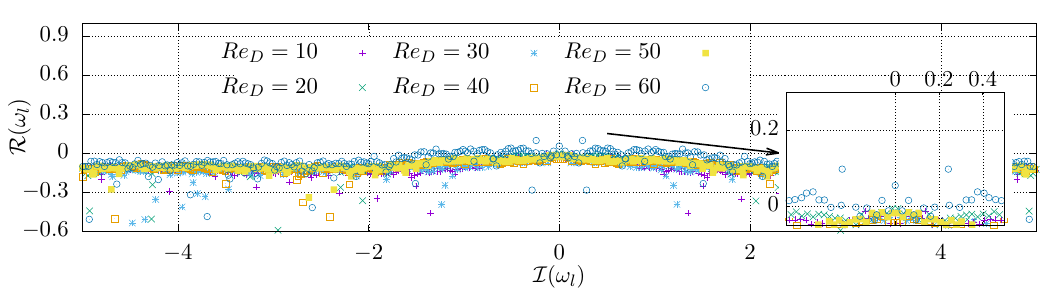}\\(b) Eigen spectra of $\mathcal{V}(\pmb{\chi})$
    \end{minipage}
    \caption{LagSAT eigenvalue spectra of (a) streamwise and (b) crossflow velocities.}
    \label{fig:cyl_ritz}
\end{figure}

The LagSAT eigenvalue spectra of the streamwise and crossflow velocity components are shown in Fig.~\ref{fig:cyl_ritz} for the Reynolds number increasing from $Re=10$ to $Re=60$ at an increment of $10$.
The modal frequencies and the growth rates are marked on, respectively, $x$ and $y$ axes, for the streamwise velocity in Fig.~\ref{fig:cyl_ritz}(a) and crossflow in Fig.~\ref{fig:cyl_ritz}(b).
Over the entire range of Reynolds number $10 \leq Re \leq 60$, the higher frequency modes ($St\gtrapprox 2$) have negative growth rates or stable responses for the both velocity components.
The low-frequency end of the spectra ($St\lessapprox 2$) have plenty of unstable modes at all Reynolds numbers for the streamwise velocity.
The inset of Fig.~\ref{fig:cyl_ritz}(a) provides a close view of the unstable eigenvalues, where three distinct branches of unstable modes are evident.
The first branch, in $0.1 < \omega_i < 0.2$, clearly shows that the modal growth rate increases with the increase of Reynolds number from $Re=10$ to $Re=50$, while the modal growth rate drastically drops for the post-critical value of $Re=60$.
The second and third branches have modal frequencies about $St\approx 0.3$ and $St\approx 0.45$, respectively, but the modal growth rates are not in a consistent order.
Notably, the LagSAT on all baseflows ($10 \leq Re \leq 60$) for the streamwise velocity engenders unstable modes, despite the fact that the flow is steady for $Re<53$.

The LagSAT eigenvalue spectra of the crossflow velocity of Fig.~\ref{fig:cyl_ritz}(b) displays several modes in the low-frequency end of the spectra ($St\lessapprox 2$) that have positive growth rate.
However, all the unstable modes belong to the post-critical Reynolds number of $Re=60$.
The growth rates of the LagSAT modes associated with the pre-critical Reynolds numbers ($10 \leq Re \leq 50$) are all negative.
The inset of Fig.~\ref{fig:cyl_ritz}(b) provides a closer view of the eigenvalues near the zero (neutral) growth rate line, clearly displaying that only the modes of baseflow with $Re=60$ are unstable.
The most unstable mode (with a highest growth rate) is a distinct mode at $St=0.243$.
Clearly, the unstable LagSAT modes of the crossflow velocity for $Re=60$ relate to the fact that the flow is absolute unstable.
Conversely, at lower Reynolds numbers, $10 \leq Re \leq 50$, all LagSAT modes of the crossflow velocity are stable.

\begin{figure}
    \centering
    \begin{minipage}{0.19\textwidth}
        \centering
        \includegraphics[width=1.0\linewidth]{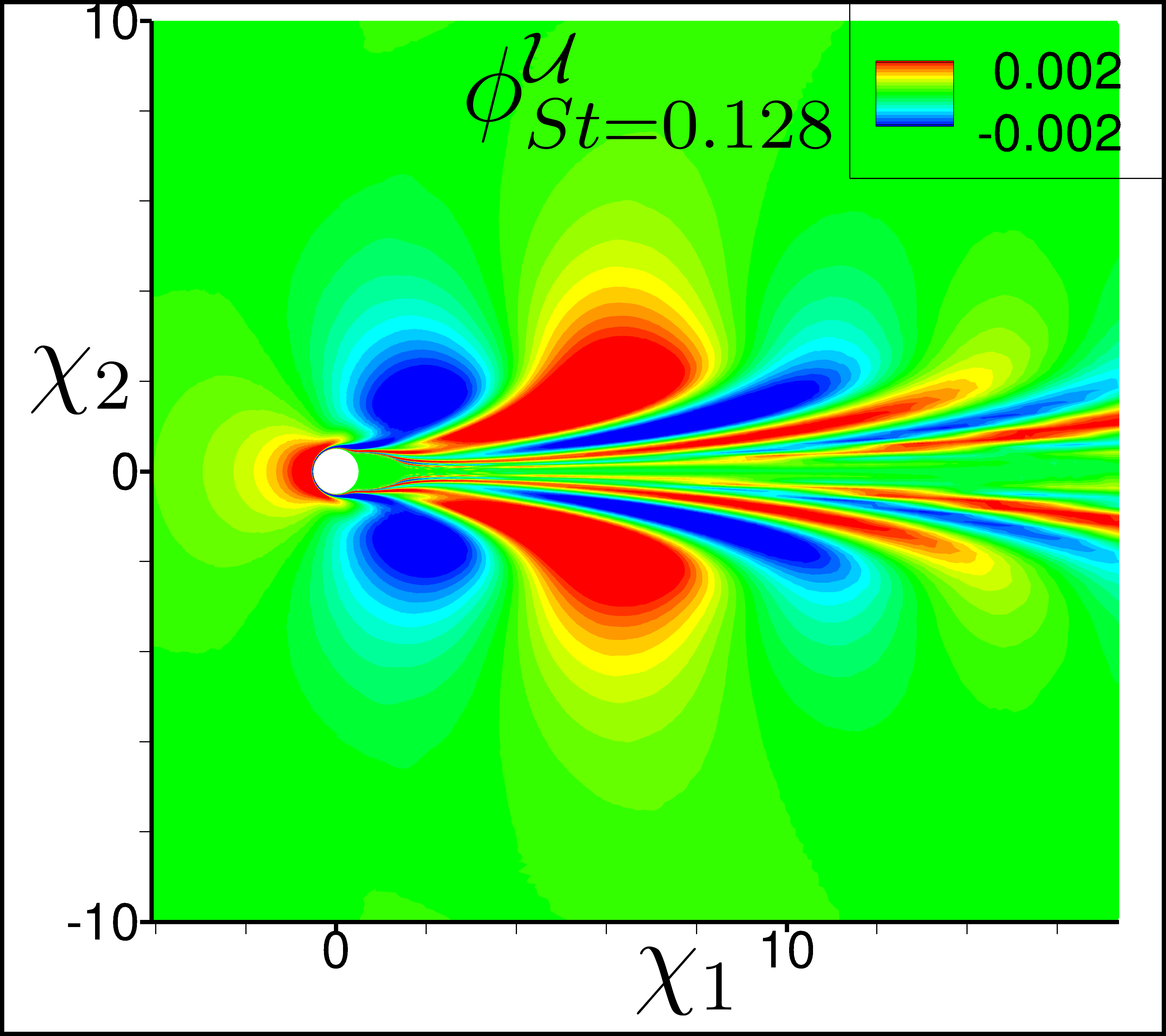}\\ (a)$Re=20; St=0.128$
    \end{minipage}
    \begin{minipage}{0.19\textwidth}
        \centering
        \includegraphics[width=1.0\linewidth]{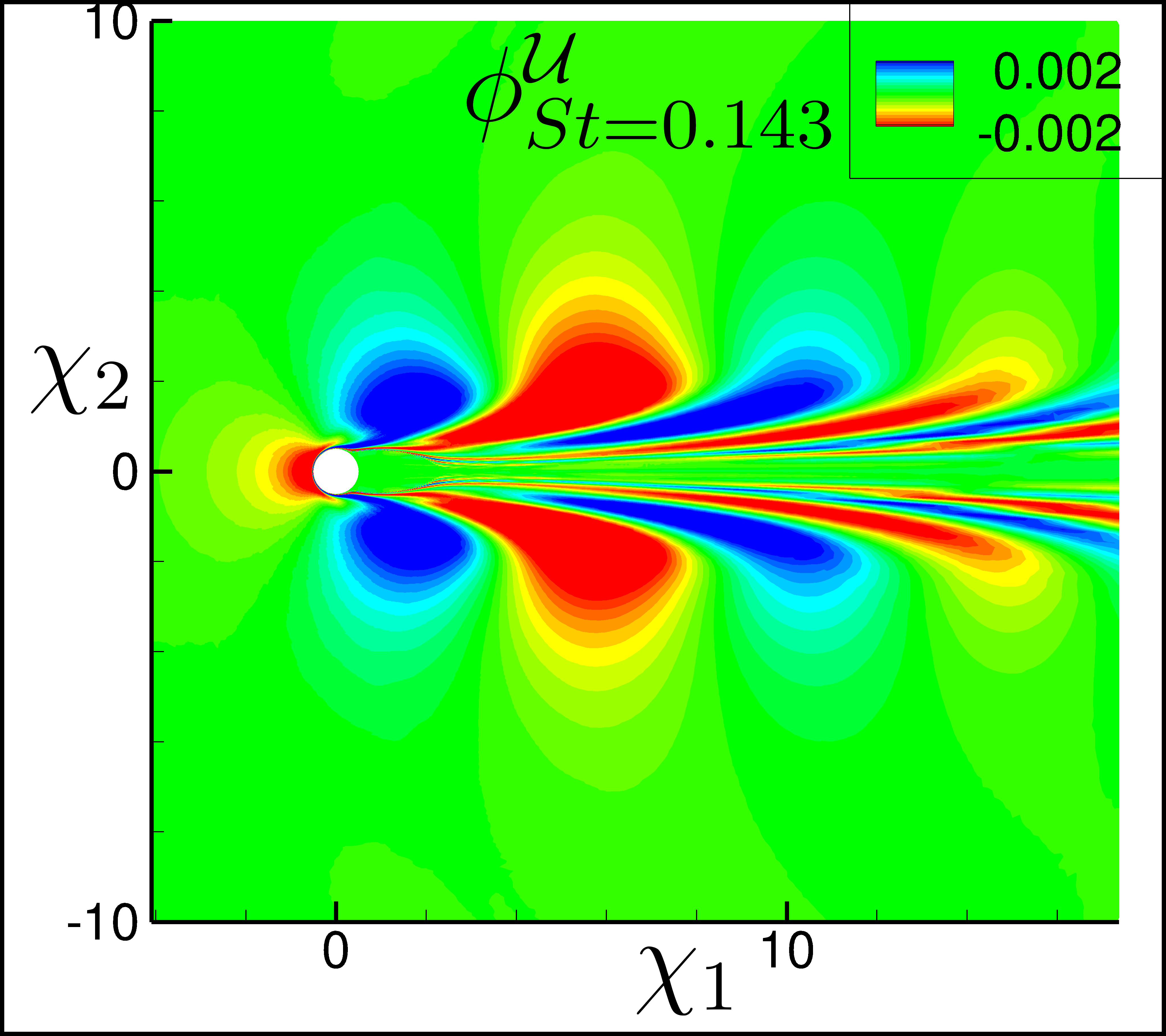}\\ (b) $Re=30; St=0.143$
    \end{minipage}
    \begin{minipage}{0.19\textwidth}
        \centering
        \includegraphics[width=1.0\linewidth]{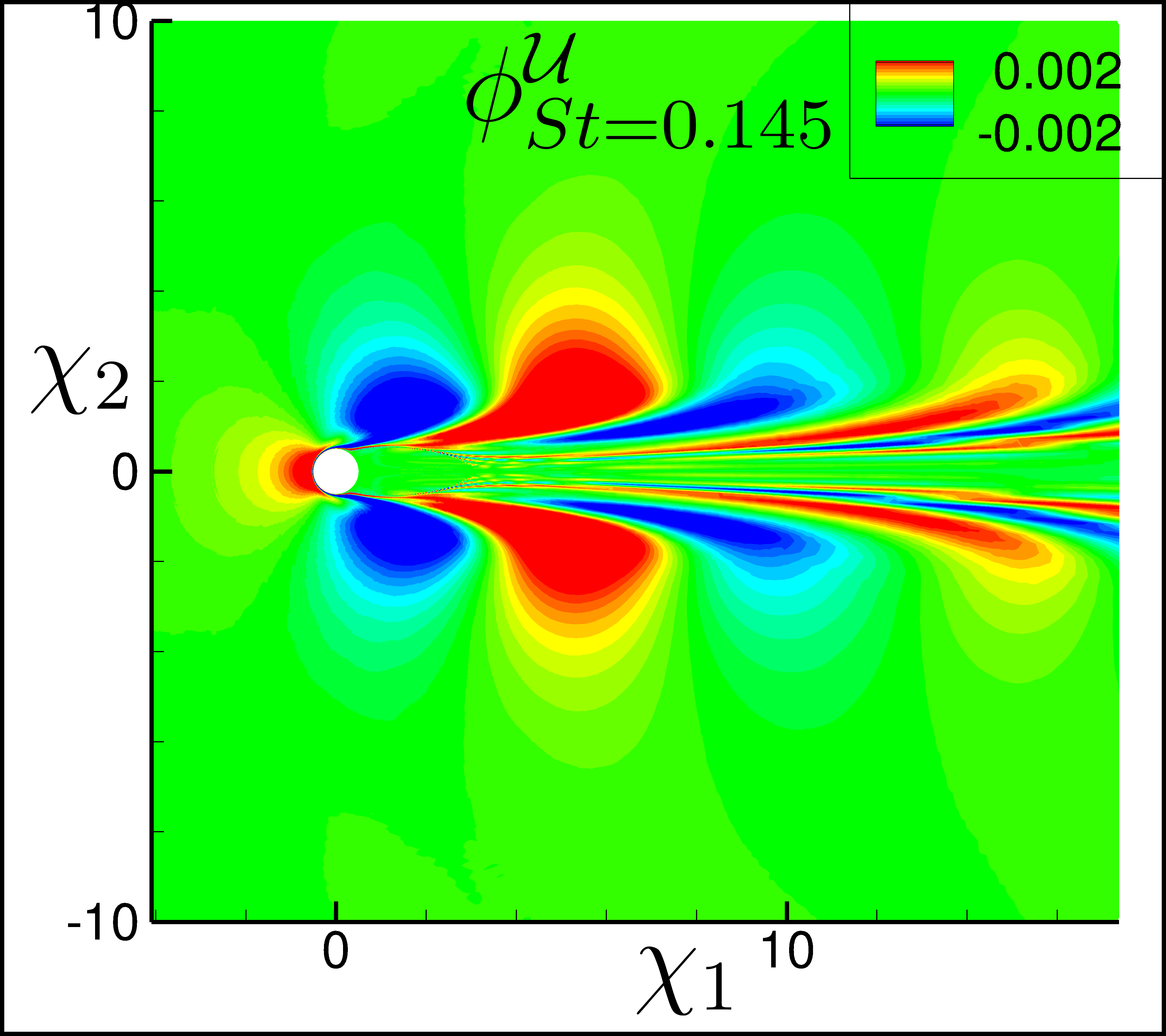}\\ (c) $Re=40; St=0.145$
    \end{minipage}
    \begin{minipage}{0.19\textwidth}
        \centering
        \includegraphics[width=1.0\linewidth]{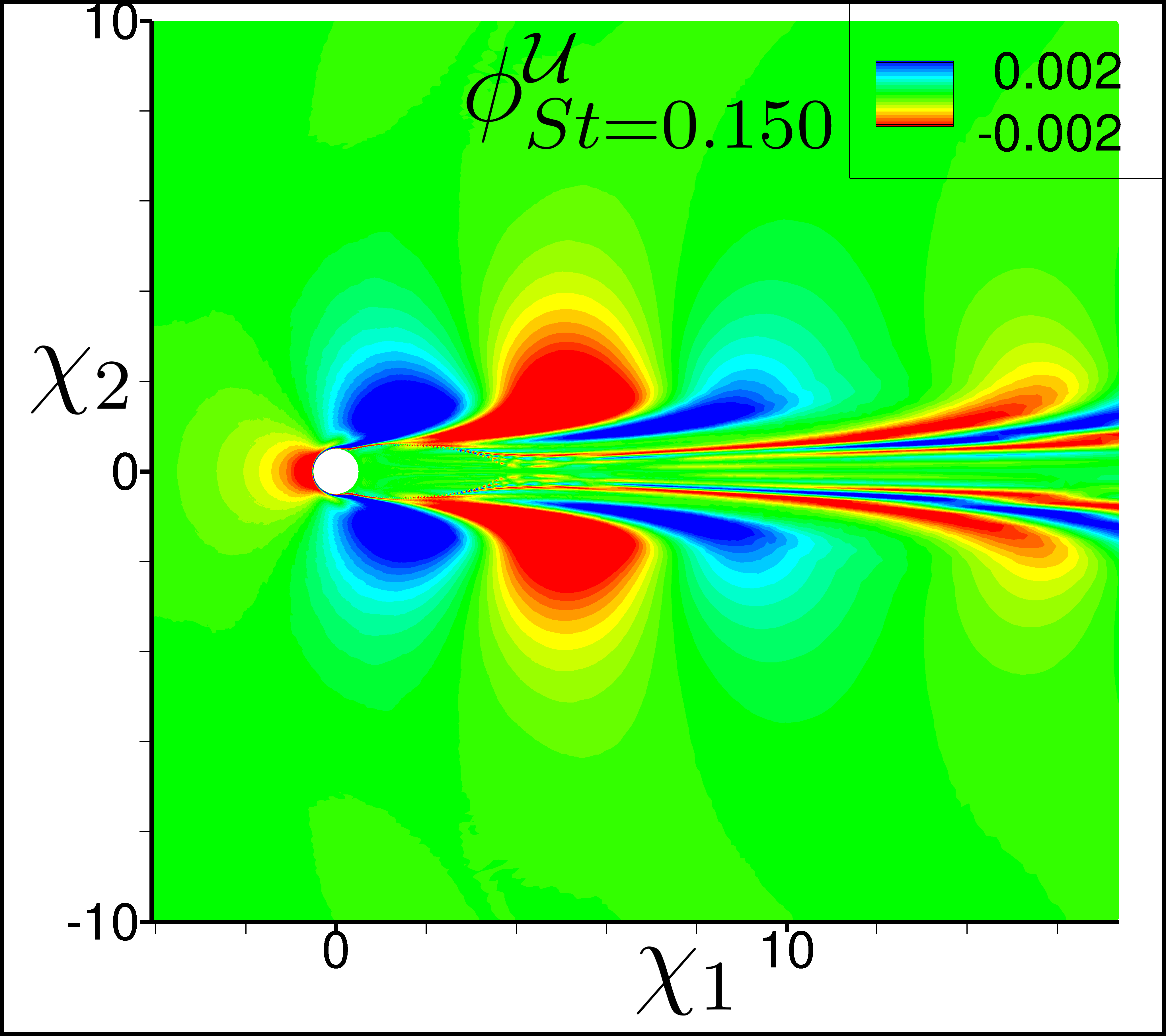}\\ (d) $Re=50; St=0.15$
    \end{minipage}
    \begin{minipage}{0.19\textwidth}
        \centering
        \includegraphics[width=1.0\linewidth]{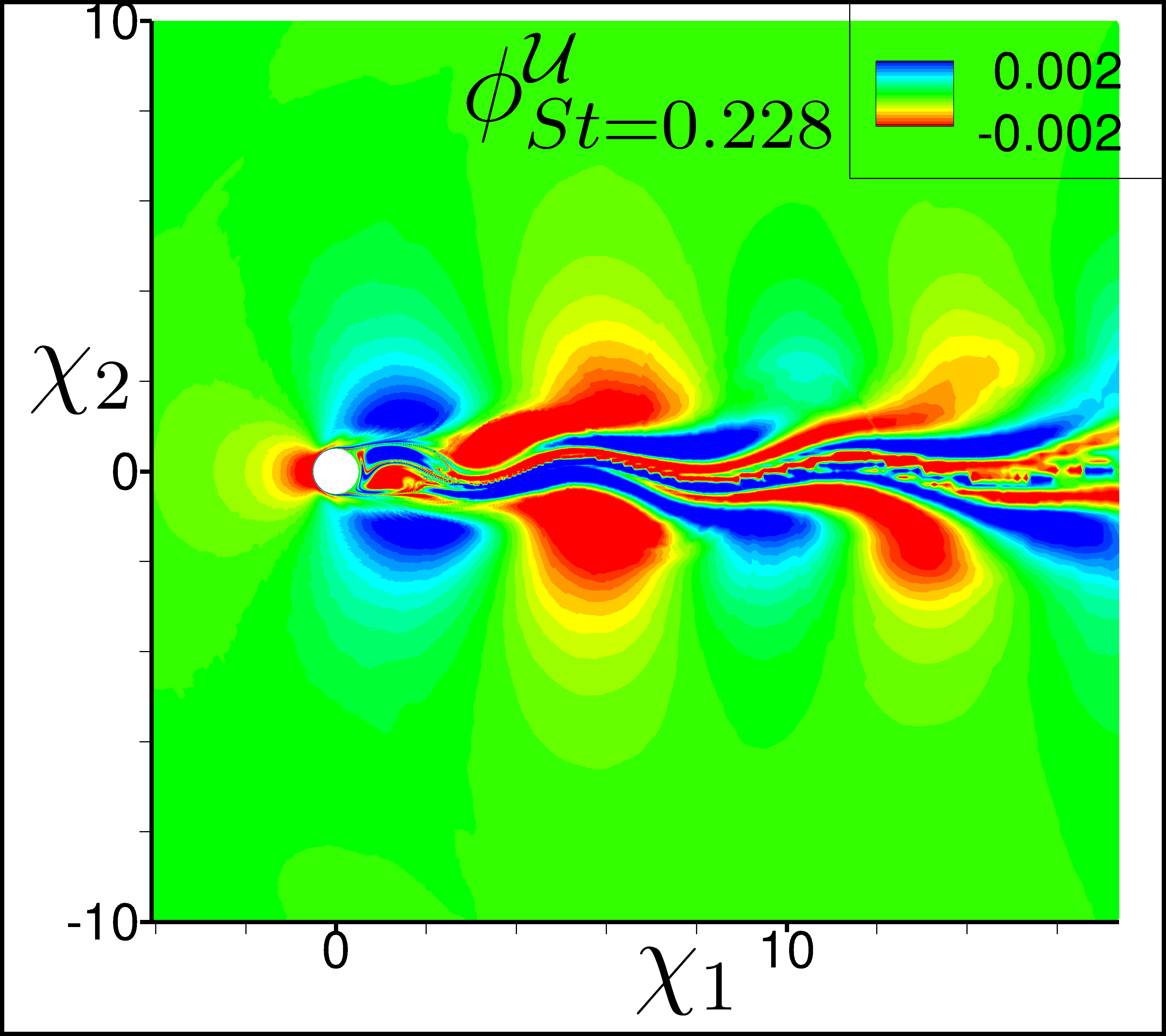}\\ (e) $Re=60; St=0.228$
    \end{minipage}
    \begin{minipage}{0.19\textwidth}
        \centering
        \includegraphics[width=1.0\linewidth]{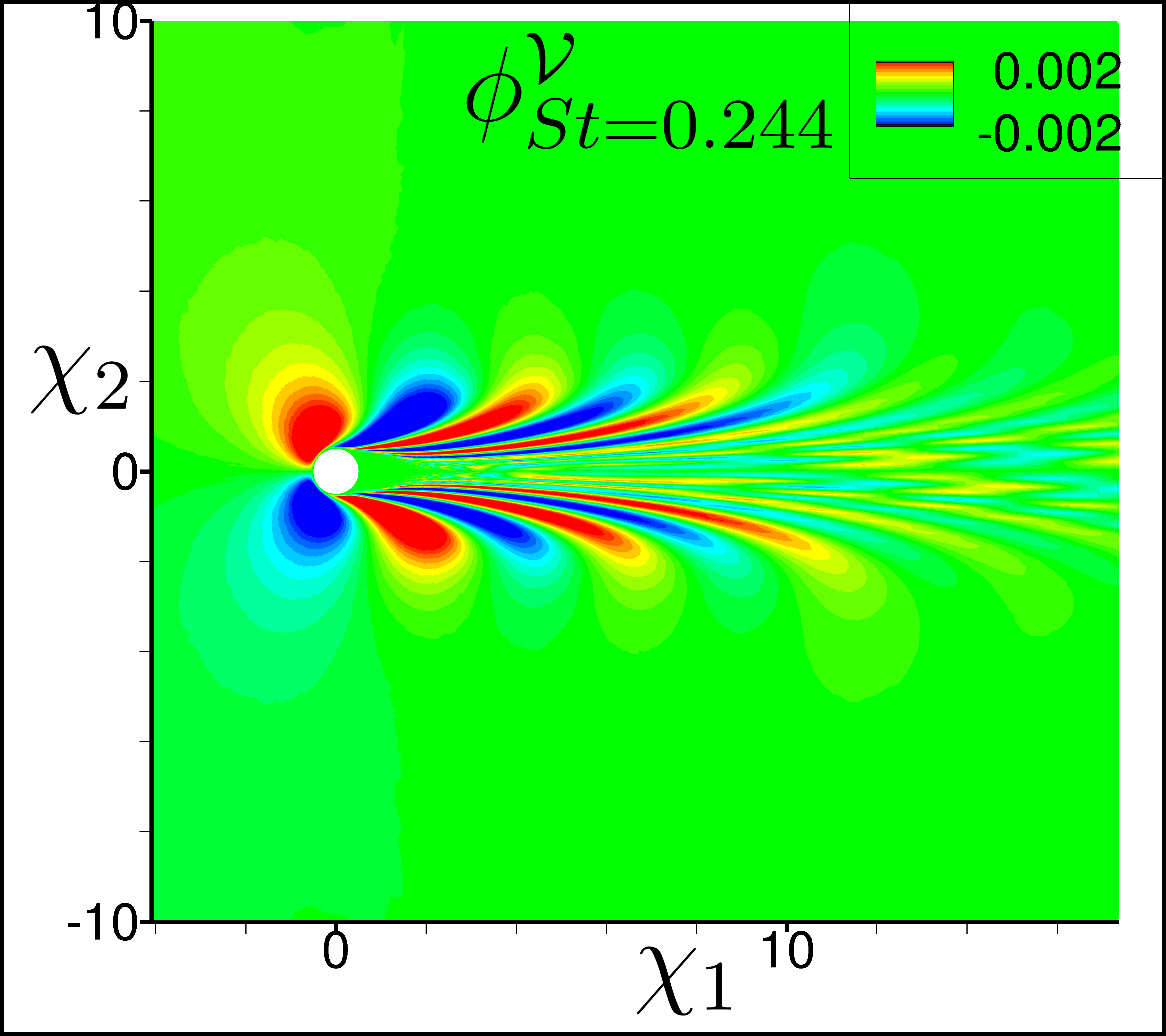}\\ (f) $Re=20; St=0.244$
    \end{minipage}
    \begin{minipage}{0.19\textwidth}
        \centering
        \includegraphics[width=1.0\linewidth]{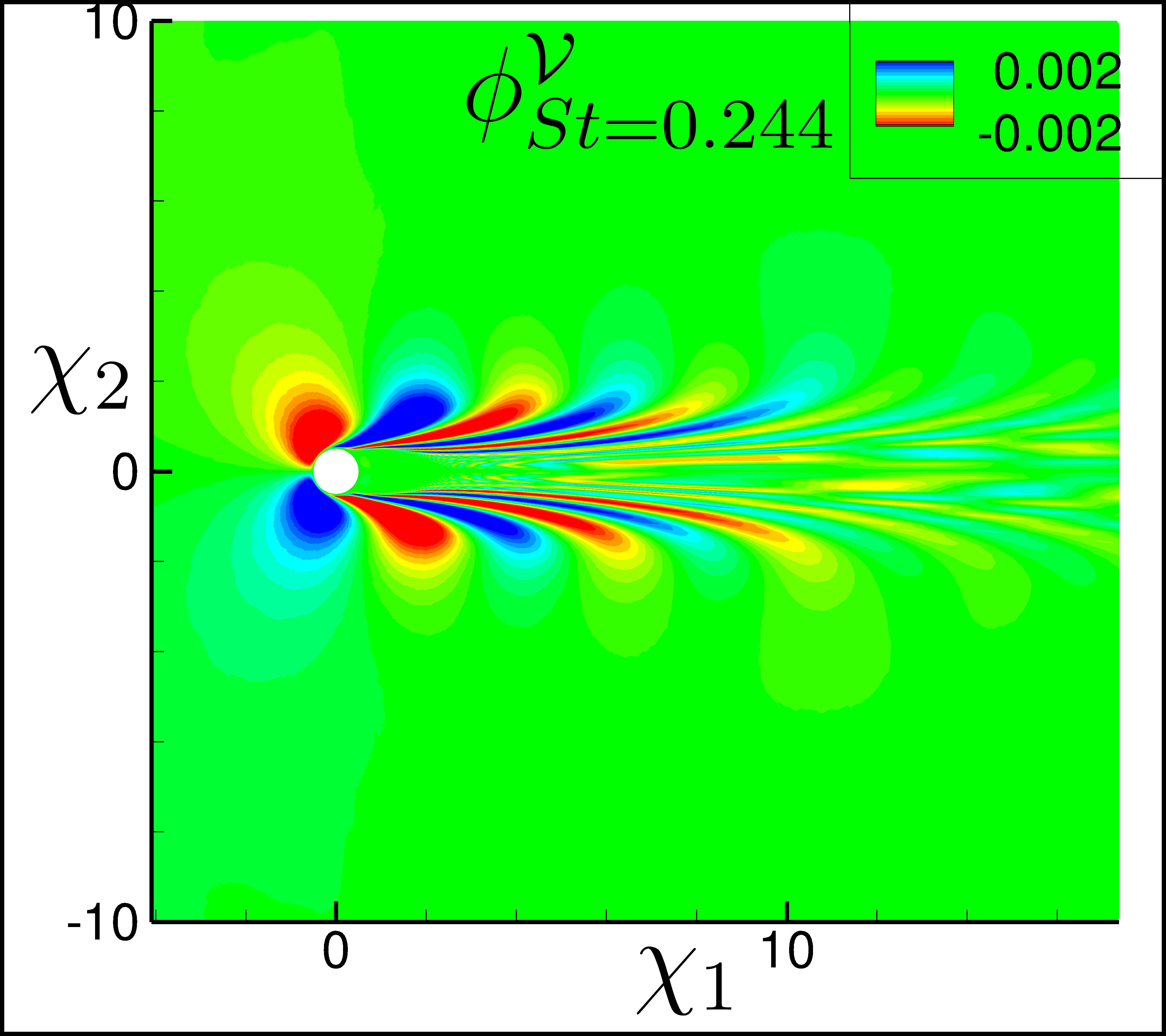}\\ (g) $Re=30; St=0.244$
    \end{minipage}
    \begin{minipage}{0.19\textwidth}
        \centering
        \includegraphics[width=1.0\linewidth]{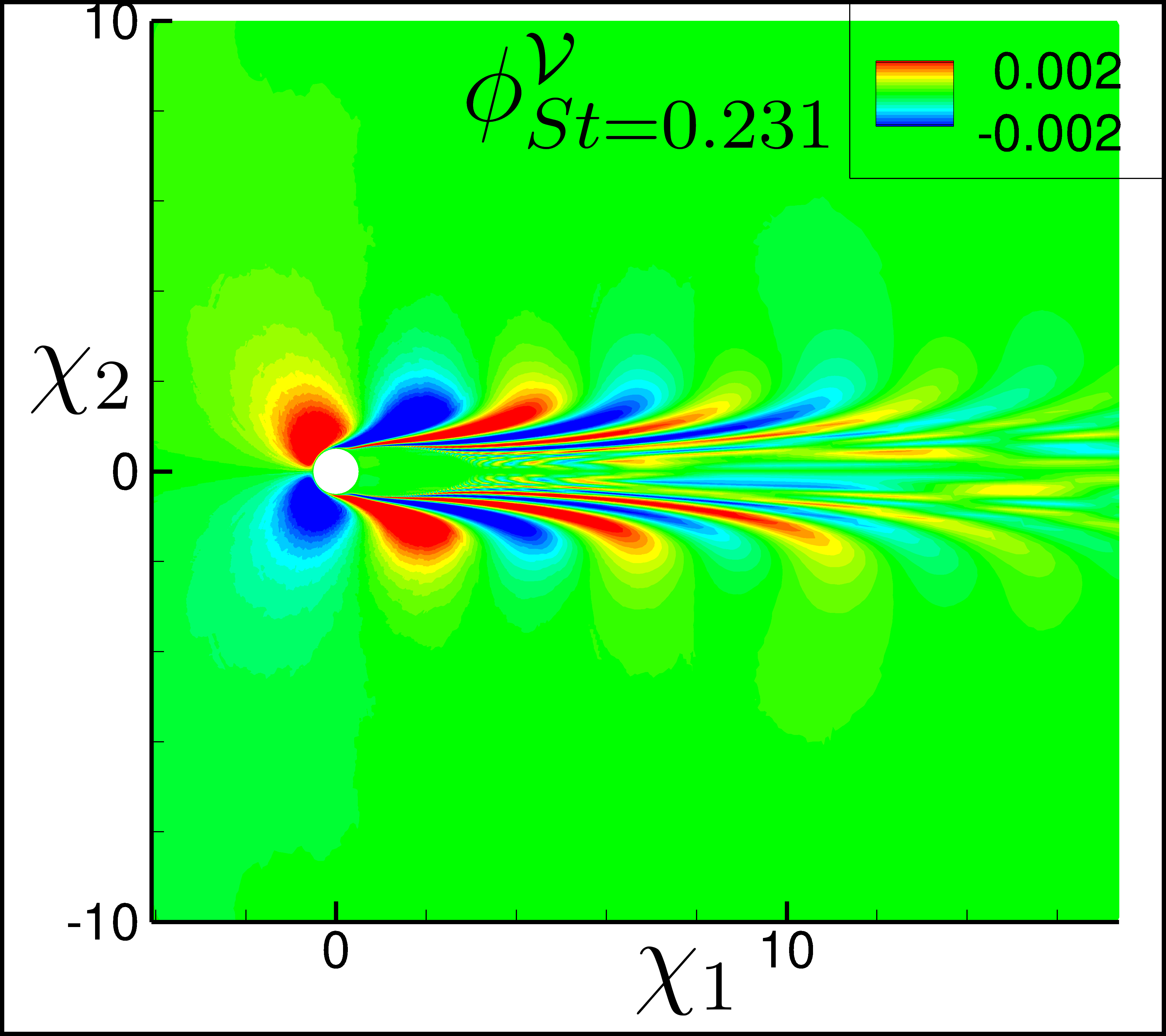}\\ (h) $Re=40; St=0.231$
    \end{minipage}
    \begin{minipage}{0.19\textwidth}
        \centering
        \includegraphics[width=1.0\linewidth]{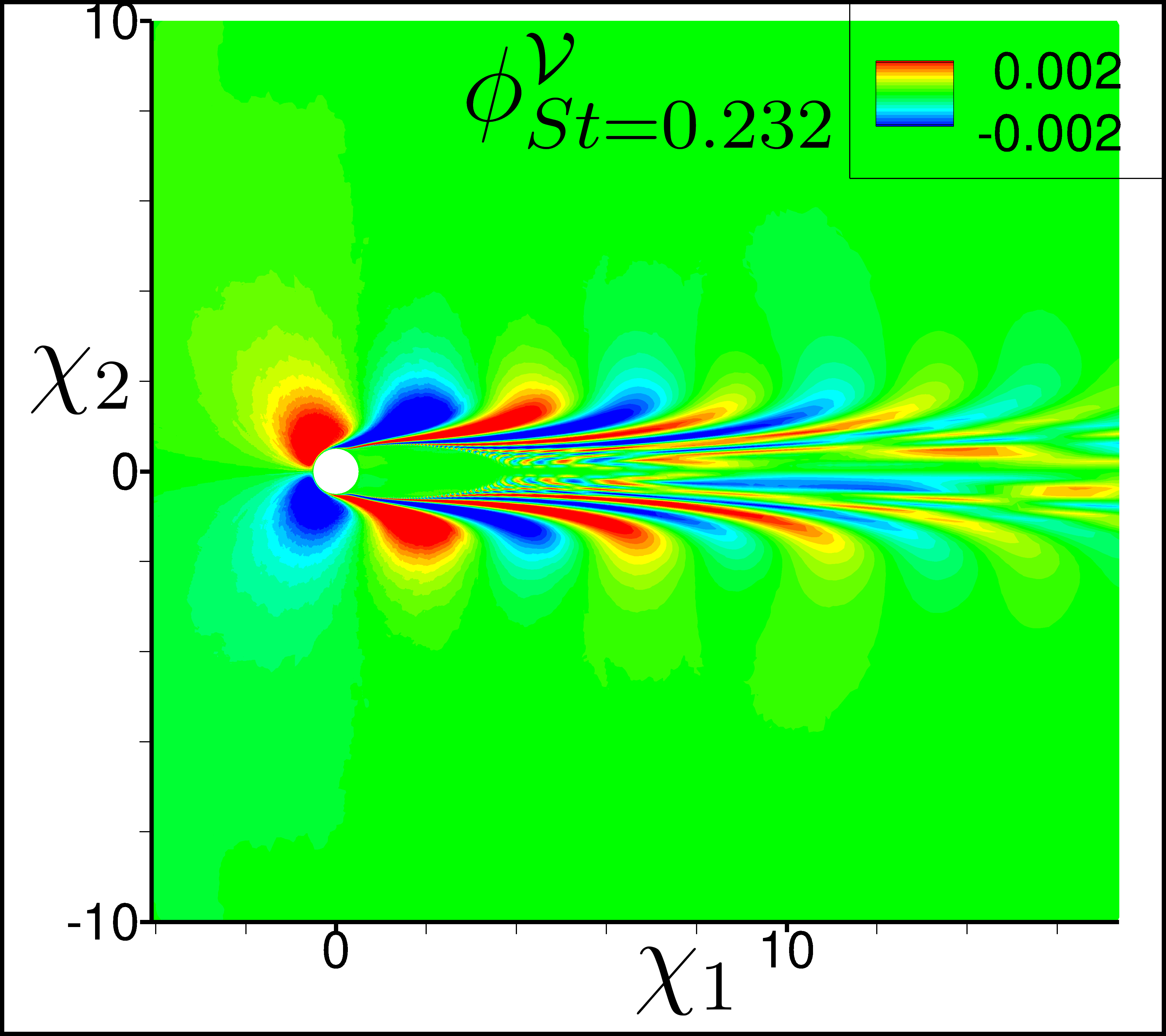}\\ (i) $Re=50; St=0.232$
    \end{minipage}
    \begin{minipage}{0.19\textwidth}
        \centering
        \includegraphics[width=1.0\linewidth]{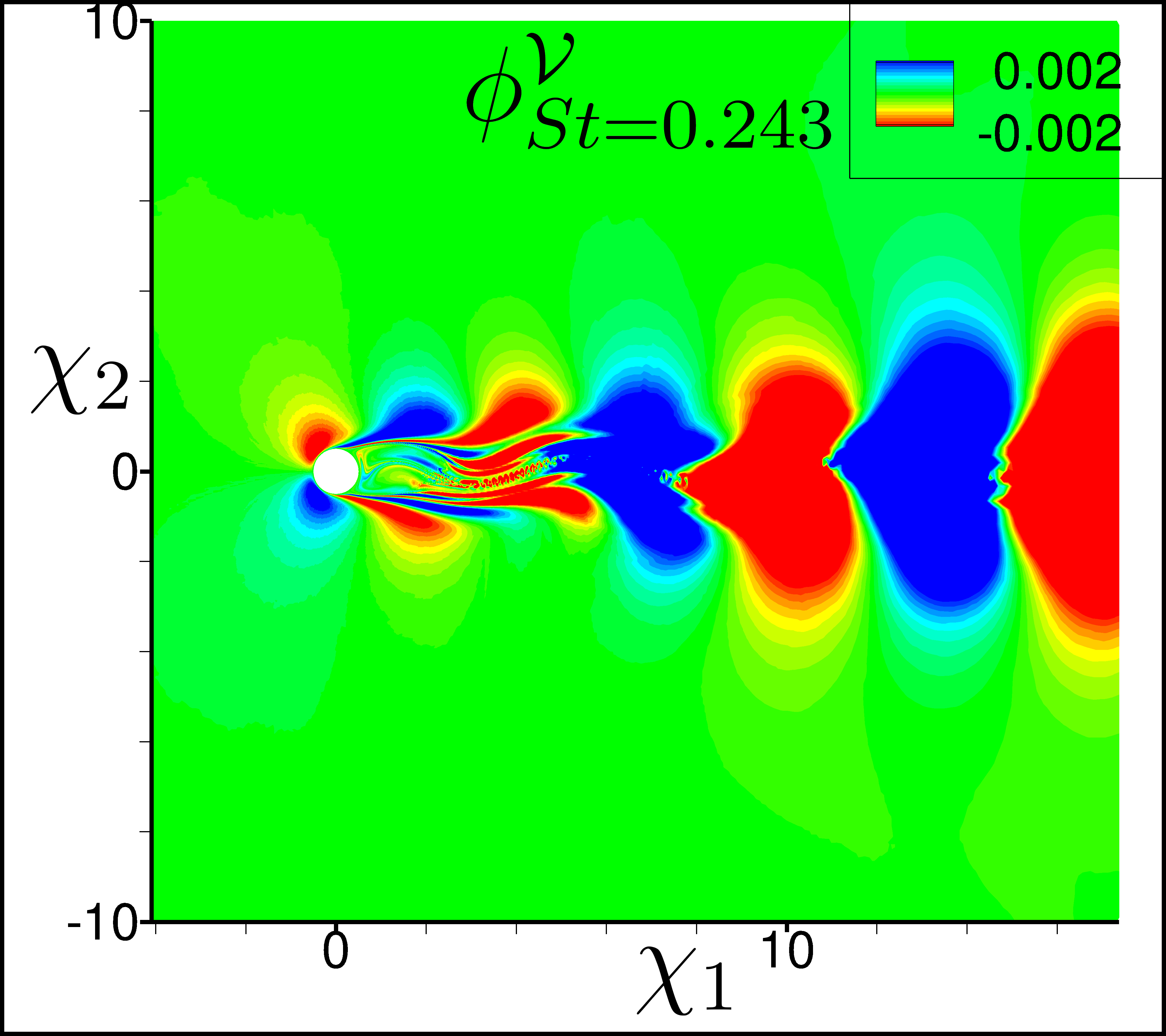}\\ (j) $Re=60; St=0.243$
    \end{minipage}
    \caption{Global LagSAT modes of the streamwise (top row) and crossflow (bottom row) velocities of compressible flow past a cylinder at $M=0.5$ for increasing Reynolds number from $Re=20$ to $Re=60$.}
    \label{fig:phi1}
\end{figure}

The global LagSAT modes of the streamwise and crossflow velocities for the flow past a cylinder at $M=0.5$ and increasing Reynolds number are presented in Fig.~\ref{fig:phi1}.
The top row displays the first branch of the unstable modes (see the inset of Fig.~\ref{fig:cyl_ritz} and corresponding discussion) of the streamwise velocity, whereas the second row exhibits the LagSAT modes with frequency close to $St=0.243$ for the crossflow velocity.
The streamwise velocity modes at $Re=20$ (Fig.~\ref{fig:phi1}a), $Re=30$ (Fig.~\ref{fig:phi1}b), $Re=40$ (Fig.~\ref{fig:phi1}c), and $Re=50$ (Fig.~\ref{fig:phi1}d) are unstable while the corresponding baseflows are steady.
On the other hand, the crossflow velocity modes at $Re=20$ (Fig.~\ref{fig:phi1}f), $Re=30$ (Fig.~\ref{fig:phi1}g), $Re=40$ (Fig.~\ref{fig:phi1}h), and $Re=50$ (Fig.~\ref{fig:phi1}i) are stable and the corresponding baseflows are steady.
For the unsteady baseflow at $Re=60$, the streamwise and crossflow LagSAT modes of Fig.~\ref{fig:phi1}(e) and Fig.~\ref{fig:phi1}(j), respectively, are unstable.

As noted before, the notion of the convective or absolute instability manifests the characteristic of baseflow to amplify a perturbation and either sweep away with the flow (convective) or retain it at the source of perturbation (absolute).
In other words, the convectively unstable flows amplify the external perturbation as it convects down with the flow, whereas the absolutely unstable flows amplify the external perturbation to exhibit self-sustained saturated oscillations~\citep{chomaz2005global}.
In the context of LagSAT, where no external perturbation is imposed on the baseflow, the inherent characteristic of the baseflow reflects through these modes.
Specifically for the flow past a cylinder configuration, the unstable modes in the streamwise direction ($x$) act as convective modes, as they grow with the flow.
On the other hand, the unstable modes in the crossflow direction ($y$), due to the flow symmetry, act as oscillatory modes.

The degree of instability of the LagSAT modes is another interesting aspect that potentially affects the nature and the onset of instability.
For instance, the increase of Reynolds number from $Re=10$ to $Re=50$ leads to a monotonic increase of the growth rate of the streamwise velocity modes from the first branch (see the inset of Fig.~\ref{fig:cyl_ritz}a).
At the onset of unsteadiness, the growth rate of the first branch mode becomes $\mathcal{R}(\omega_l)\approx 0.18$ at $Re=52$, whereas for the unsteady case of $Re=53$ the modal growth is $\mathcal{R}(\omega_l)\approx 0.07$.
Clearly, the convectively unstable modes of the streamwise velocity continues to increase the growth rate up to a threshold value before dropping at the critical Reynolds number $Re=53$.
Although the growth rates of the crossflow velocity modes are negative for the Reynolds numbers in $10 \leq Re \leq 52$, the least stable modes become more stable for decreasing Reynolds number till $Re\approx 10$.
However, further decrease of the Reynolds number reverses the trend, where the least stable modes of the crossflow velocity tend to loose their stability.

\begin{figure}
    \centering
    \begin{minipage}{1.0\textwidth}
    \centering
    \includegraphics[width=1.0\linewidth]{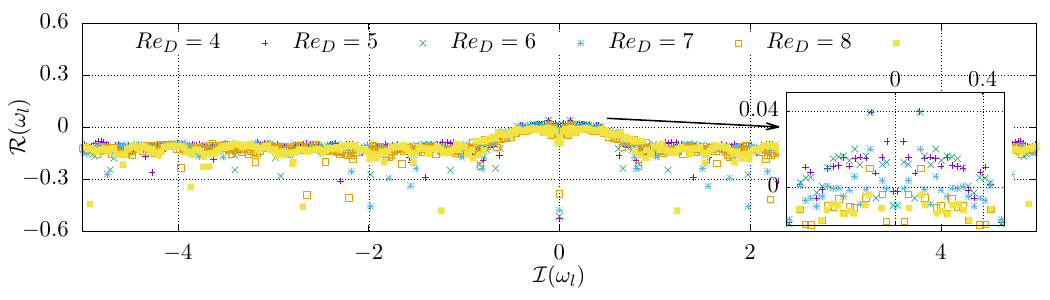}\\(a) Eigen spectra of $\mathcal{V}(\pmb{\chi})$
    \end{minipage}
    \begin{minipage}{0.192\textwidth}
        \centering
        \includegraphics[width=1.0\linewidth]{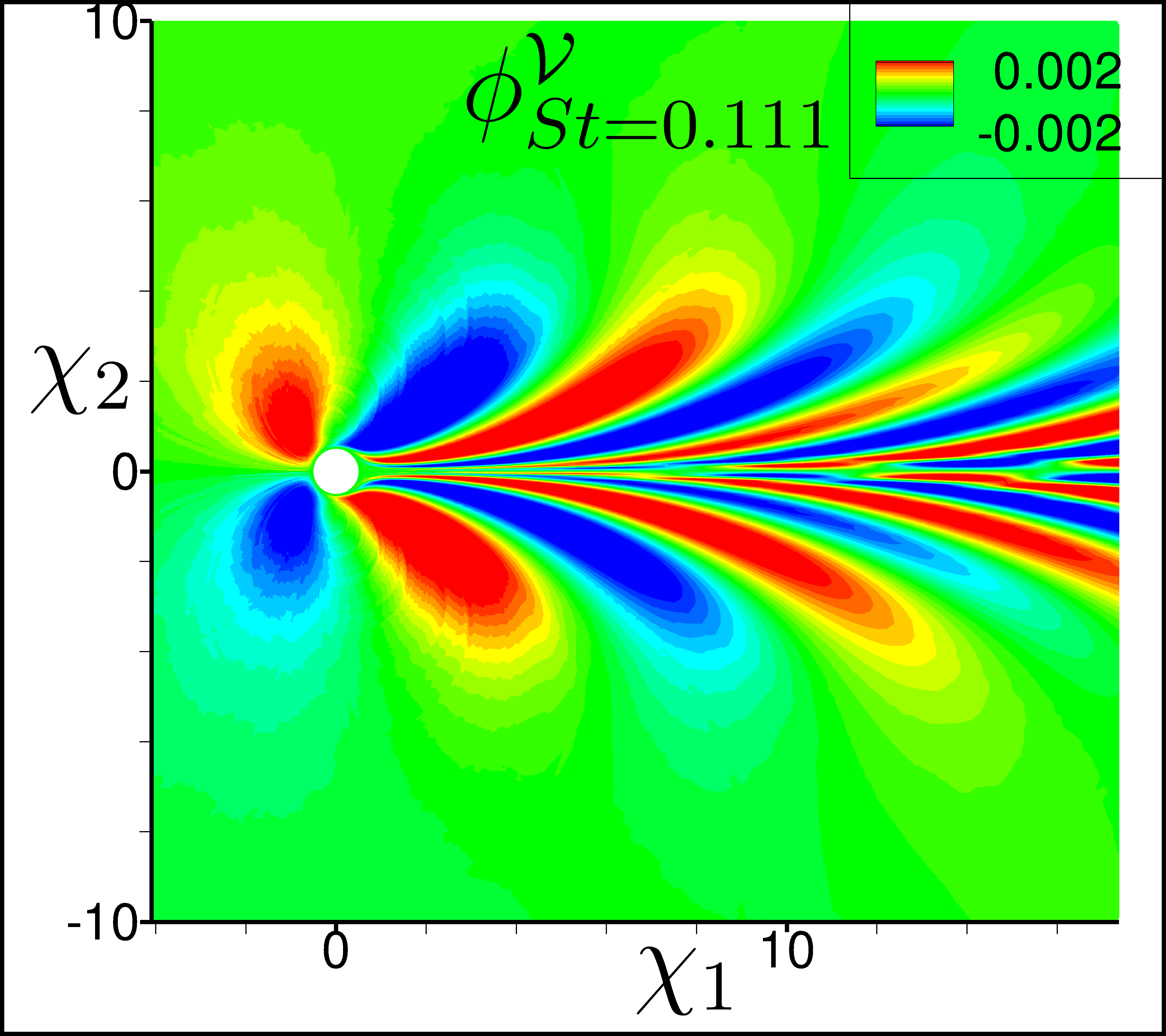}\\(b) $Re_D=4; St=0.111$
    \end{minipage}
    \begin{minipage}{0.192\textwidth}
        \centering
        \includegraphics[width=1.0\linewidth]{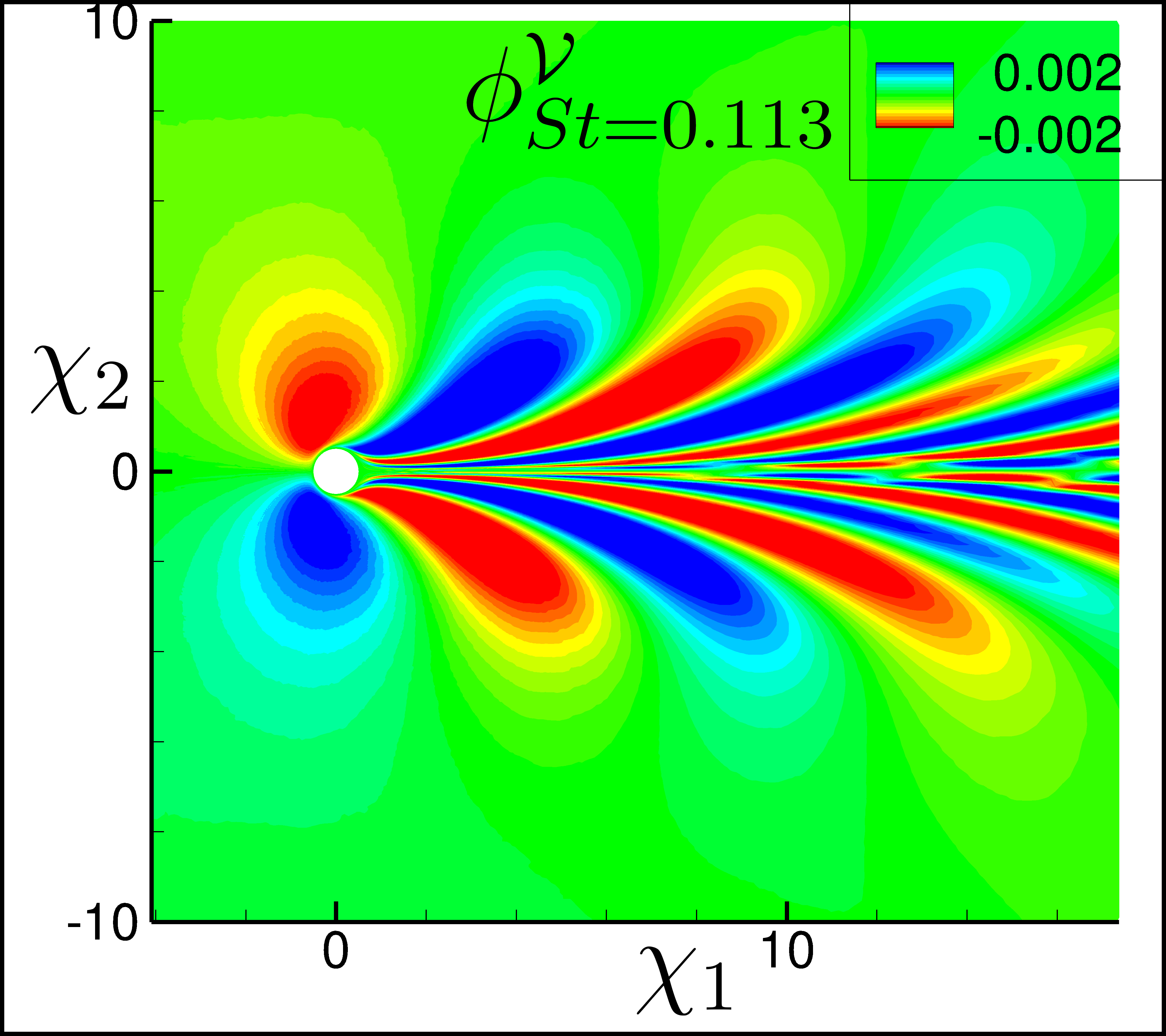}\\(c) $Re_D=5; St=0.113$
    \end{minipage}
    \begin{minipage}{0.192\textwidth}
        \centering
        \includegraphics[width=1.0\linewidth]{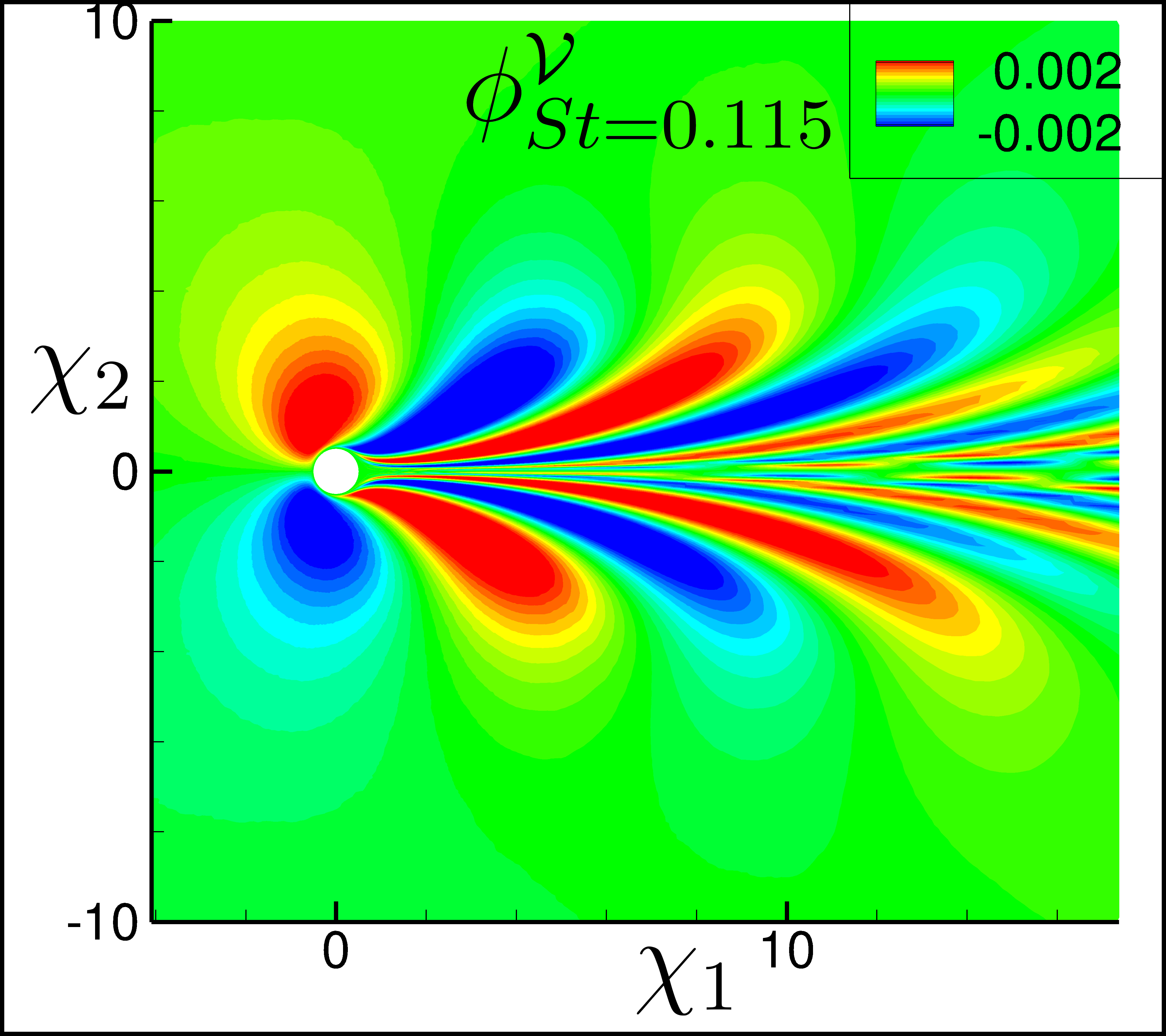}\\(d) $Re_D=6; St=0.115$
    \end{minipage}
    \begin{minipage}{0.192\textwidth}
        \centering
        \includegraphics[width=1.0\linewidth]{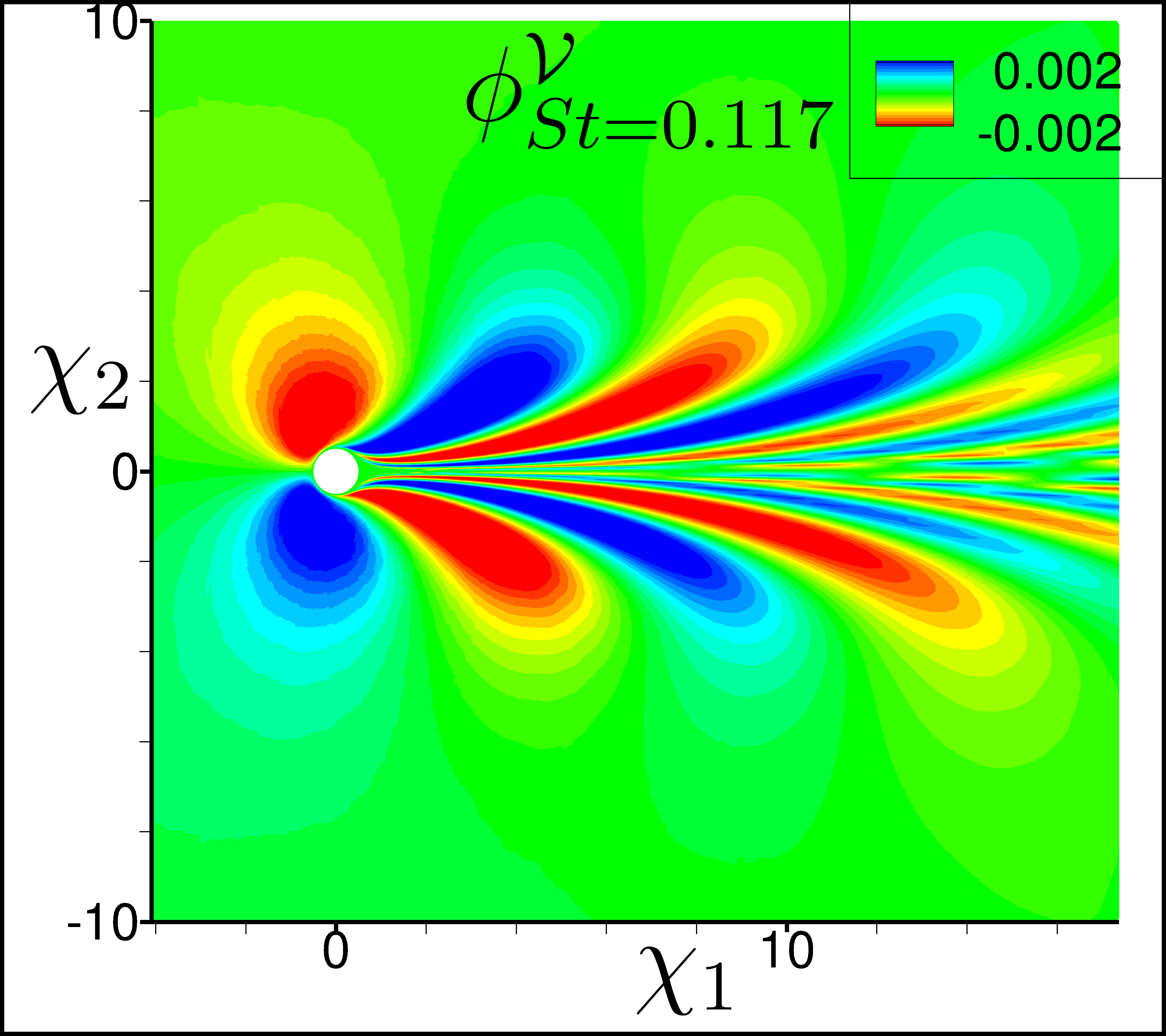}\\(e) $Re_D=7; St=0.117$
    \end{minipage}
    \begin{minipage}{0.192\textwidth}
        \centering
        \includegraphics[width=1.0\linewidth]{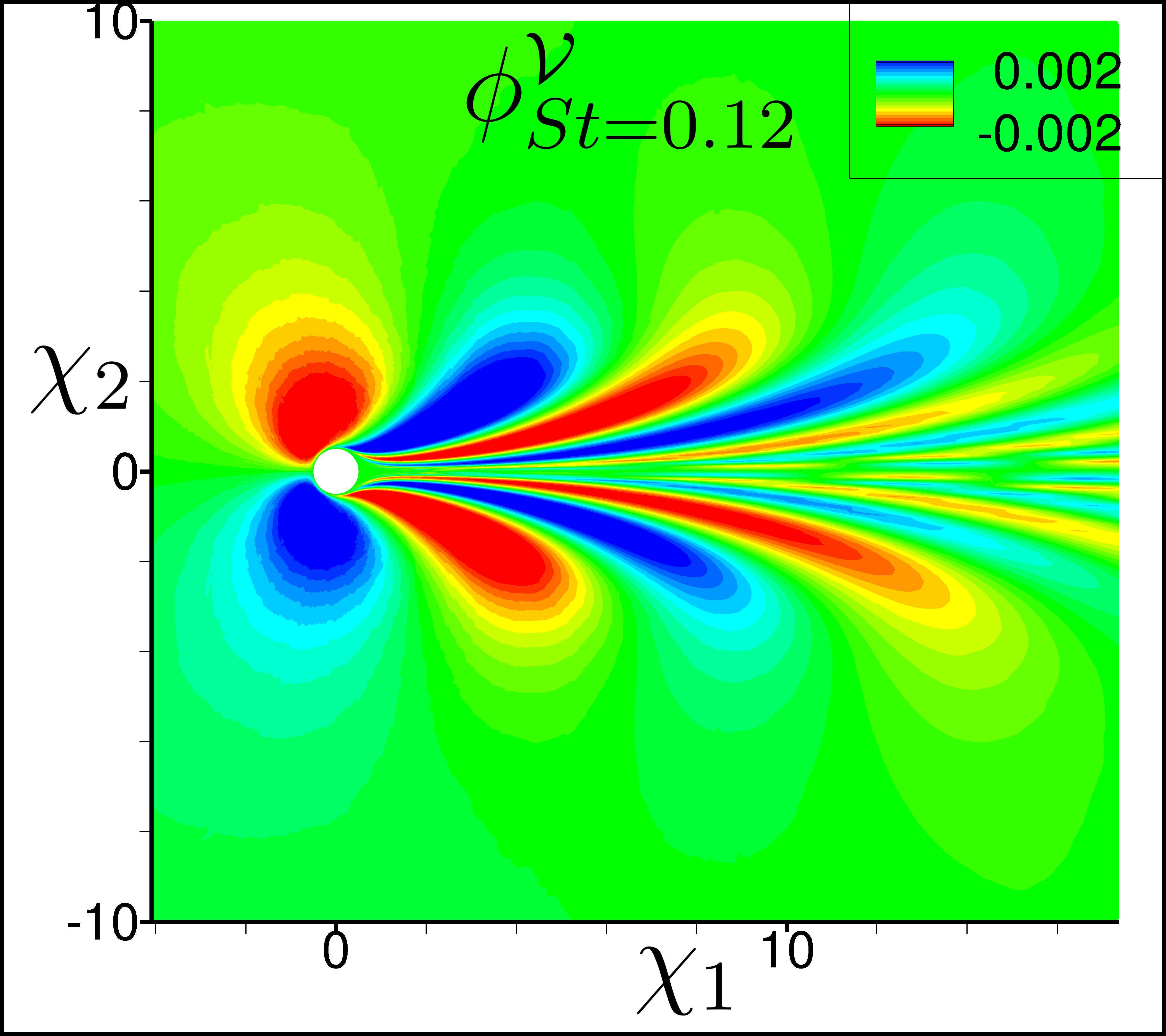}\\(f) $Re_D=8; St=0.12$
    \end{minipage}
    \caption{LagSAT eigenvalue spectra and unstable modes of crossflow velocity of compressible flow past a cylinder at $M=0.5$ and low Reynolds numbers.}
    \label{fig:cyl_ritz_Re7}
\end{figure}

To assess the stability features of the compressible ($M=0.5$) flow past a cylinder at low Reynolds numbers, we employ LagSAT in $1 \leq Re \leq 10$.
Figure~\ref{fig:cyl_ritz_Re7}(a) displays the eigen spectra associated with the LagSAT modes of the crossflow velocity for Reynolds number $Re=4$, $Re=5$, $Re=6$, and $Re=8$.
An noted before, the least stable modes become unstable for decreasing Reynolds number, at $Re=7$.
With a further decrease of Reynolds number, more number of crossflow velocity modes become unstable.
However, the magnitudes of the growth rates remain much smaller ($\mathcal{R}(\omega_l)\approx 0.04$) than the unstable modes of the streamwise velocity ($\mathcal{R}(\omega_l)\approx 0.15$), suggesting the dominance of convective phenomenon over the oscillatory nature of the instability.
This is consistent with the appearance of two steady counter-rotating recirculation regions behind the cylinder for $7< Re_D \leq 8$.
The least-stable LagSAT mode of crossflow velocity which becomes unstable for decreasing Reynolds number is shown in Figs.~\ref{fig:cyl_ritz_Re7}(b)-(f), where the Strouhal number associated with the mode is $St\approx 0.11$.
Several studies predict $Re\approx 4$ as the lower critical Reynolds number for the beginning of the convective instability by means of the linearized global/local analyses~\citep{mittal2007stabilized,monkewitz1988absolute}.
In the present case however, the flow compressibility seems to delay this transition point, similar to the global/absolute instability at $Re_D=53$.
Nonetheless, the global LagSAT precisely indicate these transition points, including the convective and absolute nature of the instability.

\begin{figure}
    \centering
    \begin{minipage}{0.48\textwidth}
    \centering
    \includegraphics[width=1.0\linewidth]{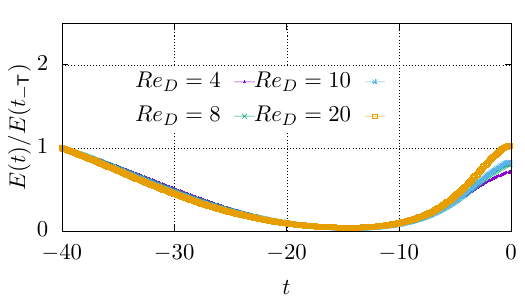}\\(a)
    \end{minipage}
    \begin{minipage}{0.48\textwidth}
    \centering
    \includegraphics[width=1.0\linewidth]{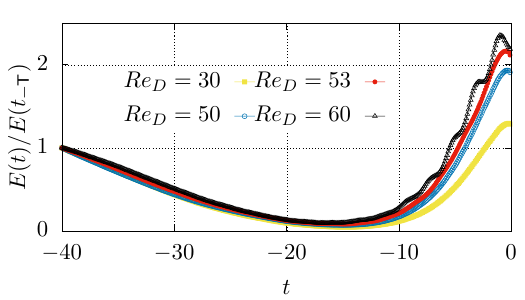}\\(b)
    \end{minipage}
    \caption{Total energy growth in LagSAT on compressible flow past a cylinder for increase Reynolds number at (a) $Re_D=4,8,10,20$, and (b) $Re_D=30,50,51,53$.}
    \label{fig:cyl_en}
\end{figure}

The total energy growth in terms of the LPOD modal energy Eq.~\ref{eq:trans-energy} in an adjoint form exhibits a transition point at $Re_D=20$, where the energy at the beginning of the Lagrangian flow map ($t=-\mathsf{T}$) increases at the identity map ($t=0$).
For all Reynolds numbers, the total energy decreases till $t\approx -15$, before increasing toward the identity map. 
For the lower values of Reynolds number $Re_D<20$, the increase of the total energy remains lower than the initial state of the flow, in a Lagrangian sense.
On the other hand, the increase of Reynolds number beyond $Re_D=20$ shows amplification of the initial energy levels, as high as a factor of $2.5$.
Note the unsteadiness introduced in the energy graphs for $Re=53$ and $Re=60$ as a results of the onset of absolute instability. 

\section{LagSAT on 2D Lid Driven Cavity} \label{sec:lagsat_ldc}

Lid-driven cavity (LDC) is a canonical flow configuration that is extensively utilized as a benchmark for numerical studies, primarily due to its simplicity.
Nonetheless, LDC exhibits a complex flow physics even in 2-dimensional setting.
The flow structure inside the cavity forms a core vortex region enclosed in a shear layer region, engendering secondary recirculation regions near the corners.
The flow remains steady at low Reynolds number, while at sufficiently higher Reynolds number it undergoes a Holf bifurcation at a critical Reynolds number~\citep{ghia1982high,shen1991hopf,ramanan1994linear,shinde2021lagrangian} to become unsteady.
The flow compressibility is known to affect this critical value of Reynolds number~\citep{bergamo2015compressible,ranjan2020robust}, by suppressing some of the dominant unstable modes from the incompressible lid-driven cavity counterpart.
The stabilization mechanisms, namely, the baroclinic torque and vorticity-dilation, are discussed in more details in~\citet{ohmichi2017compressibility}.

\begin{figure}
    \centering
    \begin{minipage}{0.192\textwidth}
        \centering
        \includegraphics[width=1.0\textwidth]{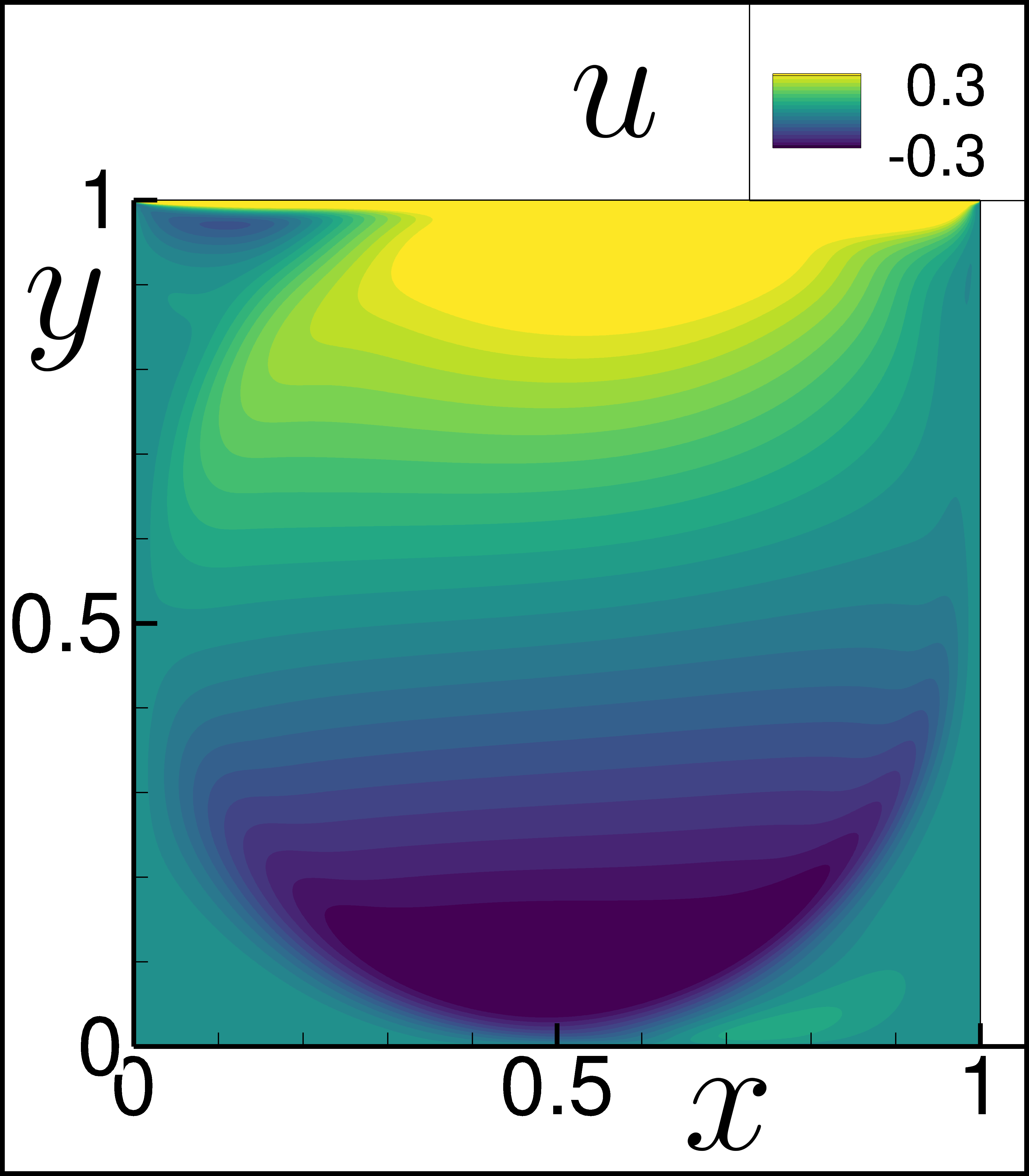}\\(a) $u(\pmb{x},t=0)$
    \end{minipage}
    \begin{minipage}{0.192\textwidth}
        \centering
        \includegraphics[width=1.0\textwidth]{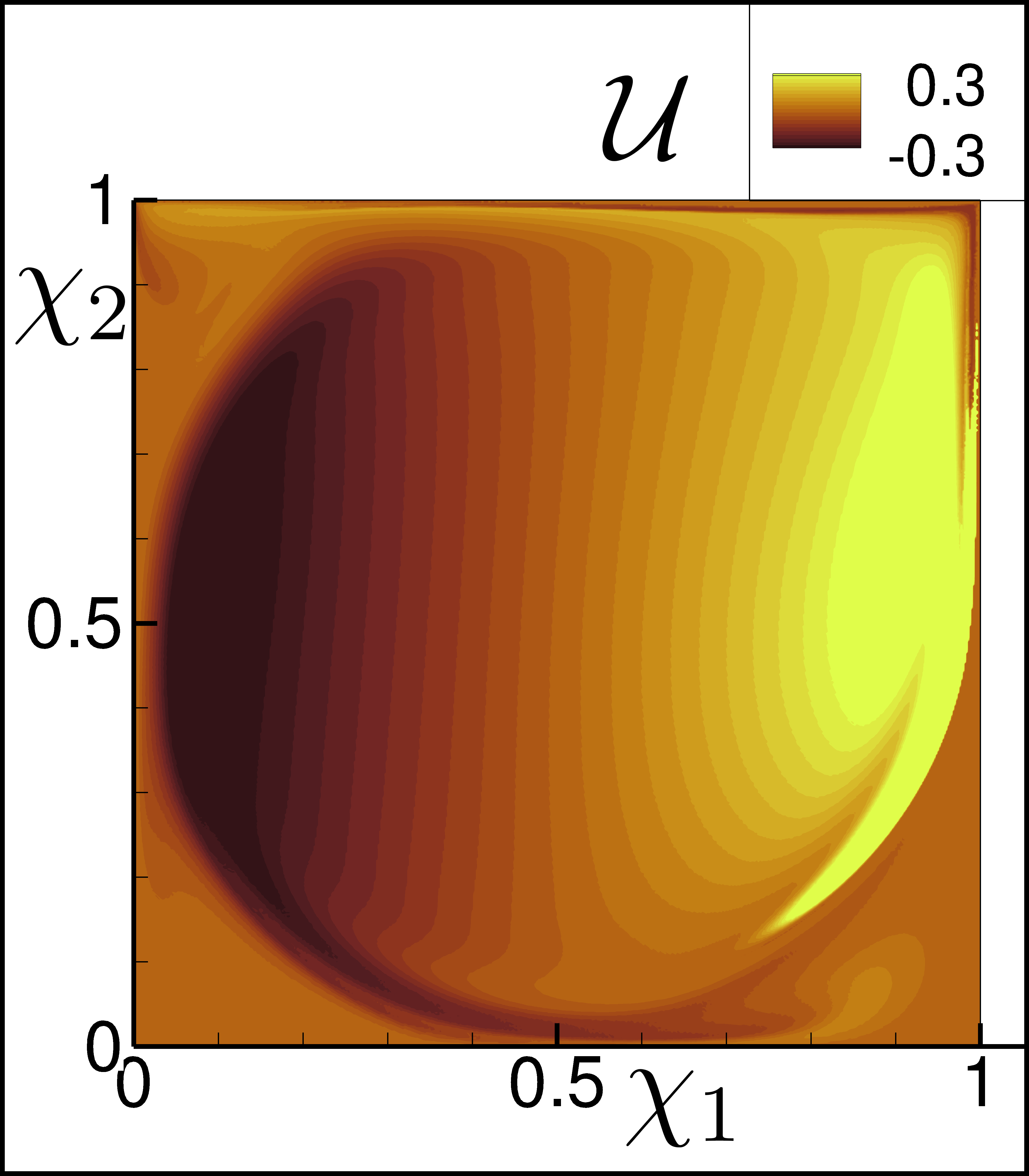}\\(b) $\mathcal{U}(\pmb{\chi},t=-2)$
    \end{minipage}
    \begin{minipage}{0.192\textwidth}
        \centering
        \includegraphics[width=1.0\textwidth]{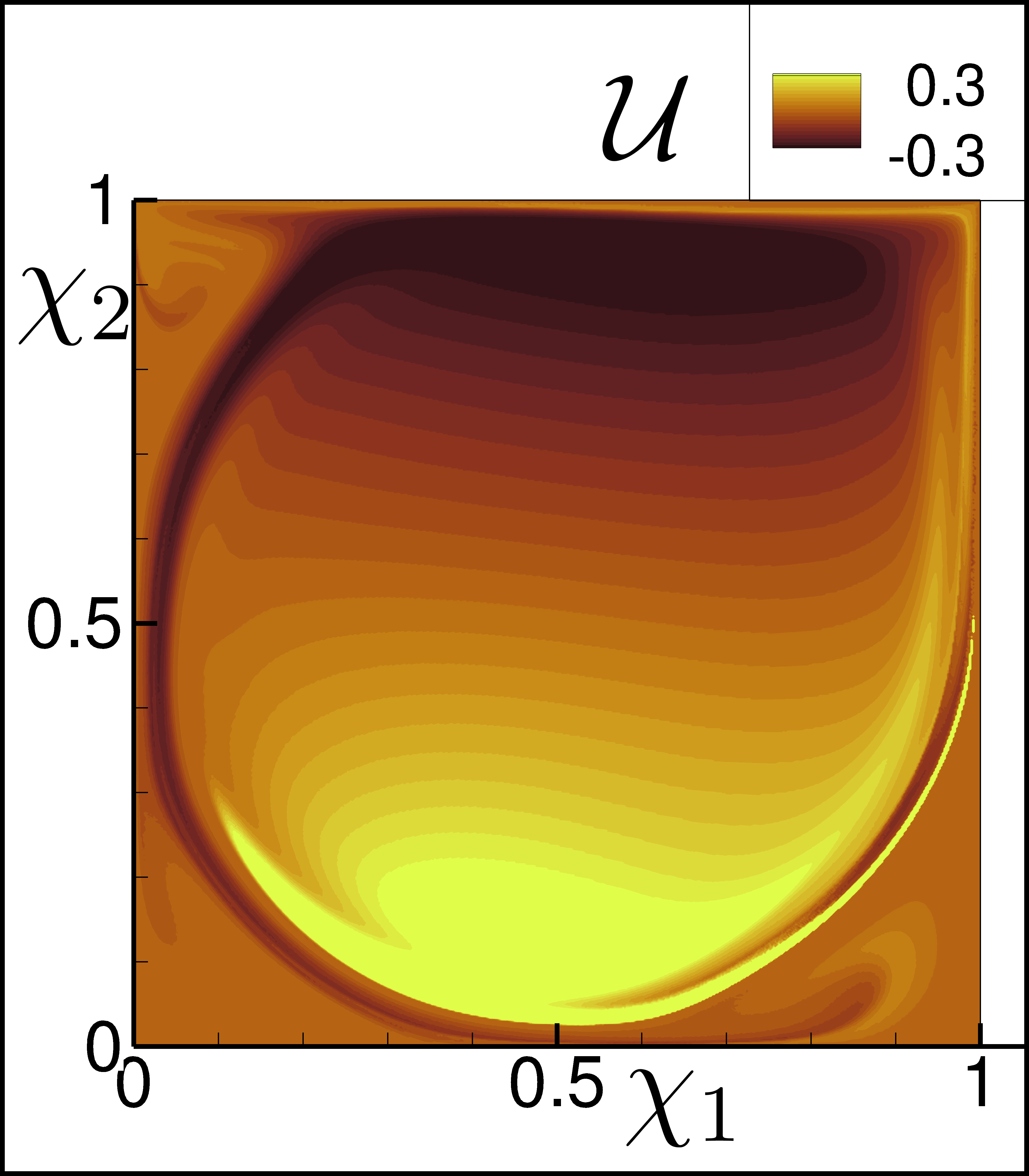}\\(c) $\mathcal{U}(\pmb{\chi},t=-4)$
    \end{minipage}
    \begin{minipage}{0.192\textwidth}
        \centering
        \includegraphics[width=1.0\textwidth]{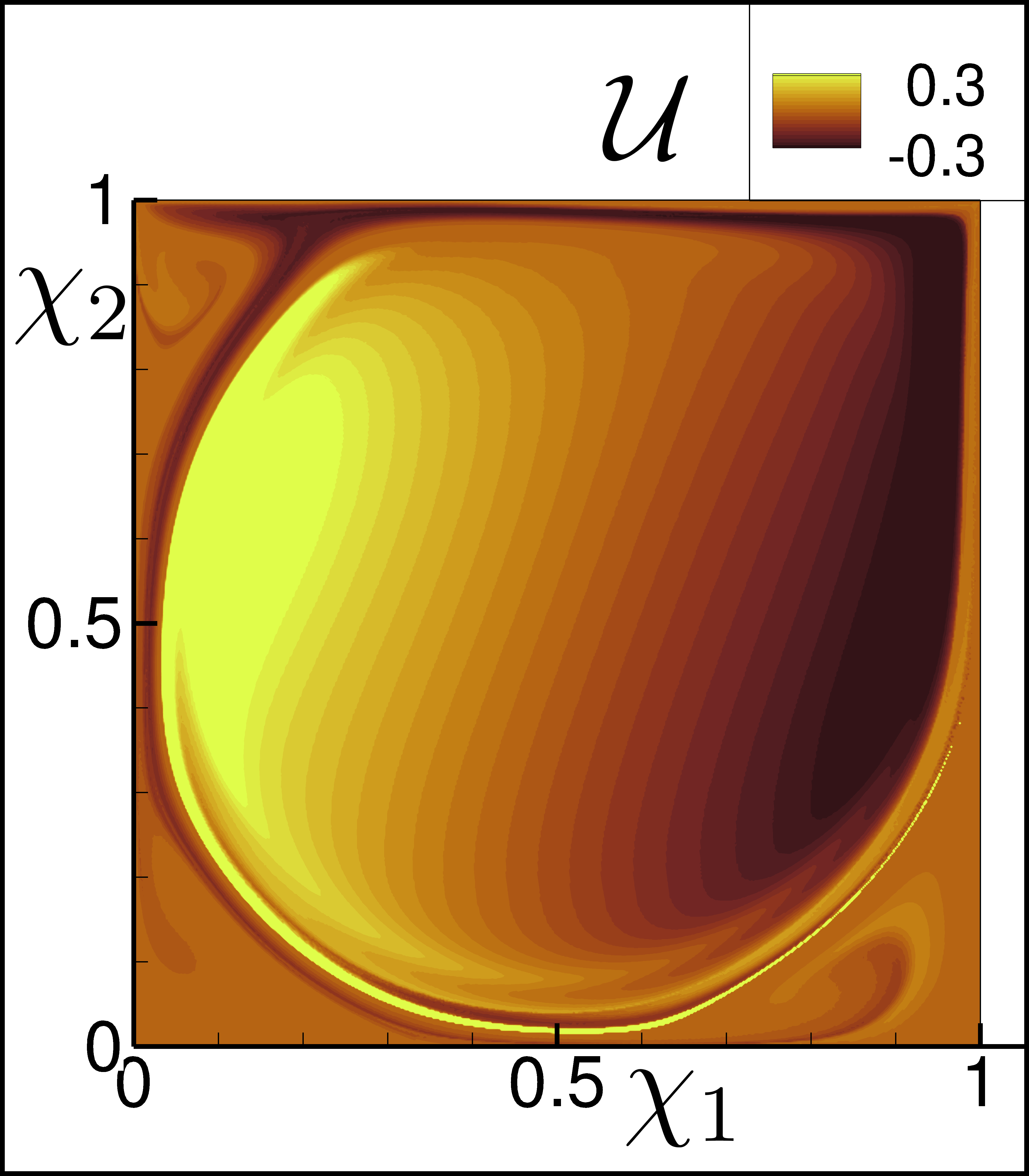}\\(d) $\mathcal{U}(\pmb{\chi},t=-6)$
    \end{minipage}
    \begin{minipage}{0.192\textwidth}
        \centering
        \includegraphics[width=1.0\textwidth]{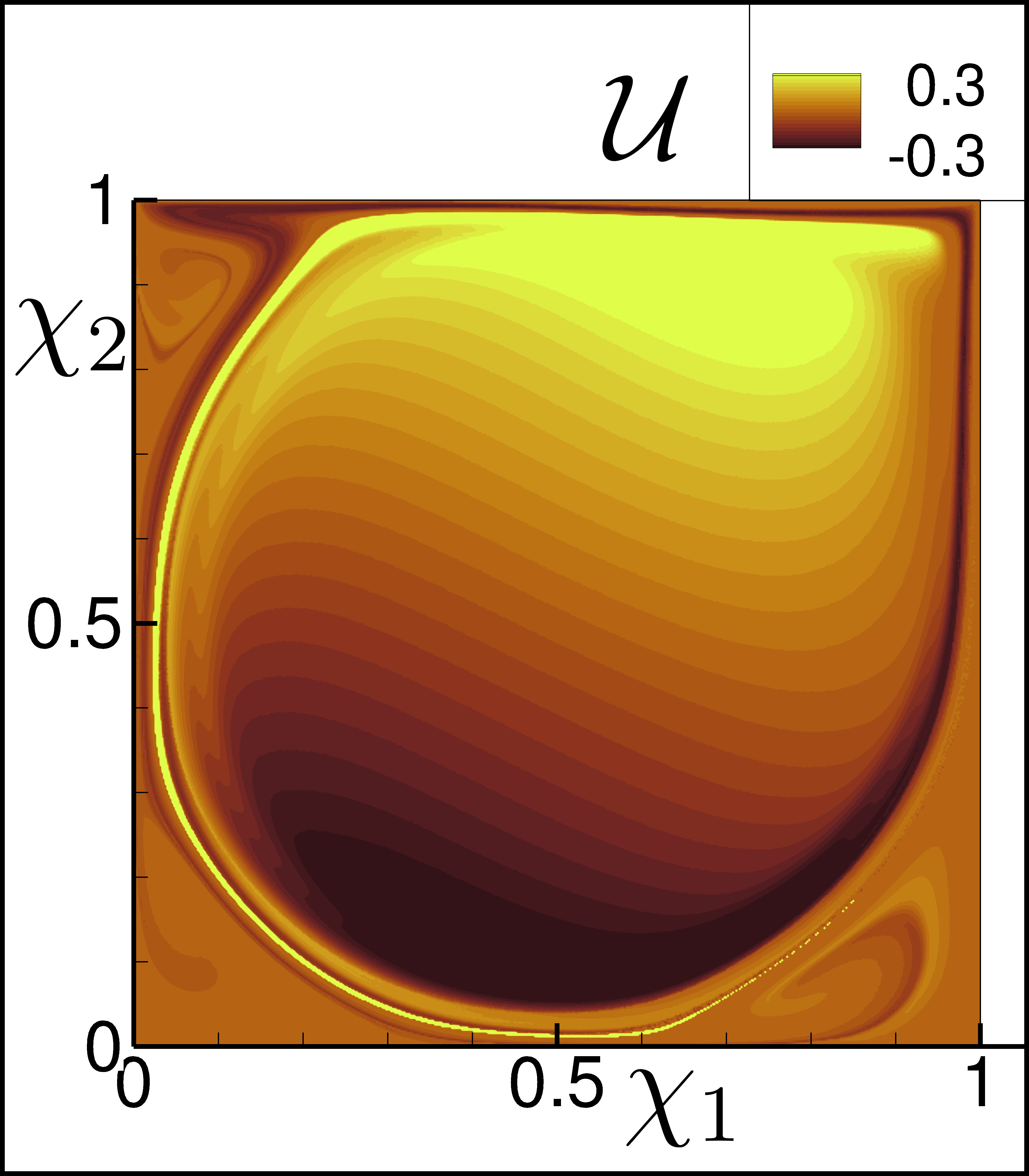}\\(e) $\mathcal{U}(\pmb{\chi},t=-8)$
    \end{minipage}
    \caption{Baseflow and select instances of Lagrangian flow map of lid drive cavity flow at $Re_L=7{,}000$ and Mach $M=0.5$.}
    \label{fig:ldc_base_lagmap}
\end{figure}

The flow configuration comprises a 2D closed cavity flow driven by the lid (top boundary of the cavity) and enclosed by the other three wall (no-slip) boundaries.
Figure~\ref{fig:ldc_base_lagmap}(a) displays a baseflow in terms of $u(\pmb{x})$ velocity component at Reynolds number of $Re_L=7{,}000$ and Mach number of $M=0.5$.
The Reynolds number here is defined in terms of the length of the square cavity $L$.
More details on the mesh independence study and numerical methodology are provided in the Appendix B of ~\citet{shinde2021lagrangian}.
In prior DNS investigations~\citep{ohmichi2017compressibility,shinde2021lagrangian}, the critical Reynolds number at Mach $M=0.5$ is reported as $Re_c\approx 10{,}500$.
To employ LagSAT on LDC, we perform a series of DNS at Reynolds numbers in $7{,}000 \leq Re_L\leq 15{,}000$, observing the Holf bifurcation and consequently the unsteady flow in LDC for $Re > 11{,}000$.
The Lagrangian flow map of the baseflow at $Re_L=7{,}000$ is displayed in Figs.~\ref{fig:ldc_base_lagmap}(b) to (e).
It comprises a total of $500$ time instances at a time interval of $0.1$, resulting in total time of $\mathsf{T=50}$, although Figs.~\ref{fig:ldc_base_lagmap}(b) to (e) display a short time horizon from the identity map, depicting the Lagrangian flow map for approximately one cycle of the least stable LagSAT mode.

\begin{figure}
    \centering
    \begin{minipage}{0.98\textwidth}
        \centering
        \includegraphics[width=1.0\textwidth]{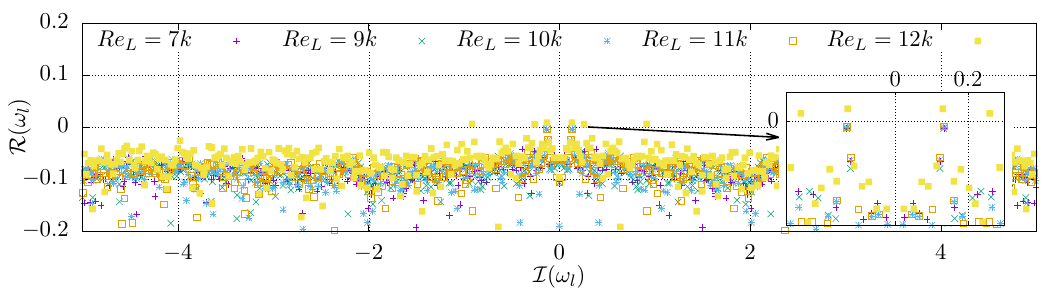}\\(a) LagSAT eigen spectra of lid-driven cavity for increasing Reynolds number.
    \end{minipage}
    \begin{minipage}{0.192\textwidth}
        \centering
        \includegraphics[width=1.0\textwidth]{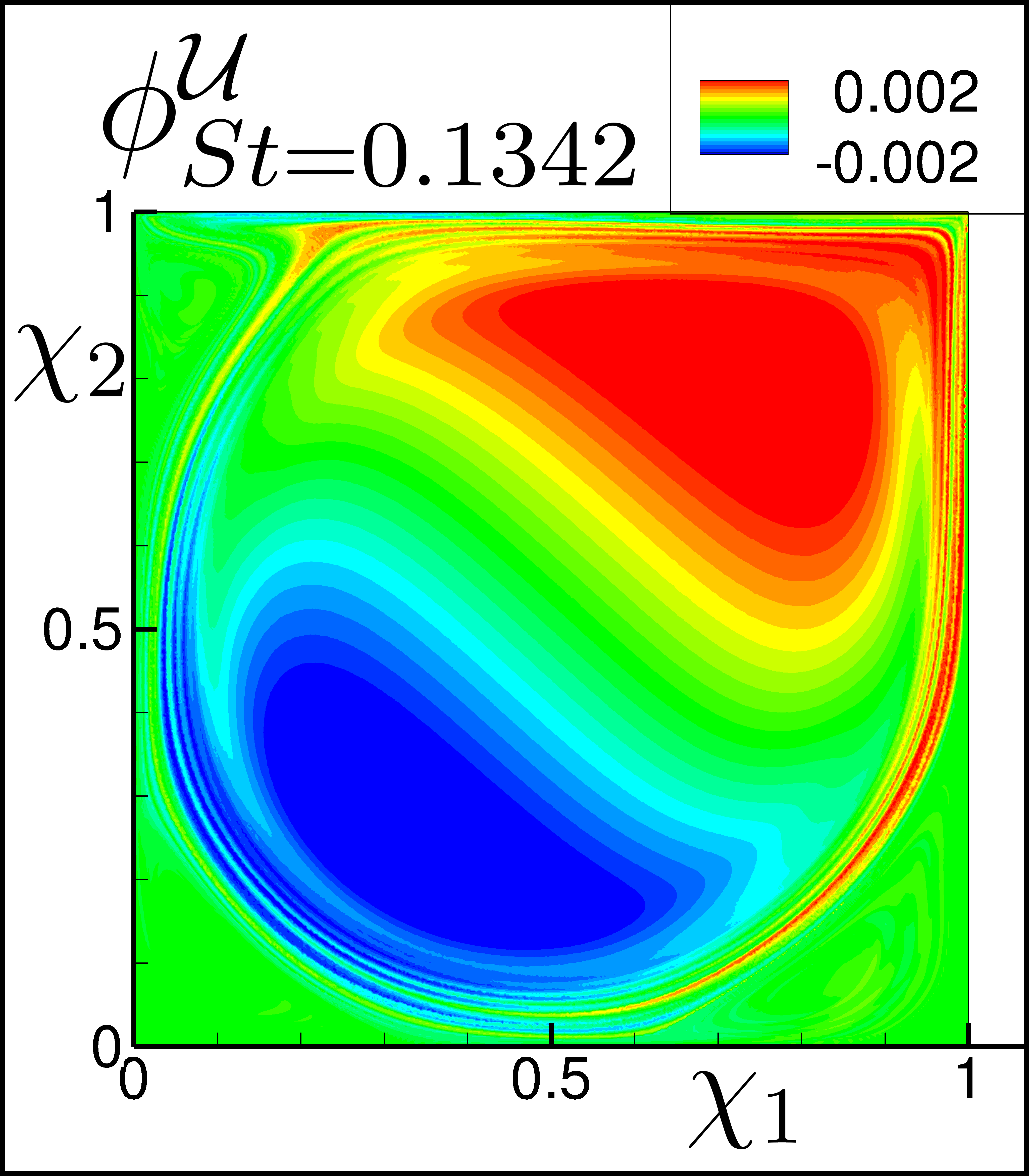}\\(b)  $Re_L=7k$ $St_L=0.1342$
    \end{minipage}
    \begin{minipage}{0.192\textwidth}
        \centering
        \includegraphics[width=1.0\textwidth]{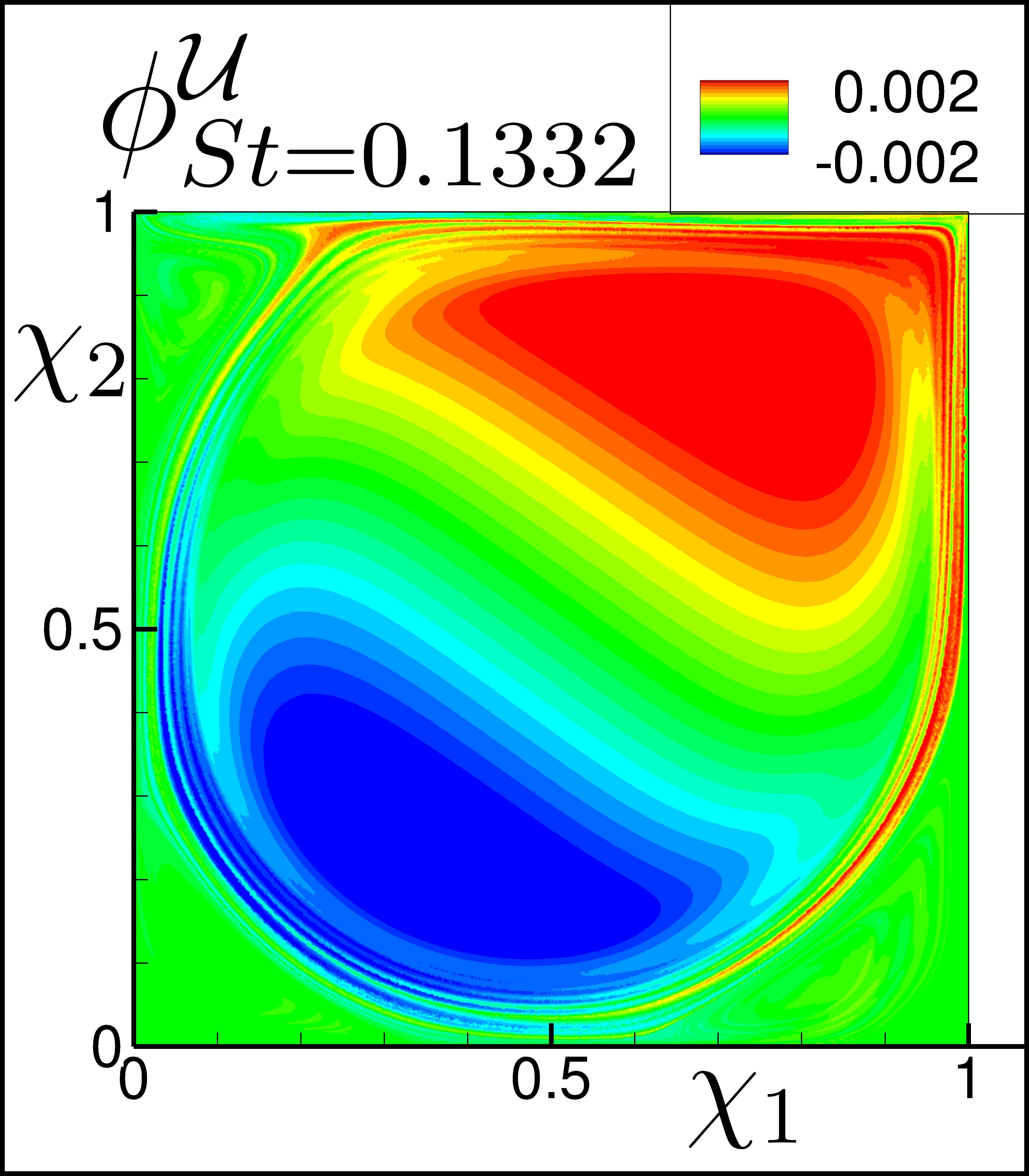}\\(c) $Re_L=9k$ $St_L=0.1332$
    \end{minipage}
    \begin{minipage}{0.192\textwidth}
        \centering
        \includegraphics[width=1.0\textwidth]{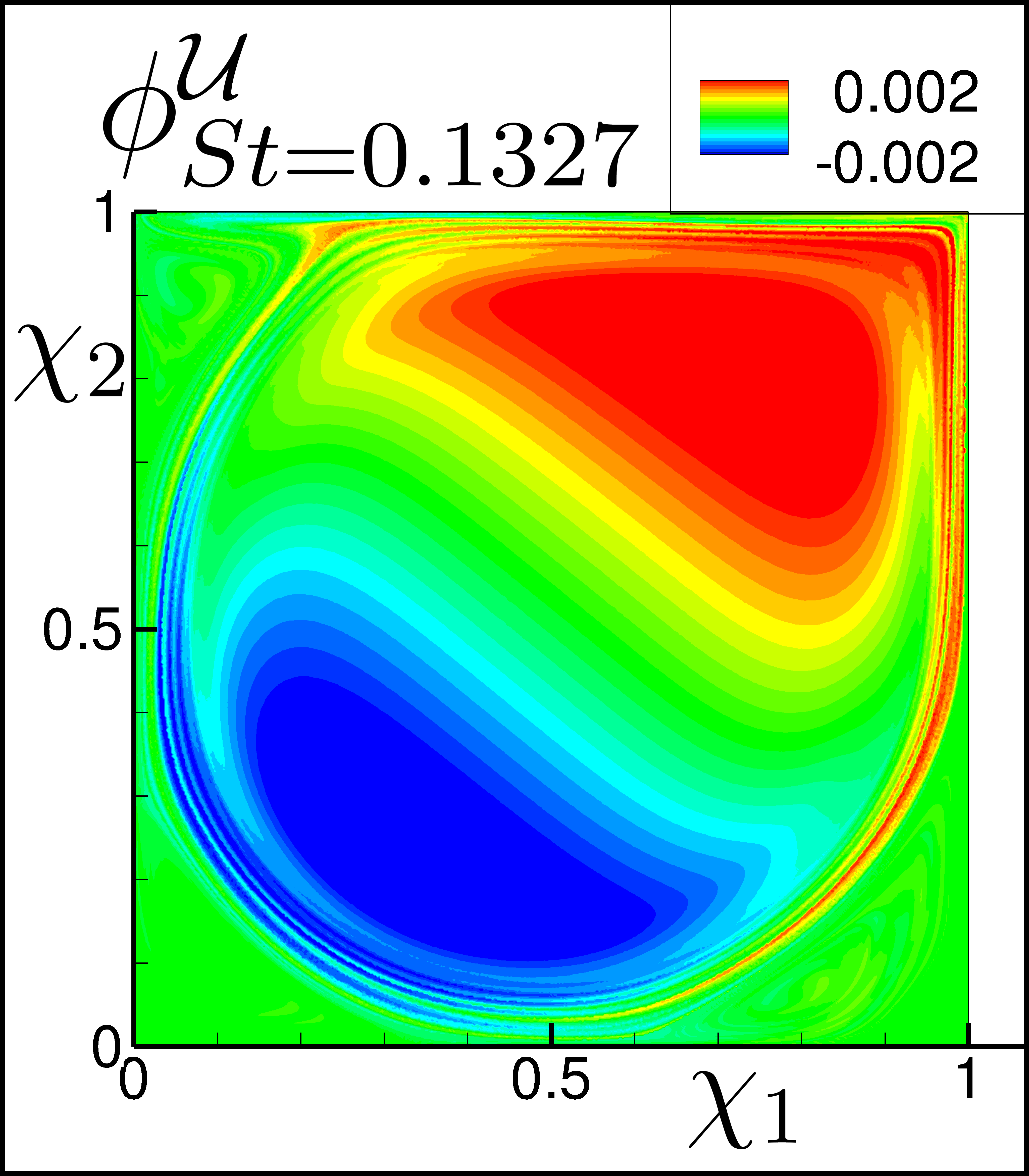}\\(d) $Re_L=10k$ $St_L=0.1327$
    \end{minipage}
    \begin{minipage}{0.192\textwidth}
        \centering
        \includegraphics[width=1.0\textwidth]{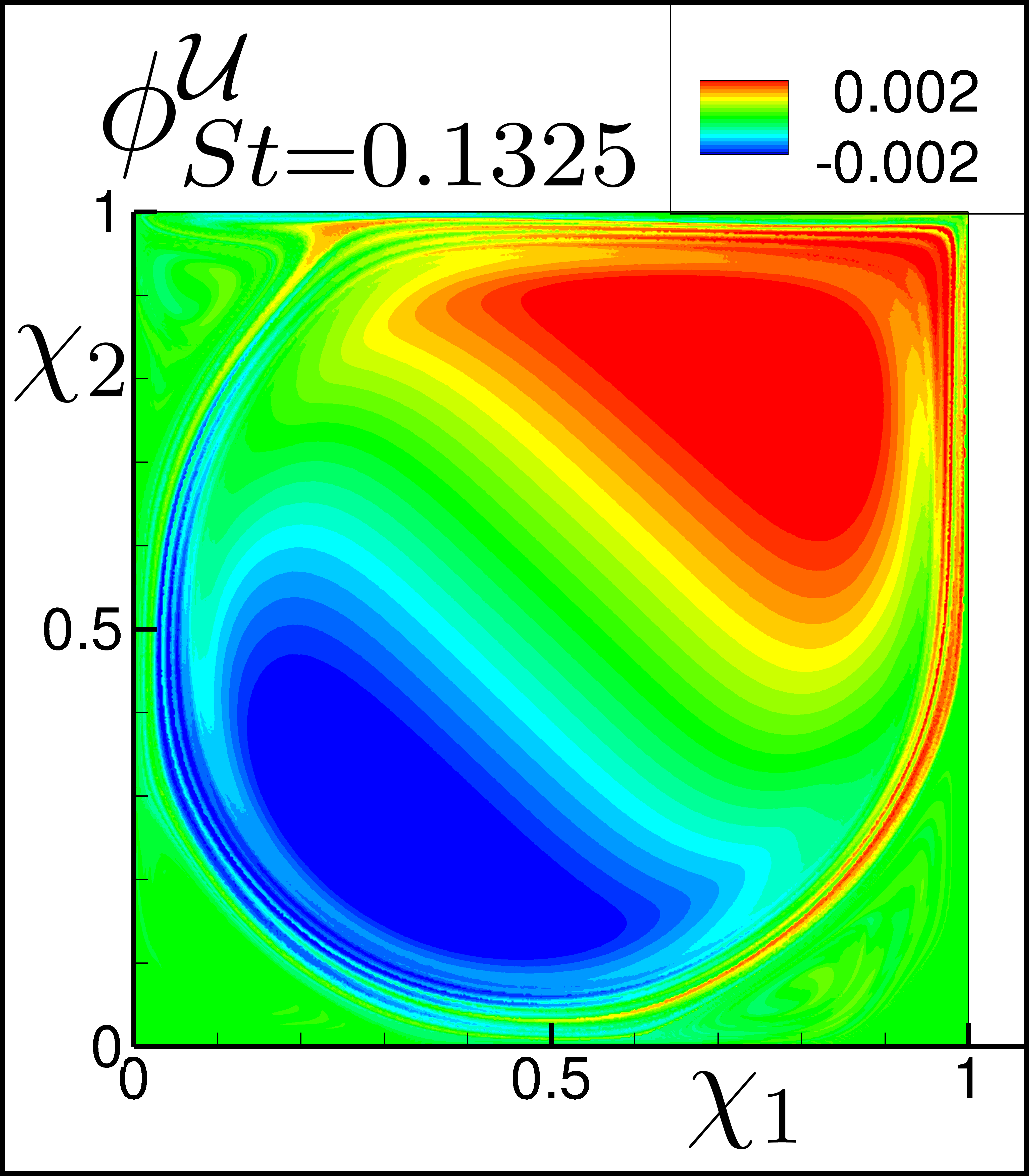}\\(e) $Re_L=11k$ $St_L=0.1325$
    \end{minipage}
    \begin{minipage}{0.192\textwidth}
        \centering
        \includegraphics[width=1.0\textwidth]{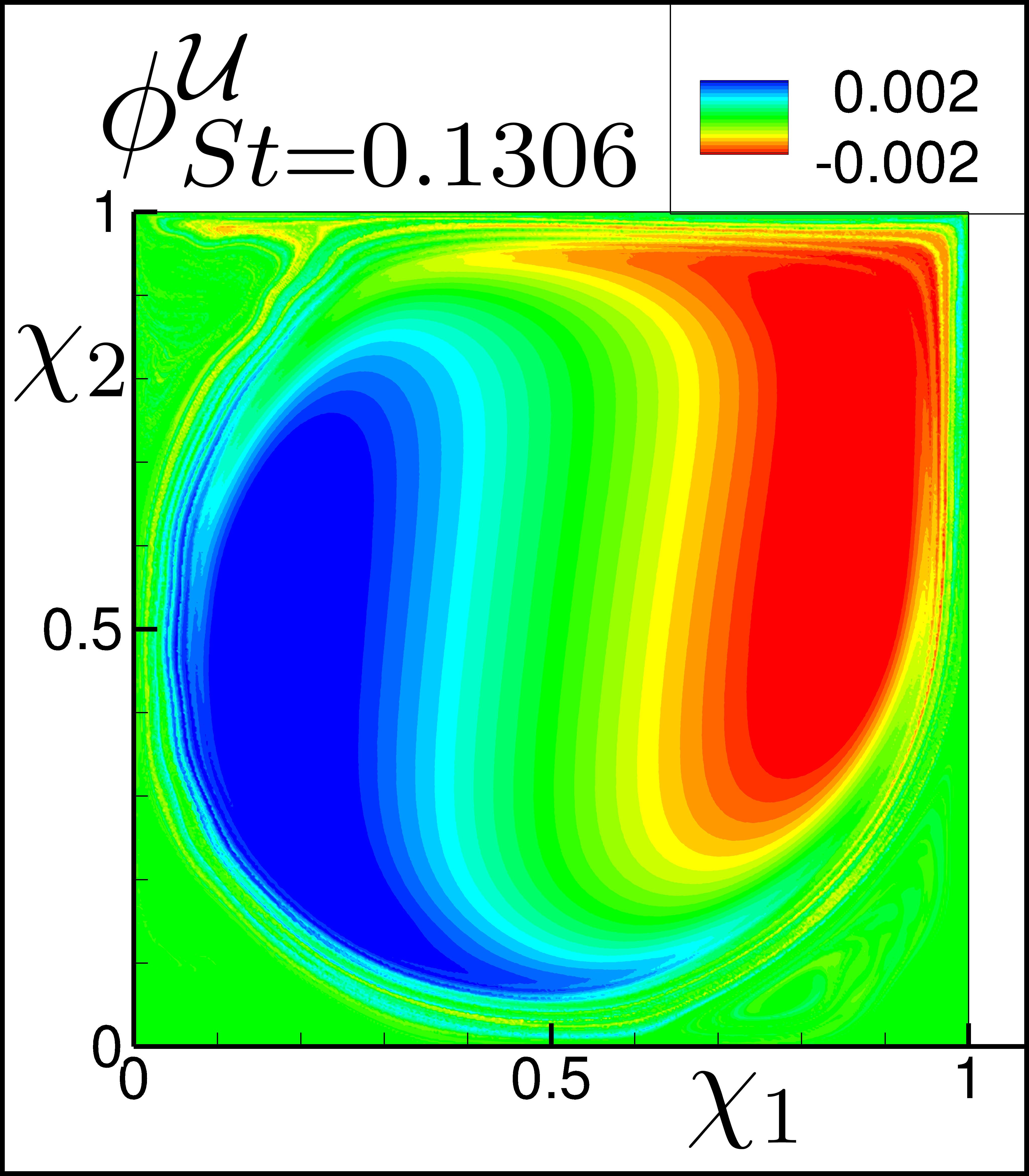}\\(f) $Re_L=12k$ $St_L=0.1306$
    \end{minipage}
    \caption{Least stable sub-critical and unstable post-critical LagSAT modes of lid-driven cavity for increasing Reynolds number from a sub-critical $Re_L=7{,}000$ to a post-critical $Re_L=12{,}000$.}
    \label{fig:ldc_phi1}
\end{figure}

The results of LagSAT on LDC are displayed in Fig.~\ref{fig:ldc_phi1} in terms of the eigen spectra and the leading instability mode.
The eigen spectra for Reynolds number $Re_L\leq 11{,}000$ show that majority of modes are largely damped (stable) whereas a few modes are marginally stable.
Nonetheless, all baseflows for $Re_L\leq 11{,}000$ are stable.
At $Re_L=12{,}000$, the flow becomes unstable with a few LagSAT modes exhibiting positive growth rate.
The least stable mode for pre-critical Reynolds numbers and the unstable mode for the post-critical Reynolds number are displayed in Figs.~\ref{fig:ldc_phi1}(b) to (f).
This unstable mode, which appears as the first unstable mode, mainly spans the inner vortex region of the lid-driven cavity, exhibiting modal frequency of $St\approx 0.13$.
Remarkably, LagSAT can extract the leading (least stable) mode at a much lower Reynolds number compared to the critical value.
The associated unsteadiness, in Lagrangian sense, of the mode marginally decreases toward the critical Reynolds number.
Note the spatial form of the post-critical mode shape (Fig.~\ref{fig:ldc_phi1}f), which appears to have evolved from the pre-critical shape of the least stable mode (Fig.~\ref{fig:ldc_phi1}b to e).

\begin{figure}
    \centering
    \begin{minipage}{0.192\textwidth}
        \centering
        \includegraphics[width=1.0\textwidth]{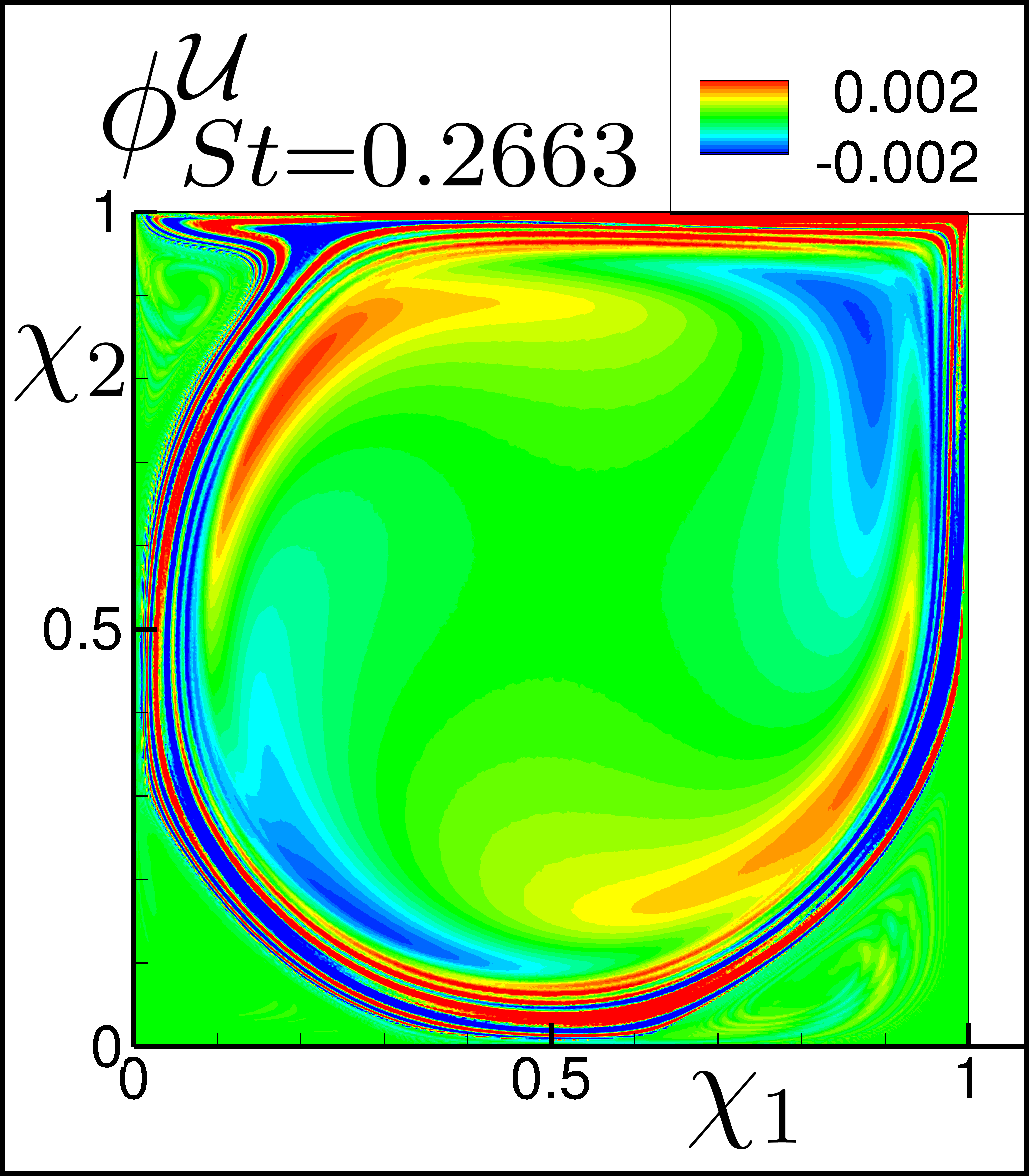}\\(b)  $Re_L=7k$ $St_L=0.2663$
    \end{minipage}
    \begin{minipage}{0.192\textwidth}
        \centering
        \includegraphics[width=1.0\textwidth]{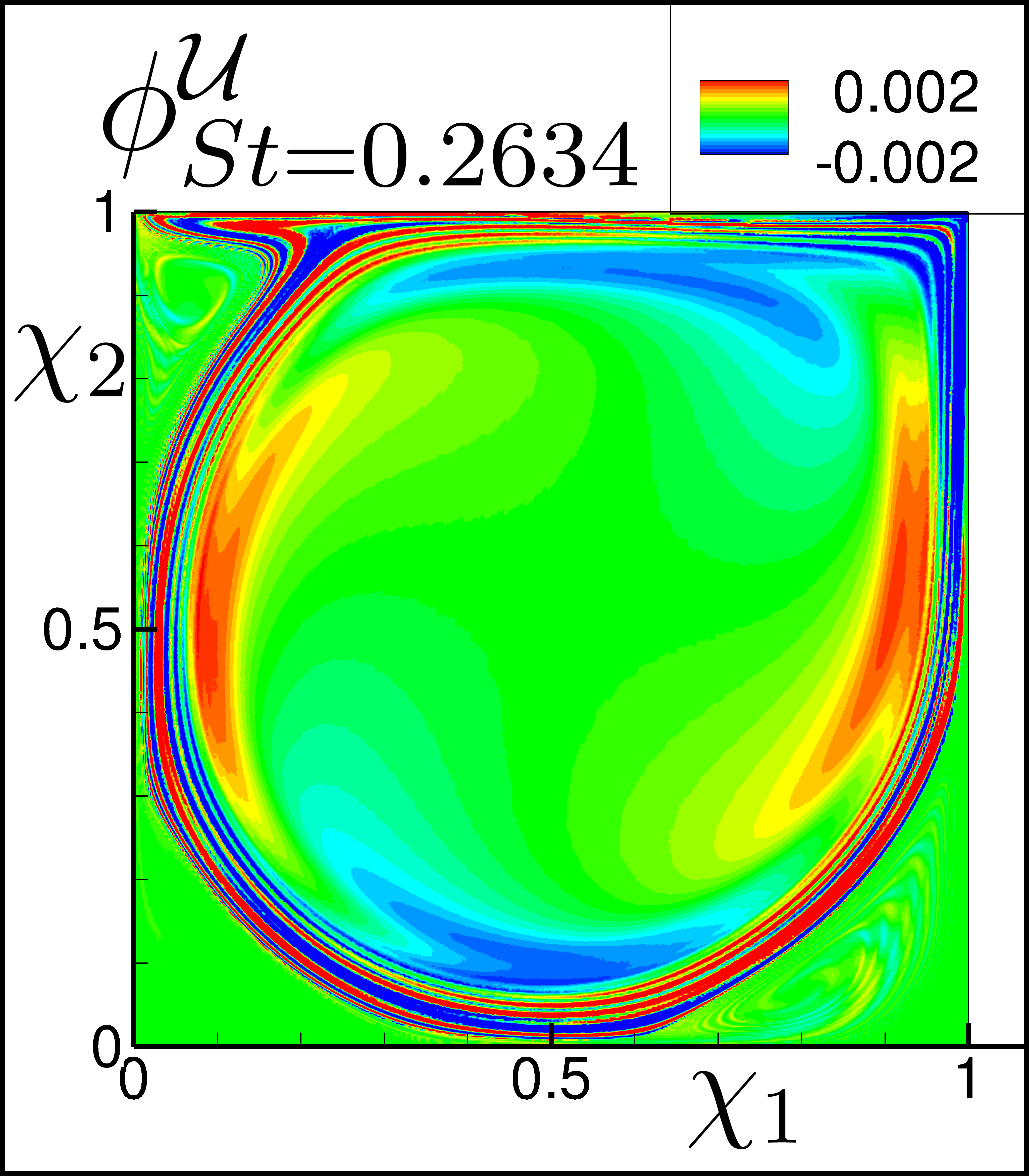}\\(c) $Re_L=9k$ $St_L=0.2634$
    \end{minipage}
    \begin{minipage}{0.192\textwidth}
        \centering
        \includegraphics[width=1.0\textwidth]{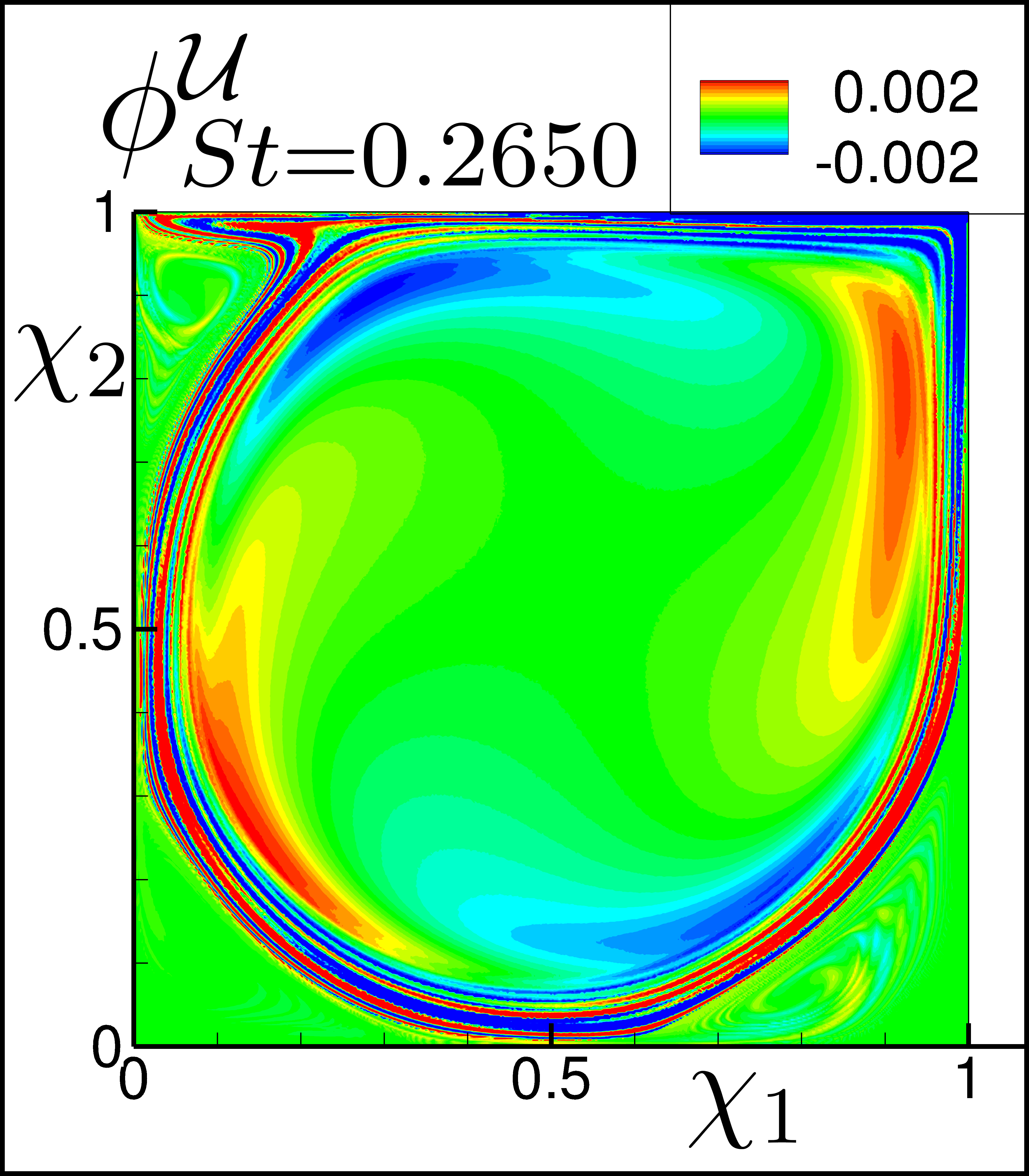}\\(d) $Re_L=10k$ $St_L=0.2650$
    \end{minipage}
    \begin{minipage}{0.192\textwidth}
        \centering
        \includegraphics[width=1.0\textwidth]{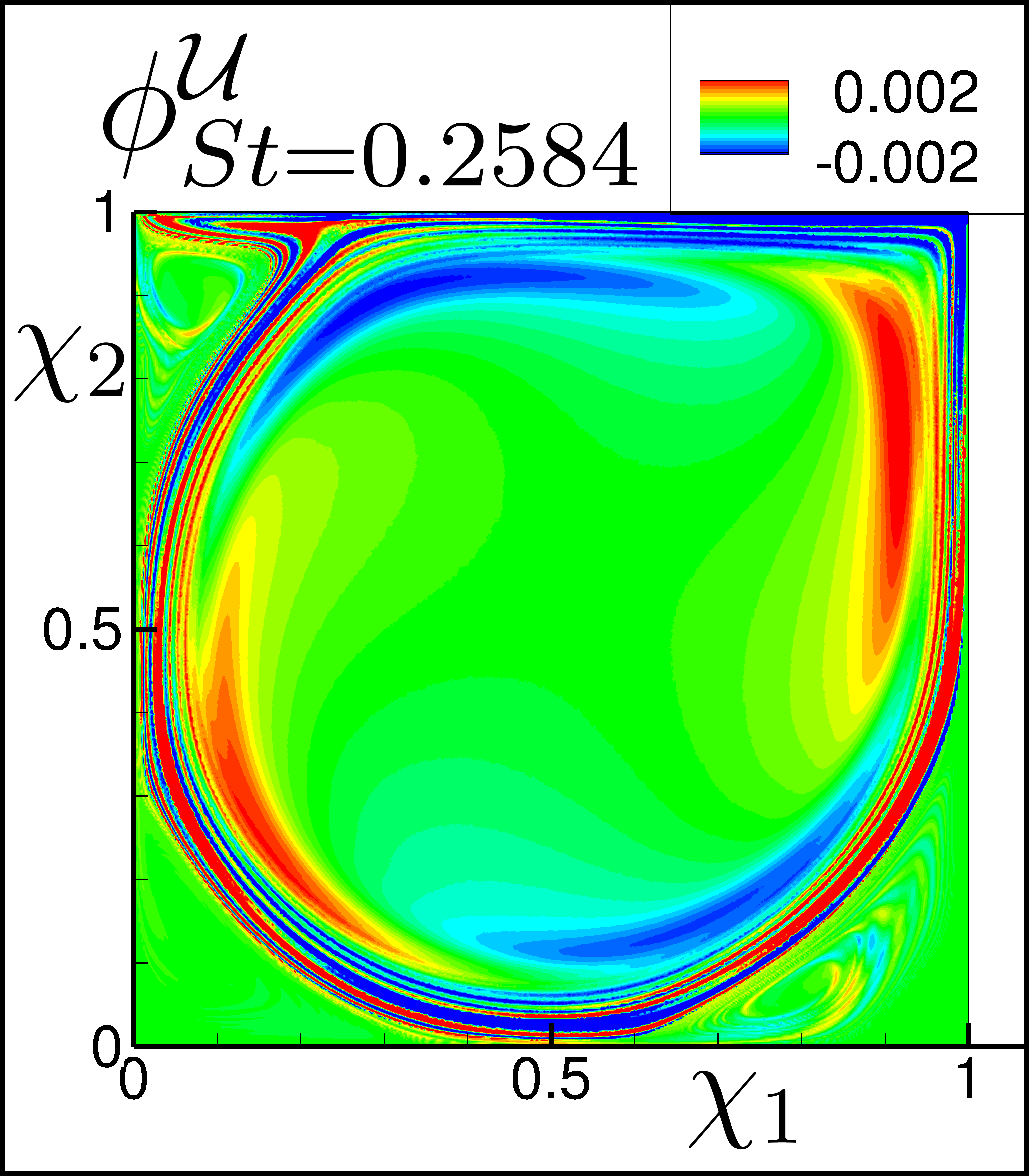}\\(e) $Re_L=11k$ $St_L=0.2584$
    \end{minipage}
    \begin{minipage}{0.192\textwidth}
        \centering
        \includegraphics[width=1.0\textwidth]{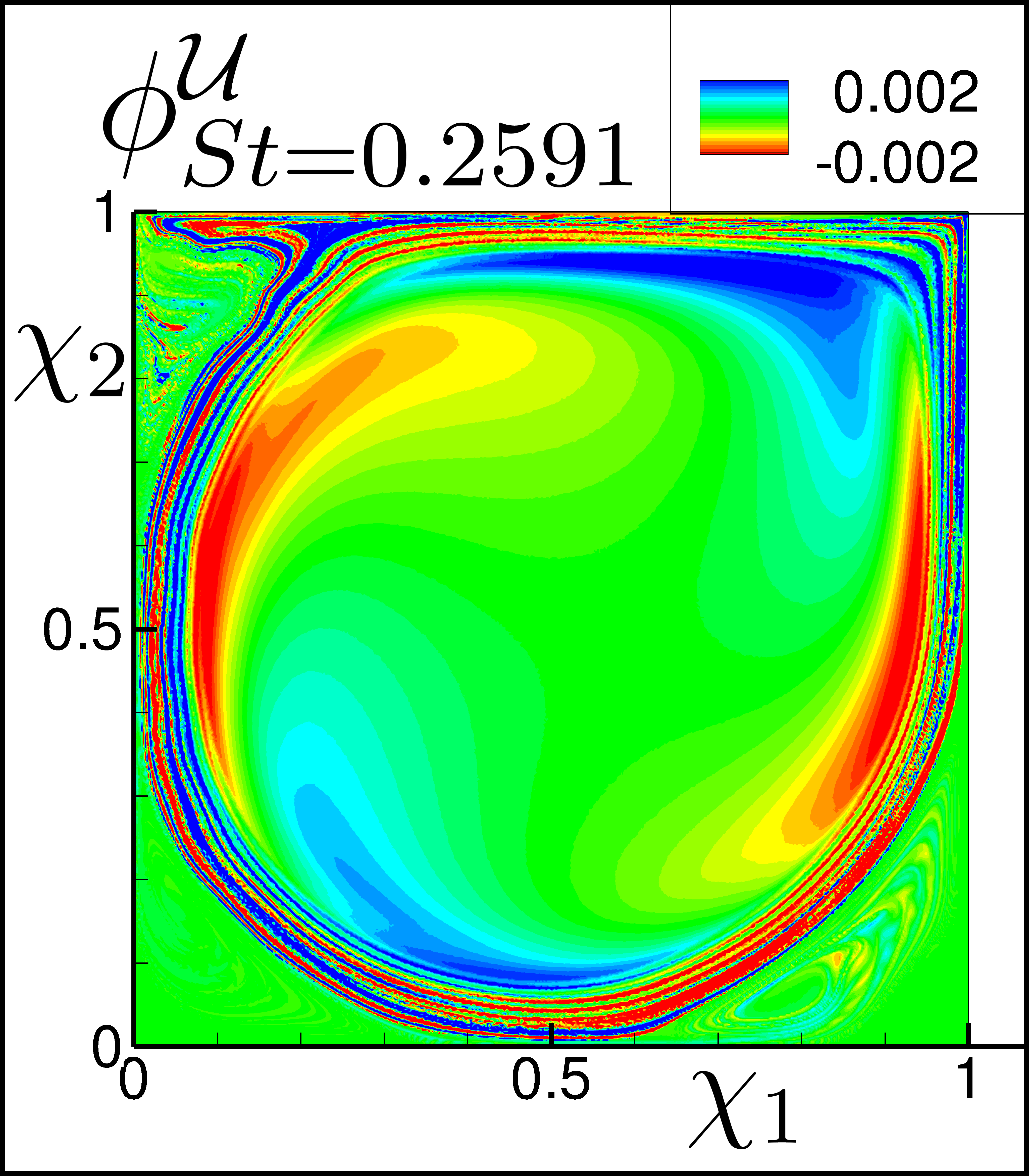}\\(f) $Re_L=12k$ $St_L=0.2591$
    \end{minipage}
    \caption{Marginally stable sub-critical and unstable post-critical LagSAT modes of lid-driven cavity for increasing Reynolds number from a sub-critical $Re_L=7{,}000$ to a post-critical $Re_L=12{,}000$.}
    \label{fig:ldc_phi2}
\end{figure}

The association of the first unstable mode and the core vortex dynamics, as performed by~\cite{shen1991hopf,ohmichi2017compressibility}, is based on the coinciding frequencies of the global unstable mode in the precritical regime and the flow oscillations in the post-critical regime.
In LagSAT framework, one can track the instability modes across the critical point, as shown in Fig.~\ref{fig:ldc_phi1}(b) to (f).
Furthermore, the spatial form of the first unstable mode remains nearly intact until the critical Reynolds number, and also to a large degree in the post-critical regime.
This is in contrast with the changing modal shape of the first unstable global mode for increasing Reynolds number, reported in~\cite{ohmichi2017compressibility}, where the unstable mode represents the core vortex dynamics at lower Reynolds number and a shear layer dynamics at higher Reynolds number.
Instead, LagSAT identifies additional unstable modes at sub/super-harmonic frequencies.
Here, two additional modes at higher harmonic frequencies of $St\approx 0.26$ and $St\approx 0.92$ also becomes unstable at $Re=12{,}000$ (Fig.~\ref{fig:ldc_phi1}a).
The second unstable mode at $St\approx 0.26$ is displayed in Fig.~\ref{fig:ldc_phi2} for increasing Reynolds number, where the spatial modal structure spans the shear layer region of the flow more than the core vortex region.
In addition, the unstable mode at $St\approx 0.92$ (not shown) also aligns more with the shear layer region compared to the vortex core, where the mode comprise smaller size and higher number of modal lobes consistent with the increase of modal frequency.

\begin{figure}
    \centering
    \includegraphics[width=1.0\linewidth]{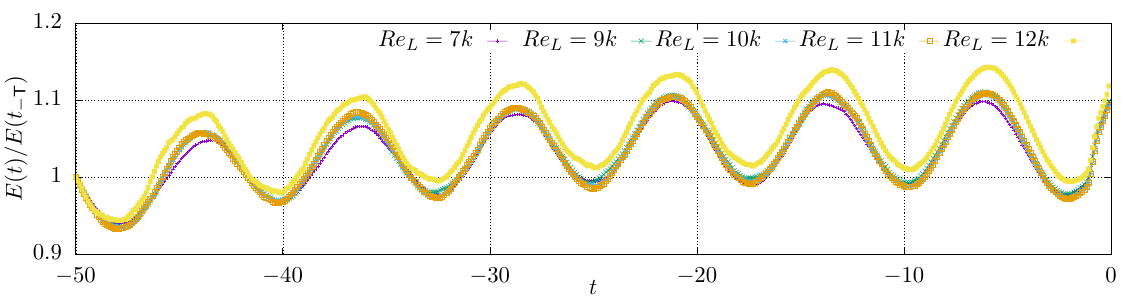}
    \caption{Total energy associated with Lagrangian flow map of lid-driven cavity for increasing Reynolds number.}
    \label{fig:ldc_en}
\end{figure}

The total energy, estimated using Eq.~\ref{eq:trans-energy}, for lid-driven cavity at various Reynolds numbers is displayed in Fig.~\ref{fig:ldc_en}.
The time evolution of the energy is nearly identical in all cases, except for the unstable case ($Re_L=12{,}000$) where the overall energy level is marginally higher (see Fig.~\ref{fig:ldc_en}).
It is also evident that the periodicity in the energy signal corresponds to the modal frequency of the least stable or unstable mode of Fig.~\ref{fig:ldc_phi1}(b).
The oscillatory form of the energy curves (Fig.~\ref{fig:ldc_en}), particularly at pre-critical Reynolds numbers, manifests in a time periodic/oscillatory flow for post-critical Reynolds numbers, suggesting the global absolute nature of the instability.

\section{Discussion and Concluding Remarks}\label{sec:concl}

Historically, flow stability has been perceived in terms of the response of a stable baseflow to external perturbation.
In contrast, the presented LagSAT elicits stability features of a baseflow without any perturbation by changing the reference frame from the Eulerian to the Lagrangian.
The Lagrangian viewpoint provides an access to the baseflow dynamics, or the dynamics that fluid parcels undergo as they travel collectively in the flow domain.
For instance, a steady {\it non-uniform} flow ({\it e.g.} a converging-diverging nozzle) does not accelerate or decelerate locally in the Eulerian reference frame, while the convective acceleration can be non-zero.
On the other hand, the fluid ``parcels" traveling with the flow ({\it i.e.}, in the Lagrangian reference frame) experience the acceleration or deceleration, exhibiting unsteady dynamics.
The originally proposed, Lagrangian modal analysis~\citep{shinde2021lagrangian} technique exploits the fact that a non-uniform Eulerian flow is unsteady in the Lagrangian frame of reference.
LagSAT utilizes some of the elements of LMA that include the adjoint form of Lagrangian DMD together with the forward time/flow implementation.

The LagSAT framework, presented in this paper, brings out the dynamics of Lagrangian flow map, considering it as a dynamical system with spacetime Lagrangian trajectories.
LagSAT finds a natural connection with Koopman operator theory~\citep{koopman1931hamiltonian,mezic2005spectral,williams2015data} via LDMD, where the Koopman observables are the Lagrangian state space variables.
In this case, the linear state space representation of the Koopman observables is a choice, while one can explore functional spaces with non-linear basis for a higher order approximation of the Koopman operator modes.
Nonetheless, the LagSAT eigenvalues/eigenvectors exhibit the stability characteristics of a dynamical system, which can be linear, nonlinear, chaotic, or turbulent.
Notably, LagSAT leads to a precise mathematical expression to calculate N-factor, where a perturbation-free N-factor of a LagSAT mode at a frequency is its growth rate ($\omega_r$) times the transient time duration ($\mathsf{T}$).
In other words, for a given perturbation that aligns with the LagSAT modal frequency and spatial form, the initial perturbation amplitude $\mathcal{A}_0$ grows to $\mathcal{A}$ following the linear modal growth rate of $\omega_r$ over time $\mathsf{T}$.

The LagSAT framework comprises three main steps: 1. build a spatio-temporal Lagrangian flow map in the adjoint setting from a given Eulerian baseflow, 2. select an appropriate region of the Lagrangian flow map for a local/global or a spatial/temporal stability analysis, and lastly 3. employ the Lagrangian modal/non-modal analyses ({\it e.g.} LDMD/LPOD) in the adjoint form but in the forward direction of the flow/time.
The intersection of the two frames of reference is the identity map, {\it i.e.}, the baseflow state at the current time $t_0$, in Lagrangian sense.
By constructing a Lagrangian flow map backward (adjoint) in time and performing the Lagrangian analyses forward in time conveniently leads to the stability features of the baseflow at the identity map.
The Lagrangian flow map forms a dynamical system whose eigenspectra and spatial modes dictate its stability, where LagSAT via LDMD linearly approximates the Koopman operator eigenspectra and modes.
The stability features include the stable/unstable spatio-temporal mode, the associated frequency/wavenumber, and the growth rate information.
Consistent with the original formulation of DMD~\citep{schmid2009dynamic,schmid2010dynamic}, the LagSAT/LDMD modes need not be orthonormal.
The non-normality of LagSAT modes can lead to the mechanism of transient energy growth~\citep{schmid2007nonmodal}, as exemplified in this article by means of, for instance, the 2D self-similar (Blasius/Falkner-Skan) boundary layer under favorable pressure gradient.

To delve into the notion of convective/absolute instabilities, LagSAT is performed on three flow configurations, namely, the self-similar Blasius/Falkner-Skan boundary layers, a 2D compressible flow past a cylinder, and a 2D compressible lid-driven cavity flow.
The hydrodynamic instabilities in these flow configurations of the following nature: convective, convective/absolute, and absolute nature, respectively.
The ability to employ LagSAT on a specific component of velocity is particularly useful to probe the nature of instability. 
For the velocity component aligned with the flow direction, LagSAT alludes to the convective nature of the instability.
On the other hand, it refers to the absolute nature of instability for the velocity component normal to the flow direction.
The open flows considered here, the self-similar boundary layers and the flow past a cylinder, manifest global stable/unstable modes those are of convective nature.
On the contrary, the lid-driven cavity flow exhibits no unstable modes until a critical Reynolds number, where the flow undergoes transition from a steady state to an unsteady state by means of a global absolute instability.
In the flow past a cylinder case, the streamwise velocity is associated with the convective instability, while the crossflow component is aligned with the oscillatory (absolute) nature of instability.
The transition point where the flow past a cylinder becomes unsteady (here at the critical Reynolds number of $Re_c=53$, Mach number $M=0.5$), some of the LagSAT modes of the crossflow component of velocity loose their stability.
It appears that the compressible flow past a cylinder at $M=0.5$ exhibits an absolute instability, which begins at the inception of two counter-rotating recirculation regions behind the cylinder at $Re\approx 8$.
The flow remains stable until the critical Reynolds number of $Re_c=53$, where the flow behind cylinder becomes unsteady.
For the low Reynolds numbers $Re\leq 7$ with no flow separation, the flow remains steady with convective instability.

An interesting feature of LagSAT is that it can be applied to the unsteady baseflow, which relieves the non-uniformity constraint due to the unsteadiness.
In this case, the formation of Lagrangian flow map must account for the time-evolving baseflow.
The ability to treat unsteady baseflows is a powerful feature of LagSAT from the flow control point of view. 
One can trace the origin of unstable modes from a post-critical regime into the pre-critical regime, as demonstrated in Fig.~\ref{fig:phi1}.
The frequency of the first least stable/unstable mode of the streamwise velocity (Fig.~\ref{fig:phi1}a-e) changes considerably for the increase of Reynolds number from $Re=20$ to $Re=60$.
In contrast, the lid-driven cavity flow exhibits much small variation in the frequency of the first/second least stable/unstable modes over a wide range of Reynolds numbers (Figs.~\ref{fig:ldc_phi1},~\ref{fig:ldc_phi2}).
To some degree, this alludes to the inaccurate prediction of the frequency of unstable modes by the conventional linear stability analysis and the issue of the choice of baseflow~\citep{juniper2014modal}.

The extension of LagSAT to post-critical unsteady/turbulent regime is realized in a straight forward manner, partly since the method is data-driven and does not involve the governing equations for the flow/perturbation and associated constraints that are typically present in the formation of convectional stability analysis techniques.
For instance, the distinction between a linear and a non-linear instability is not associated with LagSAT, unlike the LST, linear/non-linear PSE, and similar stability analysis techniques that are based on linearization of the Navier-Stokes equations.
Thus, LagSAT can be employed to investigate various mechanisms of the flow transition process, including the linear/nonlinear, transient, or bypass mechanisms~\citep{morkovin1994transition}.
Furthermore, it can be applied to any numerical/experimental flowfields irrespective of the numerical models/experimental conditions that were used to produce the baseflow.
Notably, LagSAT framework provides a mathematical expression for N-factor calculation (Eq.~\ref{eq:N-fac2}), where the modal growth rate and time-duration of the growth lead to an N-factor value for a single mode at a frequency.
Interestingly, this is a perturbation-free approach of obtaining N-factor, even though the derivation of N-factor (Eq.~\ref{eq:N-fac2}) builds on the conventional definition of N-factor comprising the initial ($\mathcal{A}_0$) and final ($\mathcal{A}$) amplitudes of perturbation.
In the same vein, LagSAT provides access to the total energy evolution over the Lagrangian flow map (from the viewpoint of dynamical system), leading to insights into the transient energy growth/decay analysis.

In this paper, LagSAT operates on the flow velocities, primarily to established the connections with the conventional stability work; however, LagSAT on the passive ({\it e.g.}, passive/active scalars, Temperature, etc.) and/or derived ({e.g.}, vorticity, eddy viscosity, etc.) variables can provide valuable physical insights.
In addition, a function space representation of Koopman observables and the selection of suitable basis may produce insightful results from the system stability and control point of view~\citep{brunton2021modern}.
The parallel implementation of LagSAT is based on a highly efficient/scalable singular/eigen-value solving algorithms in~\cite{shinde2025distributed}, making the method suitable for the global analysis of the baseflows with a large and complex domain, e.g., high-speed boundary transition geometries~\citep{choudhari2020streak,wheaton2021final}.
Furthermore, LagSAT can be used to analyze the flow over a deforming surface, in the realm of fluid structure interactions~\citep{visbal2014viscous,shinde2019transitional}, where the conventional methods generally fall short due to the domain deformation.
The Lagrangian ansatzes (LMA, LagSAT) naturally integrate the domain deformation in the mathematical frameworks.

\section*{Acknowledgment}
This work is supported by the National Aeronautics and Space Administration (NASA) under Grant No. 80NSSC24M0103.
The author sincerely thanks Dr. Meelan Choudhari (NASA Langley Research Center) for the invaluable discussions and for monitoring this effort. 
Technical exchanges with Prof. Adrian Sescu (Mississippi State University), Prof. Nathan Murray (University of Mississippi), and Dr. Paritosh Mokashi (University of Mississippi) are appreciated.
The computing resources provided by the High Performance Computing Center (HPCC) of Mississippi State University (MSU) are acknowledged.

\appendix
\section{Self-Similar Blasius and Falkner-Skan Boundary Layers} \label{sec:Blasius_baseflow}

Fluid flow in the vicinity of (no-slip) wall forms a thin region dominated by viscosity, the boundary layer region.
Following the order of magnitude analysis~\citep{prandtl1905uber}, many terms of the full Navier-Stokes equations can be neglected, leading to a much simpler set of boundary layer equations.
For a steady incompressible flow with constant viscosity and density, these equations in 2D take the following form:
\begin{equation}\label{eq:mass_pr}
    \frac{\partial u}{\partial x}+\frac{\partial v}{\partial y}=0,
\end{equation}
and 
\begin{eqnarray}\label{eq:momentum_pr}
    u\frac{\partial u}{\partial x}+v\frac{\partial v}{\partial y}&=&-\frac{1}{\rho}\frac{\partial p}{\partial x} + \nu\frac{\partial^2 u}{\partial y^2} \\
    0&=&-\frac{1}{\rho}\frac{\partial p}{\partial y},
\end{eqnarray}
as the continuity and momentum equations, respectively.
Assuming a constant pressure in the streamwise direction, {\it i.e.}, $\partial p/\partial x=0$, ~\citet{blasius1907grenzschichten} showed that the boundary layer equations have a self-similar solution.
This is due to the invariance of the equations and boundary conditions under the following transformation:
$$x\rightarrow c^2 x, \hspace{5mm} y\rightarrow cy, \hspace{5mm} u\rightarrow u, \hspace{5mm} \text{and} \hspace{5mm} v\rightarrow \frac{v}{c},$$
where $c$ is an arbitrary positive constant.
A self-similar length scale and a stream function were introduced, respectively, as
\begin{equation}\label{eq:bl_eta_psi}
\eta=\frac{y}{\sqrt{\nu x/u_e}}, \hspace{5mm} \text{and} \hspace{5mm} f(\eta)=\frac{\Psi}{\sqrt{\nu x u_e}},
\end{equation}
where $\nu$ is the kinematic viscosity.
$u_e$ and $\Psi$ stand for the outer boundary layer (edge) velocity and the stream function, respectively.
This led to the following Ordinary Differential Equation (ODE) of the boundary layer (or well known as Blasius boundary layer):
\begin{equation}\label{eq:BlasiusBL}
    2\frac{d^3 f(\eta)}{d \eta^3}+f(\eta)\frac{d^2 f(\eta)}{d \eta^2} = 0.
\end{equation}
The no-slip boundary condition, {\it i.e.}, $u=0$ and $v=0$ is employed by setting $df(\eta)/{d\eta}=0$ and $f(\eta)=0$, respectively, whereas the far field boundary condition, {\it i.e.}, $u=u_e$ becomes $df(\eta)/{d\eta}=1$.

\begin{figure}
    \centering
    \begin{minipage}{0.40\textwidth}
        \centering
        \includegraphics[width=1.0\textwidth]{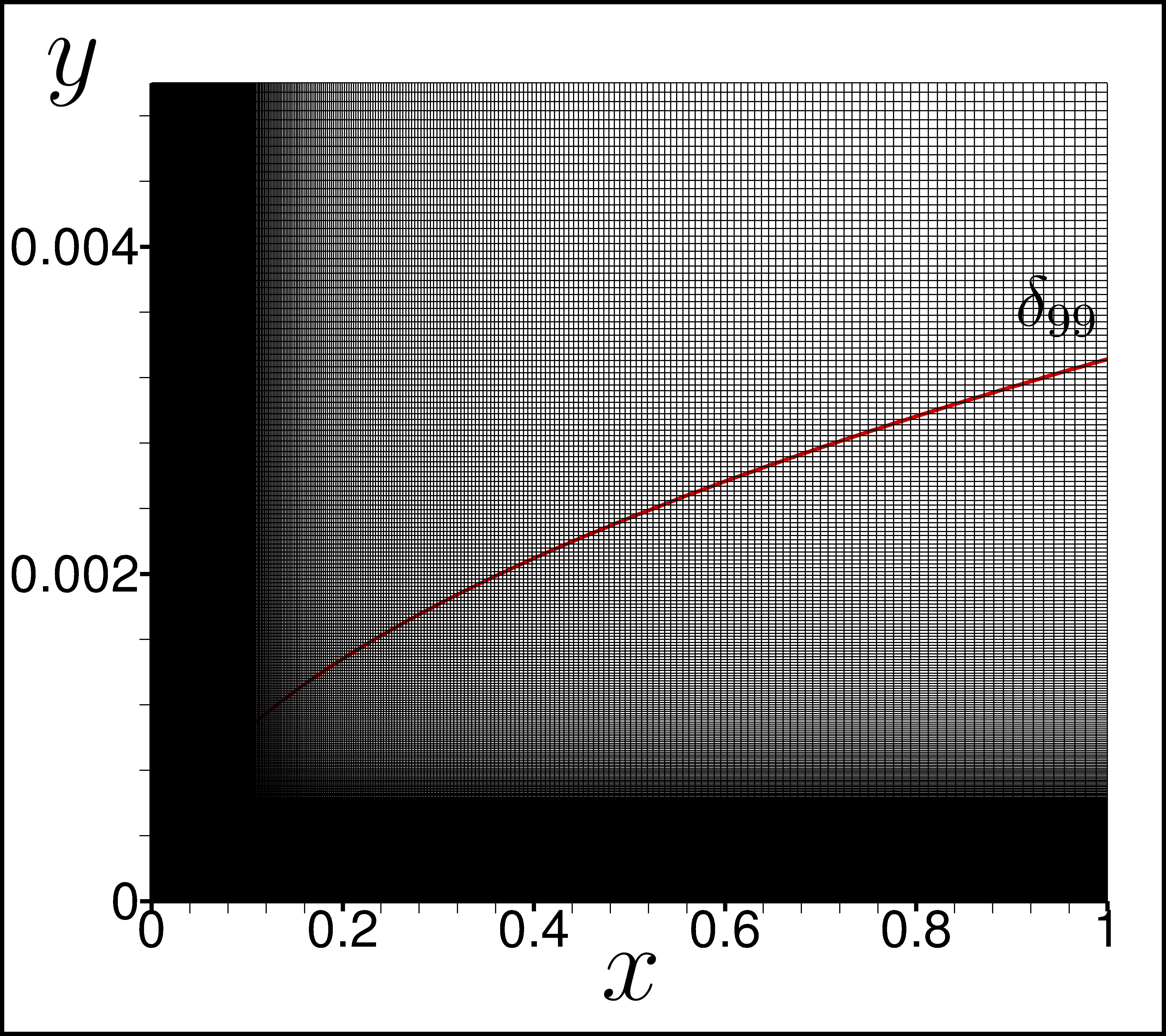}\\(a) $(N_x\times N_y) = (1000\times 1000)$
    \end{minipage}
    \begin{minipage}{0.58\textwidth}
        \centering
        \includegraphics[width=0.85\textwidth]{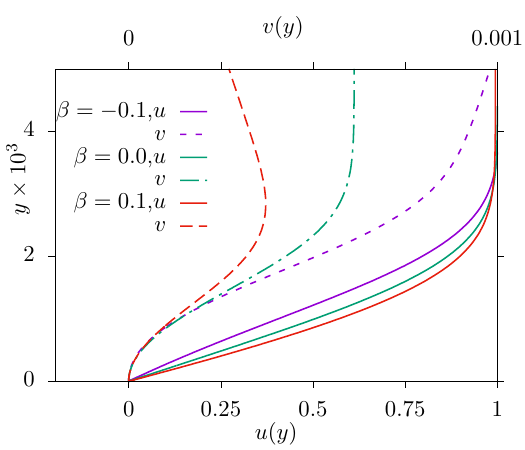}\\(b) $u(y)$, $v(y)$ at $x=0.91$
    \end{minipage}
    \caption{Computational mesh and velocity profiles of the self-similar boundary layer profiles (Blasius and Falkner-Skan) for $\beta=[-0.1,0.0,0.1]$. (a) Computational mesh - $N_x=1000,N_y=1000$, (b) Blasius and Falkner-Skan velocity profiles under favorable/adverse pressure gradient.}
    \label{fig:Blas_Falk_Skan}
\end{figure}

The numerical solution to the third-order non-linear ODE, the Blasius boundary layer, is obtained by utilizing the shooting method.
To impose the condition at the far field, different trials for the slope values are performed until the far field condition is matched.
An off-the-shelf python ODE solver functionality, odeint, is used.
It is known to employ Fortran based ODEPACK solvers.
For the set convergence criteria, namely, the maximum error of $10^{-8}$ and the maximum number of $500$ iterations, the solver typically converges within $\lessapprox 30$ iterations.
A sample number mesh is displayed in Fig.~\ref{fig:Blas_Falk_Skan}(a), comprising $1000\times 1000$ mesh points in the streamwise and wall-normal directions.
The mesh is created using a logarithmic function of the distance from the leading edge, producing a highly refined mesh in the leading edge region both the streamwise and wall-normal directions.
Thus, the boundary layer typically comprises over $90\%$ of the total number grid points, as shown in Fig.~\ref{fig:Blas_Falk_Skan}(a) with reference to the boundary layer thickness curve - $\delta_{99}$.
The complete solution to the Blasius boundary layer is presented in Fig.~\ref{fig:Blasius_xy} in terms of the 2D $u(x,y)$ profile (in Fig.~\ref{fig:Blasius_xy}a), 1D $u(y)$ and $v(y)$ profiles (in Fig.~\ref{fig:Blasius_xy}b), and 1D $u(x)$ profile (in Fig.~\ref{fig:Blasius_xy}c).
As noted before, the freestream conditions were conveniently selected to encompass the appropriate Reynolds number range.

A generalized form of the Blasius boundary layer was derived by~\cite{falkneb1931lxxxv} by retaining the streamwise pressure gradient term and modeling it as:
\begin{equation}\label{eq:falk-skan}
    -\frac{1}{\rho}\frac{\partial p}{\partial x} = u_e\frac{du_e}{dx}.
\end{equation}
The boundary layer edge velocity, $u_e$, was obtained by means of an inviscid flow over a wedge with an angle $\beta$ as:
\begin{equation}\label{eq:axm}
    u_e (x) = cx^m,
\end{equation}
where $c$ is an arbitrary field and $m$ is a dimensionless power.
The wedge angle $\beta$ and $m$ are related by:
\begin{equation}\label{eq:beta_m}
    \beta = \frac{2m}{m+1}.
\end{equation}
Thus, the pressure gradient term of Eq.~\ref{eq:falk-skan} takes the following form:
\begin{equation}\label{eq:press_grad_term}
    -\frac{1}{\rho}\frac{d p}{d x} = m\frac{u_e^2}{x}.
\end{equation}
The self-similar length scale and stream function of the Falkner-Skan boundary layer in a parameterized form can be given as:
\begin{equation}\label{eq:fs_eta_psi}
 \eta = y\sqrt{\frac{u_e}{(2-\beta)\nu x}} \hspace{5mm}\text{and}\hspace{5mm} f(\eta)= \frac{\psi}{\sqrt{(2-\beta)u_e \nu x}}
\end{equation}
Incorporating the pressure gradient term, and the scaling parameters above lead to the following Falkner-Skan equation:
\begin{equation}\label{eq:fs}
        \frac{d^3 f(\eta)}{d \eta^3}+f(\eta)\frac{d^2 f(\eta)}{d \eta^2}+\beta\left[ 1-\left(\frac{df(\eta)}{d \eta}\right)^2 \right]=0,
\end{equation}
which can be solved by employing essentially the same set of boundary conditions and numerical procedure that are used for the Blasius boundary layer.

The effect of pressure gradient via the wedge angle parameter $\beta$ on the self-similar boundary layer is shown in Fig.~\ref{fig:Blas_Falk_Skan}(b) for three values of $\beta=[-0.1,0.0,0.1]$.
The changes in the streamwise and wall-normal velocities are evident.
The selected $\beta$ values represent both the adverse ($\beta=-0.1$) and favorable ($\beta=0.1$) pressure gradient effects.
As expected, the Falkner-Skan profile returns to the zero pressure gradient case of Blasius boundary layer for $\beta=0$.
The smaller values of $\beta$ are conveniently chosen to ensure the same numerical treatment and accuracy of the solution.
The numerical solution at the far field is sensitive to the initial guess of solution, and may exhibit a solution divergence for ill-suited initial slopes.
In addition, the ODE solver may not converge with a desired accuracy for large negative values of $\beta$ (for instance, $\lessapprox -0.2$) representing a strong adverse pressure, or large positive values of $\beta$ (for instance, $\gtrapprox 1.0$) representing a strong favorable pressure.
Nonetheless, informed initial guesses and tracking of errors can help improve the operating range of the $\beta$ parameter.

\section{Numerical Sensitivity of LagSAT} \label{sec:Num_Sense}

The LagSAT procedure involves gathering of a number of snapshots that represent the Lagrangian flow map.
For LagSAT on Eulerian steady baseflows, an arbitrary time duration ($\mathsf{T}$) can be employed, where a finite sized computational/experimental domain or a physically relevant time duration can be adopted to perform the analysis.
For time dependent baseflows, such as the periodic or unsteady flows, LagSAT can be performed in a transient setting for a shorter time duration.
Nonetheless, the results of LagSAT depend of the choice of numerical parameters, namely, the space, time, and/or frequency discretizations.
Assuming the Lagrangian flow map is well resolved in terms of the Nyquest criterion both in the space and time, the LagSAT results generally depend on the number of snapshots utilized.
The number of snapshots for a fixed time-duration correspond to the spectral resolution at higher frequencies in the LagSAT analysis.
Here we present the numerical sensitivity of LagSAT results to 1) the number of snapshots at a fixed time-step, 2) the time-step at a fixed number of snapshots, and 3) the number of snapshots at a fixed time-duration.

\begin{figure}
    \centering
    \begin{minipage}{0.24\textwidth}
    \centering
    \includegraphics[width=1.0\linewidth]{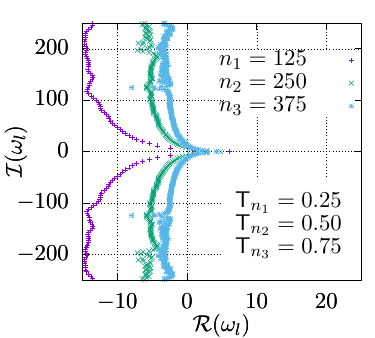}\\(a) Fixed $\delta t$ eigen-spectra
    \end{minipage}
    \begin{minipage}{0.24\textwidth}
    \centering
    \includegraphics[width=1.0\linewidth]{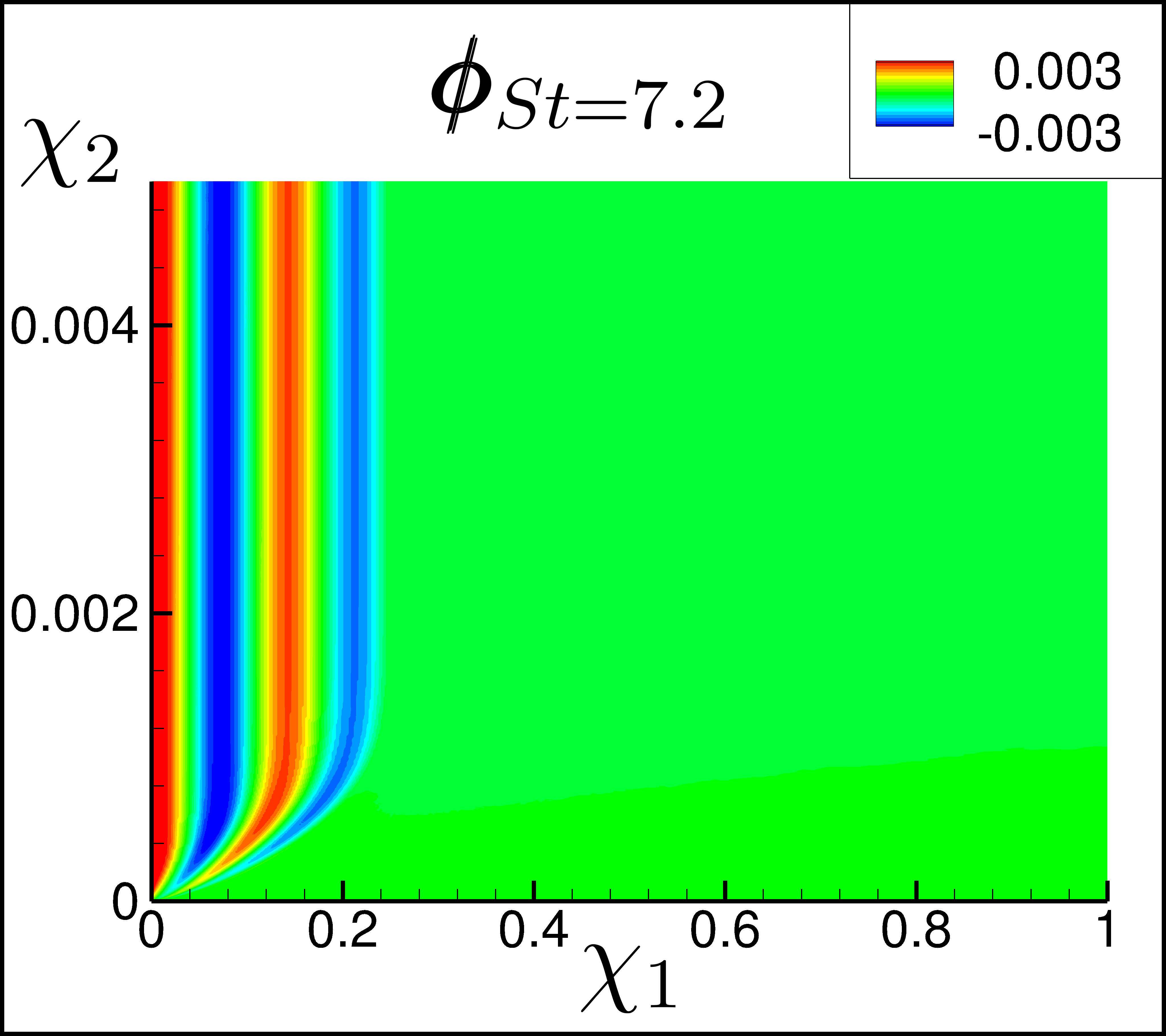}\\(b) $n_1=125$, $\mathsf{T}=0.25$
    \end{minipage}
    \begin{minipage}{0.24\textwidth}
    \centering
    \includegraphics[width=1.0\linewidth]{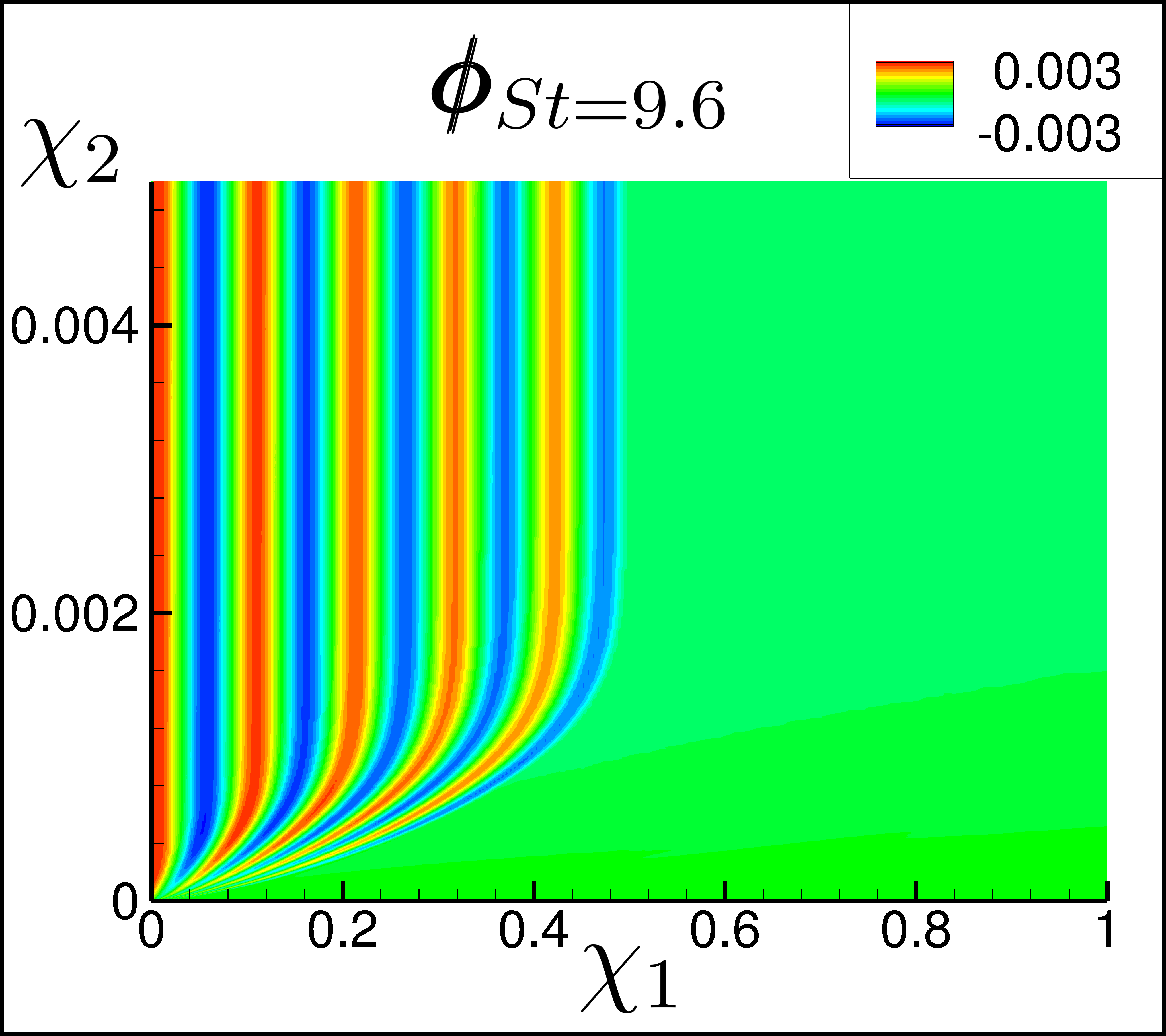}\\(c) $n_2=250$, $\mathsf{T}=0.50$
    \end{minipage}
    \begin{minipage}{0.24\textwidth}
    \centering
    \includegraphics[width=1.0\linewidth]{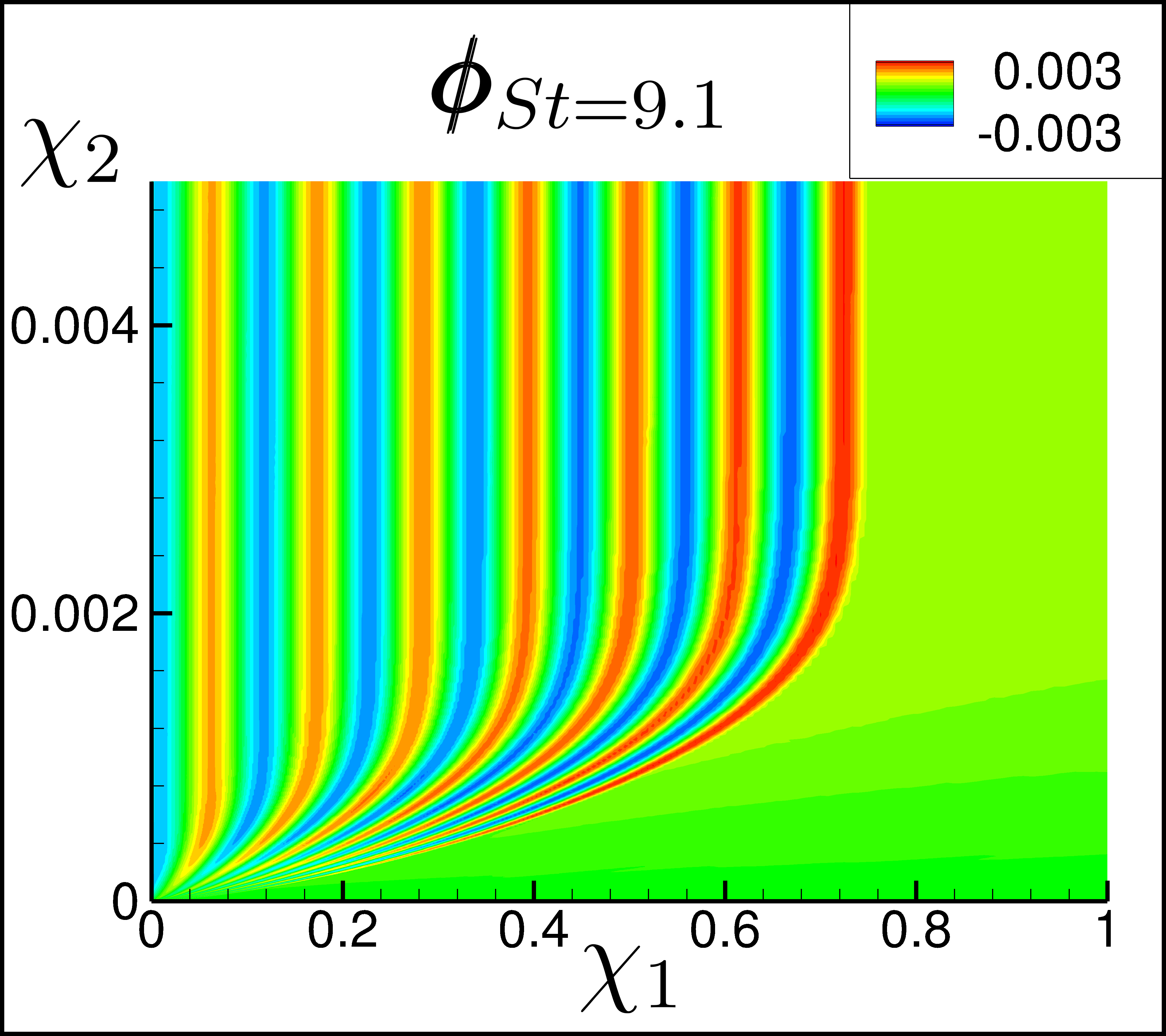}\\(d) $n_3=375$, $\mathsf{T}=0.75$
    \end{minipage}
    \begin{minipage}{0.24\textwidth}
    \centering
    \includegraphics[width=1.0\linewidth]{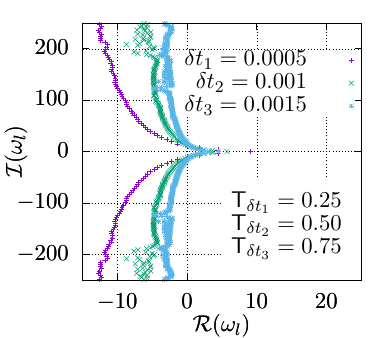}\\(e) Fixed $n$ eigen-spectra
    \end{minipage}    
    \begin{minipage}{0.24\textwidth}
    \centering
    \includegraphics[width=1.0\linewidth]{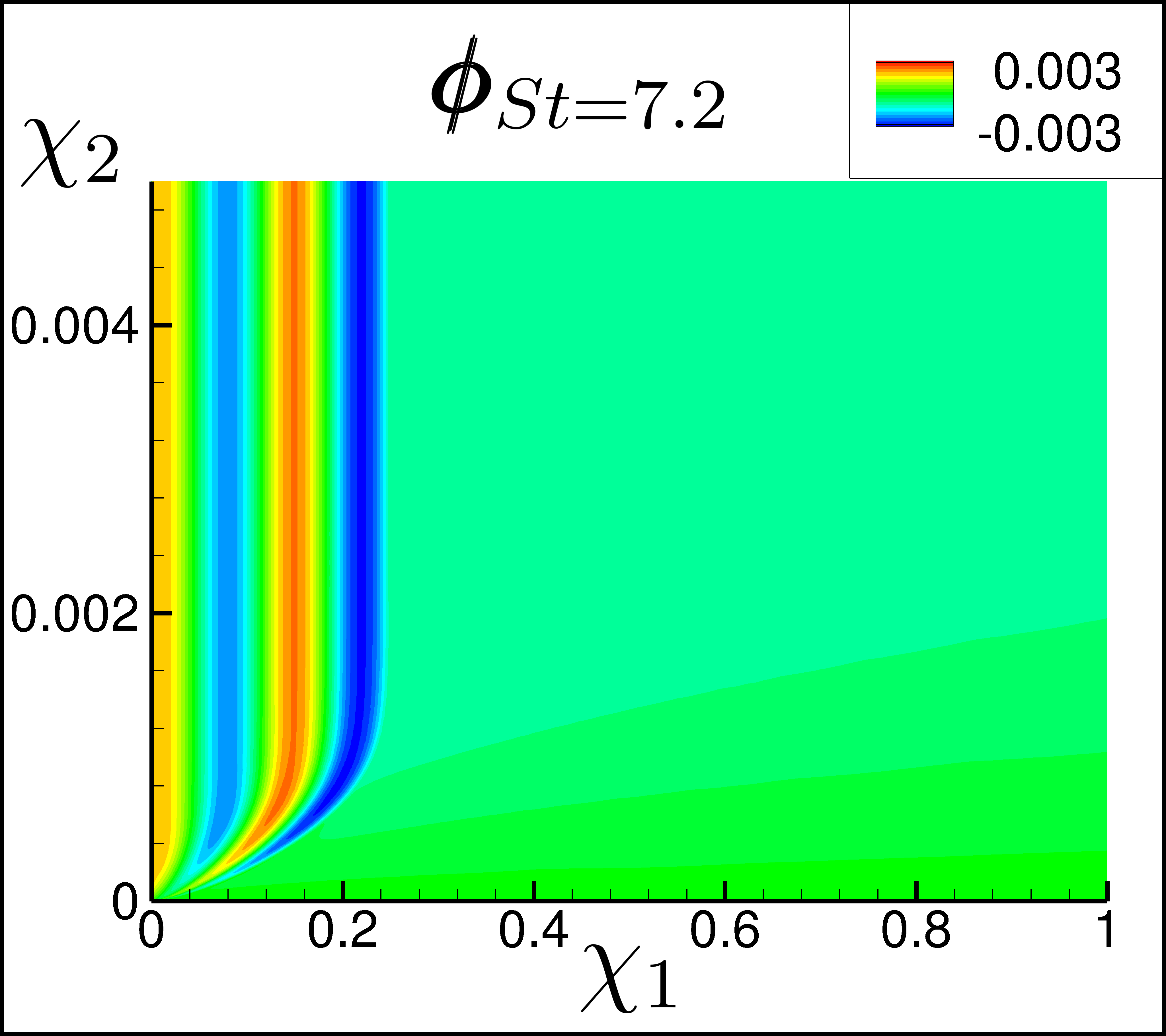}\\(f) $\delta t_1=0.0005$, $\mathsf{T}=0.25$
    \end{minipage}        
    \begin{minipage}{0.24\textwidth}
    \centering
    \includegraphics[width=1.0\linewidth]{figs/aldmd_St9d6_beta0d0T0d50.png}\\(g) $\delta t_2=0.001$, $\mathsf{T}=0.50$
    \end{minipage}    
    \begin{minipage}{0.24\textwidth}
    \centering
    \includegraphics[width=1.0\linewidth]{figs/aldmd_St9d1_beta0d0T0d75.png}\\(h) $\delta t_3=0.0015$, $\mathsf{T}=0.75$
    \end{minipage}
    \caption{Effect of time-duration in terms of the number of snapshots and time-step size on LagSAT modes. (a) Eigen spectra at constant time-step and LagSAT modes: (b) $\phi_{St=7.2}$ for $\mathsf{T}=0.25$, (c) $\phi_{St=9.6}$ for $\mathsf{T}=0.50$, and (d) $\phi_{St=9.1}$ for $\mathsf{T}=0.75$. (e) Eigen spectra at constant number of snapshots and LagSAT modes: (f) $\phi_{St=7.2}$ for $\mathsf{T}=0.25$, (g) $\phi_{St=9.6}$ for $\mathsf{T}=0.50$, and (h) $\phi_{St=9.1}$ for $\mathsf{T}=0.75$.}
    \label{fig:num_nt_delta}
\end{figure}

The sensitivity analysis of LagSAT is presented in terms of the 2D Blasius boundary layer at the spatial resolution of $1000 \times 1000$ using non-uniform mesh points, as discussed in Appendix~\ref{sec:Blasius_baseflow} and in the results section (Sec.~\ref{sec:lagsat_BlaFalSka}).
The streamwise extent and freestream velocity that are used for the 2D Blasius boundary layer are, respectively, $L=1$ and $u_\infty=1$.
Thus, the time duration of $\mathsf{T}=1.0$ approximately corresponds to the flow residence time, allowing Lagrangian modes to evolve over the entire computational domain, $\mathcal{D}$.
For smaller values of $\mathsf{T}$, the Lagrangian flow evolves over corresponding shorter flow distances; this is consistent with the spatio-temporal nature of the LagSAT analysis.
As noted before, the time duration $\mathsf{T}$ is a physical rather than numerical parameter. 
It dictates the time/space horizon of the LagSAT analysis.
Nonetheless, the numerical sensitivity of the results can be compared for a fixed value of time duration, even over a short time period.

Figure~\ref{fig:num_nt_delta} presents LagSAT modes that are evolving for increasing time duration with values $\mathsf{T}=[0.25,0.50,0.75]$.
The increase of $\mathsf{T}$ can be achieved by keeping the time-step constant and increasing the number of snapshots, and vice a versa.
The eigen spectra for the fixed time-step $\delta t=0.002$ cases are displayed in Fig.~\ref{fig:num_nt_delta}(a), while the representative LagSAT modes are shown in Figs.~\ref{fig:num_nt_delta}(b), (c), and (d) for the time durations of $\mathsf{T}=0.25$, $\mathsf{T}=0.50$, and $\mathsf{T}=0.75$, respectively.
In the second row (of Fig.~\ref{fig:num_nt_delta}), the eigen spectra for the fixed number of snapshot $\delta t=500$ cases are displayed in Fig.~\ref{fig:num_nt_delta}(e), and the representative LagSAT modes are shown in Figs.~\ref{fig:num_nt_delta}(f), (g), and (h) for the respective time durations.
The modal streamwise extent from the leading edge increases with the time duration, and consequently the local Reynolds number $Re_\chi$ for the LagSAT analysis.
Accordingly, the modal growth rates appear to increase for the increase of time duration $\mathsf{T}$, transitioning these modes toward lesser stability.
For a particular time duration, the effect of time-resolution (or equivalently the number of snapshots) can be compared between the two rows of Fig.~\ref{fig:num_nt_delta}.
The specific values of modal growth rate and N-factor for these cases are included in Table~\ref{tab:N-fac}.
In general, the larger values of time step and/or lower values of the number of snapshots appear to introduce errors in the LagSAT analysis, where a finer time-step and a higher number of snapshots lead to converging results in terms of the modal growth rate and N-factor values.

\begin{table}
    \centering
    \begin{tabular}{cccccccc}
        \hline
        Case \# & Timestep  & Total Snapshots & Total time  & Strouhal & Growth rate & n-factor \\
        \hline
        & ($\delta t$)  & ($n$) & ($\mathsf{T}$) & ($St$) & ($\omega_r$) & ($n=\omega_r \mathsf{T}$) \\ [1.ex]
        \hline
        0. & 0.002 & 500 & 1.0 & 9.8 &  1.0252 & 1.0252 \\[.5ex]
        1. & 0.002 & 125 & 0.25 & 7.2 & -2.3934 & -0.5983\\
        2. & 0.002 & 250 & 0.50 & 9.6 & -0.6470 & -0.3235\\
        3. & 0.002 & 375 & 0.75 & 9.1 &  0.4310 & 0.3233\\[.5ex]
        4. & 0.0005 & 500 & 0.25 & 7.2 & 1.7876 & 0.4469\\
        5. & 0.001 & 500 & 0.50 & 9.6 &  0.4498 & 0.2249\\
        6. & 0.0015 & 500 & 0.75 & 9.1 & 0.7301 & 0.5476\\[.5ex]
        7. & 0.008 & 125 & 1.0 & 9.9 & -0.2186 & -0.2186 \\
        8. & 0.004 & 250 & 1.0 & 9.9 & 0.4468 & 0.4468 \\
        9. & 0.00267 & 375 & 1.0 & 9.8 & 0.9038 & 0.9038\\
        \hline
    \end{tabular}
    \caption{Effect of time-step size, number of snapshots, and total time duration on n-factor}
    \label{tab:N-fac}
\end{table}

\begin{figure}
    \begin{minipage}{0.24\textwidth}
    \centering
    \includegraphics[width=1.0\linewidth]{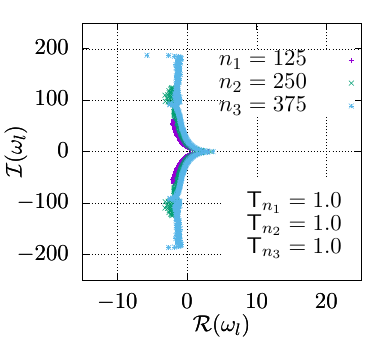}\\(a) Fixed $\mathsf{T}$ eigen-spectra
    \end{minipage}
    \begin{minipage}{0.24\textwidth}
    \centering
    \includegraphics[width=1.0\linewidth]{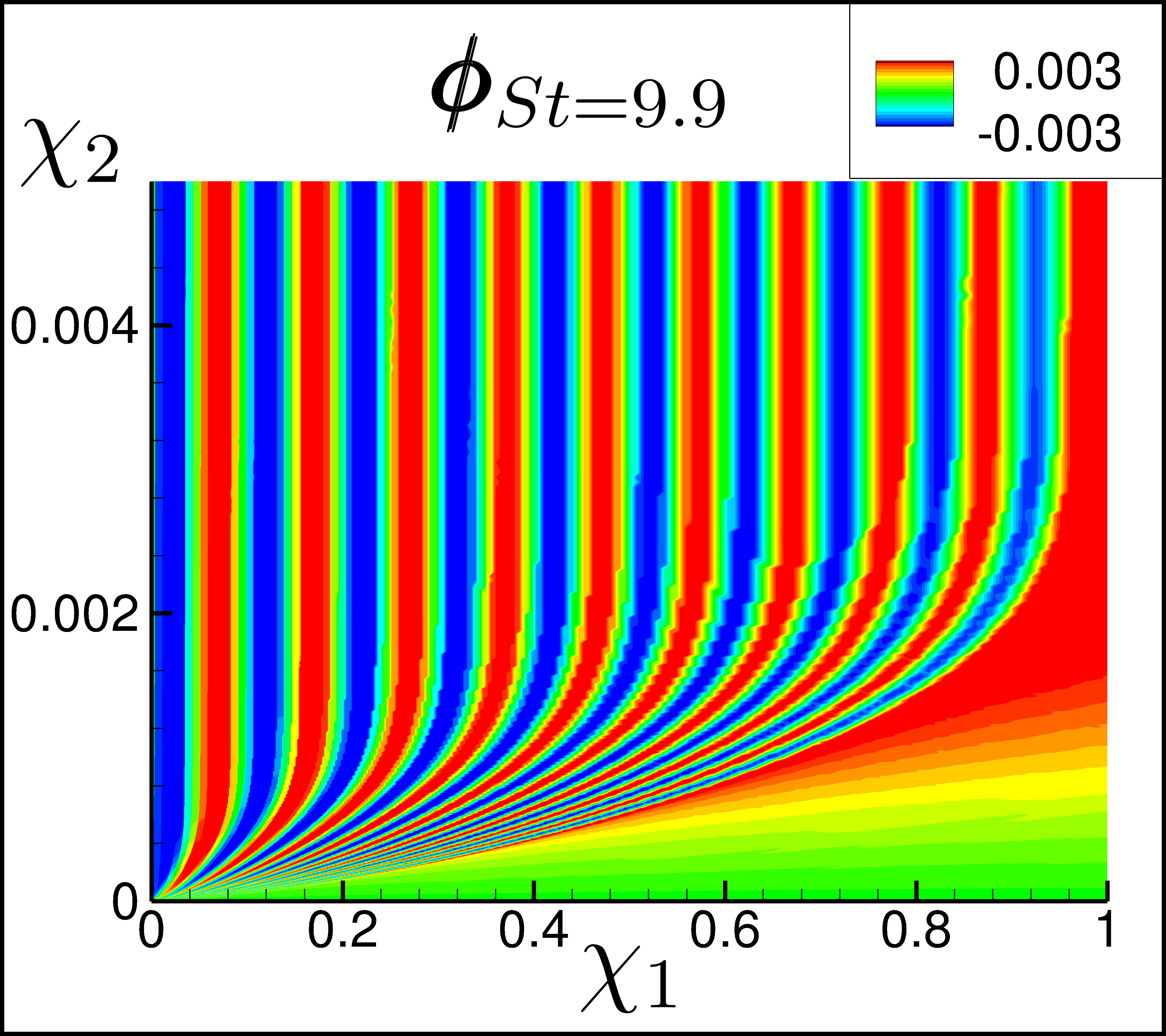}\\(b) $n_1=125$, $\delta t_1=0.008$
    \end{minipage}
    \begin{minipage}{0.24\textwidth}
    \centering
    \includegraphics[width=1.0\linewidth]{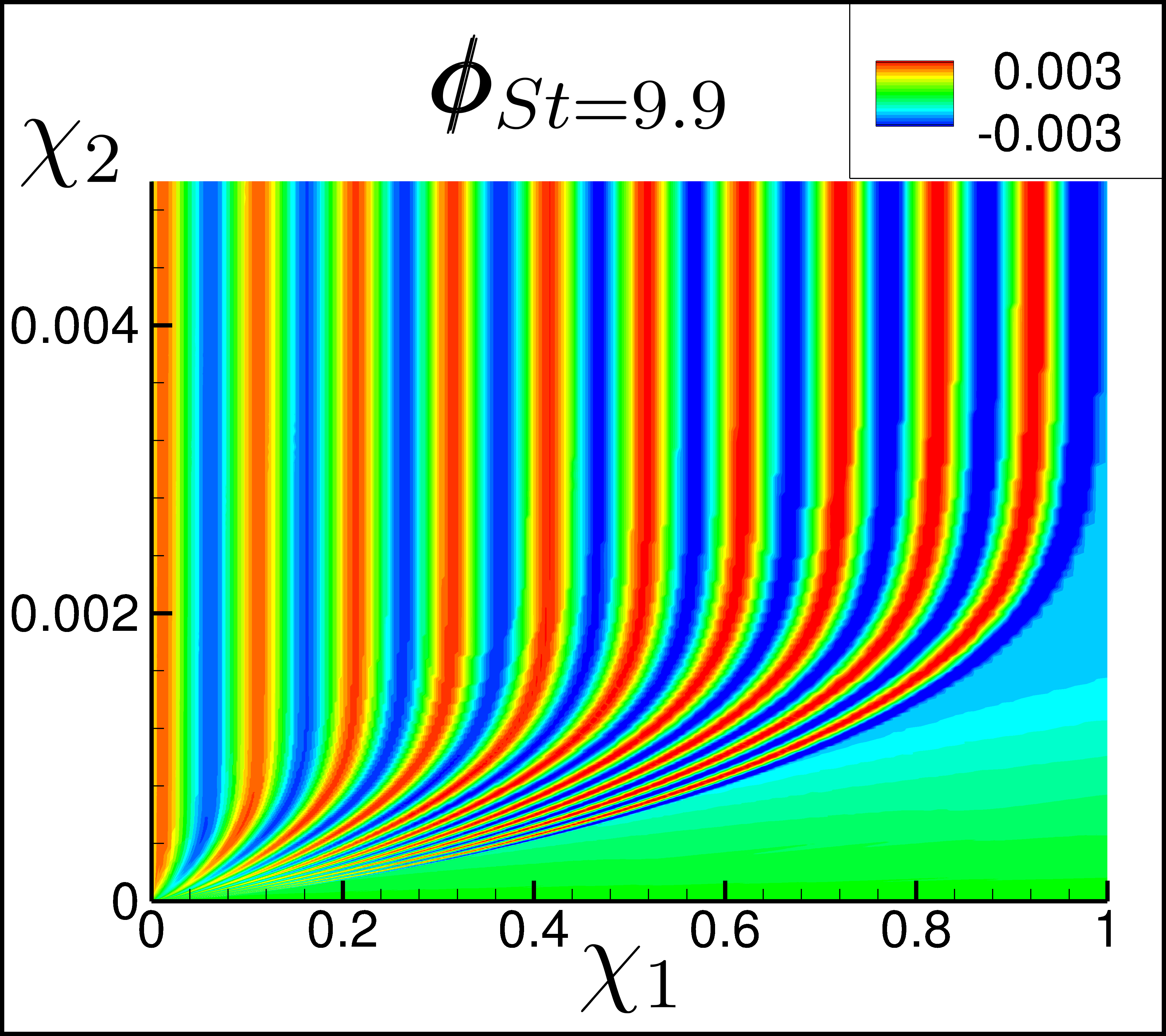}\\(c) $n_2=250$, $\delta t_2=0.004$
    \end{minipage}
    \begin{minipage}{0.24\textwidth}
    \centering
    \includegraphics[width=1.0\linewidth]{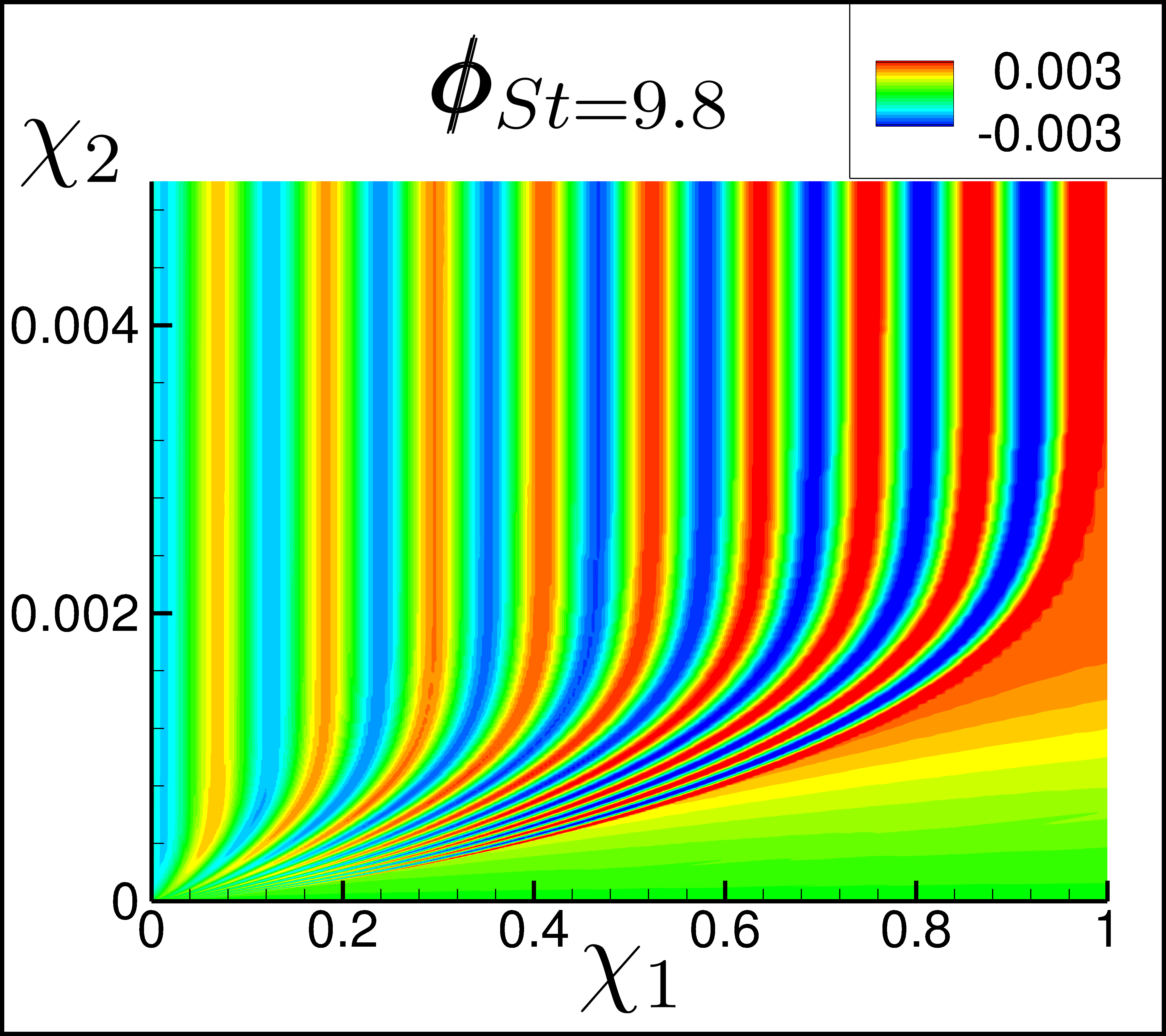}\\(d) $n_1=375$, $\delta t_1=0.00267$
    \end{minipage}
    \caption{Effect of number of snapshots and time-step size on LagSAT modes. (a) Eigen spectra at constant time-duration and LagSAT modes: (b) $\phi_{St=9.9}$ for $\mathsf{T}=0.25$, (c) $\phi_{St=9.6}$ for $\mathsf{T}=0.50$, and (d) $\phi_{St=9.1}$ for $\mathsf{T}=0.75$.} 
    \label{fig:num_const}
\end{figure}

For the fixed time duration ($\mathsf{T}=1.0$), the effect of time-step and number of snapshots on the modal growth rate and n-factor is consistent with the small time duration cases (Table~\ref{tab:N-fac}).
In general, the LagSAT modes and the eigen-spectra tend to convergence for increasing the number of snapshots and/or decreasing timestep size.
The eigen-spectra and representative LagSAT mode (at $St=9.8$) are shown in Fig.~\ref{fig:num_const}, where the LagSAT mode at a lower resolution of $\delta t=0.008$ and $n=125$ produce negative growth rate/n-factor.
In contrast, at higher resolution cases of Fig.~\ref{fig:num_const}(c) (with $\delta t=0.004$ and $n=250$) and Fig.~\ref{fig:num_const} ( with $\delta t=0.00267$ and $n=375$) converge to a growth rate $\omega_r\approx 1.0$ and n-factor value of $n \approx 1.0$ for this mode ($\phi_{St=9.8}$).

\bibliographystyle{jfm}
\bibliography{jfm-instructions}

\end{document}